\documentclass[onecolumn,showpacs,preprintnumbers,amsmath,amssymb,superscriptaddress,10pt,aps,prl,longbibliography]{revtex4-2}

\usepackage{graphicx}
\usepackage{amssymb,amsmath,amsfonts,bm,hyperref}
\usepackage{color}
\usepackage{ulem}
\usepackage{bbm}
\usepackage{graphicx}
\usepackage{animate}
\usepackage{dcolumn}
\usepackage{mathtools}
\usepackage{pifont}

\def\I{\uppercase\expandafter{\romannumeral 1}}
\def\II{\uppercase\expandafter{\romannumeral 2}}
\def\III{{\uppercase\expandafter{\romannumeral 3}}}
\def\IV{{\uppercase\expandafter{\romannumeral 4}}}
\def\V{{\uppercase\expandafter{\romannumeral 5}}}
\def\VI{{\uppercase\expandafter{\romannumeral 6}}}
\def\VII{{\uppercase\expandafter{\romannumeral 7}}}
\def\i{\lowercase\expandafter{\romannumeral 1}}
\def\ii{\lowercase\expandafter{\romannumeral 2}}
\def\iii{{\lowercase\expandafter{\romannumeral 3}}}
\def\iv{{\lowercase\expandafter{\romannumeral 4}}}
\def\v{{\lowercase\expandafter{\romannumeral 5}}}
\def\vi{{\lowercase\expandafter{\romannumeral 6}}}
\def\vii{{\lowercase\expandafter{\romannumeral 7}}}
\def\nn{\nonumber\\}

\def\k{\mathbf{k}}
\def\G{\mathbf{G}}
\def\Q{\mathbf{Q}}
\def\kt{\widetilde{\mathbf{k}}}
\def\q{\mathbf{q}}
\def\g{\mathbf{g}}
\def\qt{\widetilde{\mathbf{q}}}
\def\nn{\nonumber \\}
\def\hc{\hat{c}}
\def\hcd{\hat{c}^{\dagger}}
\def\hd{\hat{d}}
\def\hdd{\hat{d}^{\dagger}}

\newcommand{\ket}[1]{\vert #1 \rangle}
\newcommand{\bra}[1]{\langle #1 \vert}
\newcommand{\bracket}[2]{\langle #1 \vert #2 \rangle}

\newcommand{\bk}{\bm{k}}
\newcommand{\bq}{\bm{q}}
\newcommand{\btk}{\widetilde{\bm{k}}}
\newcommand{\btq}{\widetilde{\bm{q}}}
\newcommand{\br}{\bm{r}}
\newcommand{\cop}{\hat{c}}
\newcommand{\dop}{\hat{d}}

\begin{document}

\title{Synergistic correlated states and nontrivial topology in  coupled graphene-insulator heterostructures}
\author{Xin Lu}
\affiliation{School of Physical Science and Technology, ShanghaiTech University, Shanghai 201210, China}
\author{Shihao Zhang}
\affiliation{School of Physical Science and Technology, ShanghaiTech University, Shanghai 201210, China}
\author{Yaning Wang}
\affiliation{Shenyang National Laboratory for Materials Science, Institute of Metal Research, Chinese Academy of Sciences, Shenyang, China}
\author{Xiang Gao}
\affiliation{State Key Laboratory of Quantum Optics and Quantum Optics Devices, Institute of Opto-Electronics, Shanxi University,
030006 Taiyuan, China}
\affiliation{Collaborative Innovation Center of Extreme Optics, Shanxi University, 030006 Taiyuan, China}
\author{Kaining Yang}
\affiliation{State Key Laboratory of Quantum Optics and Quantum Optics Devices, Institute of Opto-Electronics, Shanxi University,
030006 Taiyuan, China}
\affiliation{Collaborative Innovation Center of Extreme Optics, Shanxi University, 030006 Taiyuan, China}
\author{Zhongqing Guo}
\affiliation{School of Physical Science and Technology, ShanghaiTech University, Shanghai 201210, China}
\author{Yuchen Gao}
\affiliation{Collaborative Innovation Center of Quantum Matter, Beijing 100871, China}
\affiliation{State Key Lab for Mesoscopic Physics and Frontiers Science Center for Nano-Optoelectronics, School of Physics, Peking University, Beijing 100871, China}

\author{Yu Ye}
\affiliation{Collaborative Innovation Center of Quantum Matter, Beijing 100871, China}
\affiliation{State Key Lab for Mesoscopic Physics and Frontiers Science Center for Nano-Optoelectronics, School of Physics, Peking University, Beijing 100871, China}

\author{Zheng Vitto Han}
\affiliation{State Key Laboratory of Quantum Optics and Quantum Optics Devices, Institute of Opto-Electronics, Shanxi University,
030006 Taiyuan, China}
\affiliation{Collaborative Innovation Center of Extreme Optics, Shanxi University, 030006 Taiyuan, China}

\author{Jianpeng Liu}
\email{liujp@shanghaitech.edu.cn}
\affiliation{School of Physical Science and Technology, ShanghaiTech University, Shanghai 201210, China}
\affiliation{ShanghaiTech Laboratory for Topological Physics, ShanghaiTech University, Shanghai 201210, China}

\bibliographystyle{apsrev4-2}

\begin{abstract} 
In this work, we study the synergistic correlated  states in two distinct types of interacting electronic systems coupled by interlayer Coulomb interactions. We propose that this scenario can be realized in a  type of Coulomb-coupled graphene-insulator heterostructures with gate tunable band alignment. We find that, by virtue of the interlayer Coulomb coupling between the interacting electrons in the two layers, electronic states that cannot be revealed in either individual layer  would emerge in a cooperative and synergistic manner. 
Specifically, as a result of the band alignment, charge carriers can be transferred between graphene and  the substrate under the control of gate voltages, which can yield a long-wavelength electronic crystal at the surface of the substrate. This electronic crystal exerts a superlattice Coulomb potential on the Dirac electrons in graphene, which generates subbands with reduced non-interacting Fermi velocity. As a result, $e$-$e$ Coulomb interactions within graphene would play a more important role, giving rise to a gapped Dirac state at the charge neutrality point, accompanied by interaction-enhanced Fermi velocity.
Moreover, the superlattice potential can  give rise to topologically nontrivial subband structures which are tunable by superlattice's constant and anisotropy. Reciprocally, the electronic crystal formed in the substrate can be substantially stabilized in such coupled bilayer heterostructure  by virtue of  the cooperative interlayer Coulomb coupling.
We further perform high-throughput first principles calculations to identify a number of promising insulating materials as candidate substrates for graphene to demonstrate these effects. 

\end{abstract} 

\maketitle

\section{Introduction}
Graphene hosts two-dimensional (2D) massless Dirac electrons with linear dispersions and nontrivial Berry phases around two inequivalent $K$ and $K'$ valleys in the Brillouin zone (BZ) \cite{novoselov_science2004,castroneto_rmp2009}. 
Such linear dispersions and topological properties of Dirac cones bestow various intriguing single-particle physical properties to graphene including the relativistic Landau levels, the Klein tunneling effects, and the nontrivial edge states, etc. \cite{castroneto_rmp2009}. Besides, low-energy Dirac fermions in graphene also exhibit distinct $e$-$e$ interaction effects \cite{kotov_rmp2012}, such as the interaction-enhanced Fermi velocity \cite{nair_science2008,guinea-np10}, the gap opening at the charge neutrality point \cite{jung_prb2011,Tang_science2018,trushin_prl2011}, and even chiral superconductivity when the Fermi level locates at the van Hove singularity \cite{chubukov-chiral-graphene-np12}.

Insulating transition metal oxides (TMOs) and transition metal chalcogenides (TMCs) have also stimulated significant research interests over the past few decades due to the diverse correlated phenomena discovered in these systems such as Mott insulator \cite{Imada-mott-RMP-1998}, excitonic insulator \cite{wu-wte2-np22,cobden-wte2-np22}, and various complex symmetry-breaking states \cite{tokura-science00,balents-review14}.  Under charge dopings, these insulating TMOs and/or TMCs may show 
more intriguing correlated states including unconventional superconductivity \cite{lnw-rmp06,Sigrist-uncSC-RMP-1991,wte2-supercond-science18} and long-wavelength charge density wave \cite{Frano-CDWmanganate-NatMat-2016}.

An open question is what would happen if two types of distinct interacting many-electron systems, i.e., the interacting Dirac fermions in graphene and the correlated electrons in (slightly) charge doped TMO and/or TMC insulators, are integrated into a single platform. Especially, how the mutual couplings would affect the interacting electronic states in both systems. Inspired by recent pioneering experiments in CrOCl-graphene \cite{wang_arxiv2021}, 1T-TaS$_2$-graphene \cite{Altvater-GrTaS2-arxiv-2022}, and CrI$_3$-graphene \cite{Tseng-CrI3-arxiv22} heterostructures, here we propose that such a scenario (of interacting Dirac fermions coupled with the correlated electrons in charge doped TMO/TMC insulators) can be realized in  graphene-insulator heterostructures with gate tunable band alignment. In this work, we show that, by virtue of the interlayer Coulomb coupling between the interacting electrons in the two layers, intriguing correlated physics that cannot be seen in either individual layer would emerge in a cooperative and synergistic manner in such band-aligned graphene-insulator heterostructures.

\begin{figure*}[htb]
    \centering
    \includegraphics[width=0.75\textwidth]{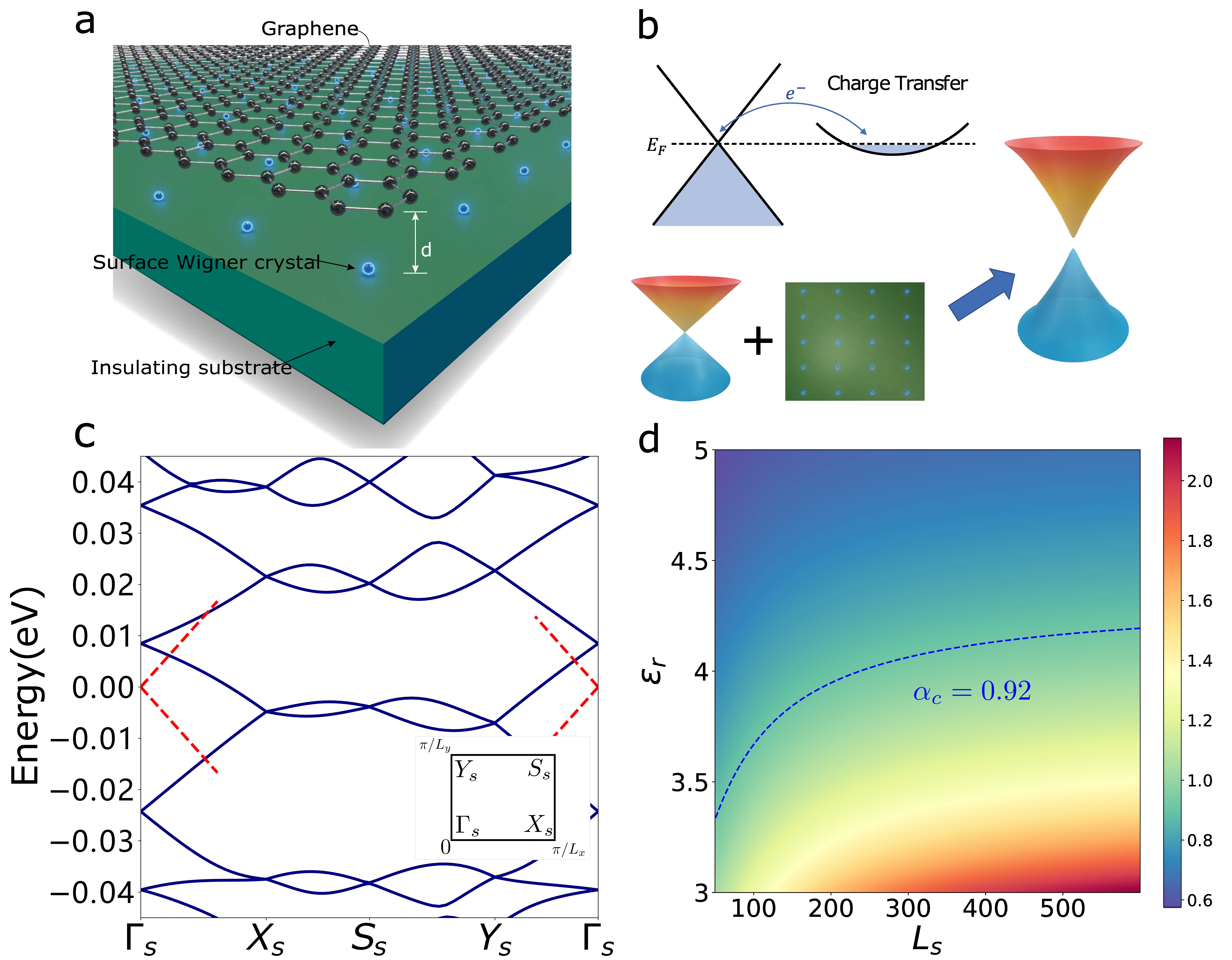}
    \caption{(a) Cartoon illustration of a monolayer graphene supported by an insulating substrate with long-wavelength charge order (blue dots), with an interlayer distance $d$.  (b) Schematic of charge transfer in a band-aligned graphene-insulator heterostructure and its effects on the Dirac dispersion. (c) shows the non-interacting band structure by blue solid lines with $r=1.2$ and $L_s = 600\,$\AA. The red dashed lines represent the non-interacting Dirac cones in free-standing graphene. The
inset marks the high-symmetry points in the
superlattice Brillouin zone.  (d) shows the calculated effective fine structure constant $\alpha(L_s,\epsilon_r)$, where the dashed line marks the critical value $\alpha_c\approx 0.92$.}
    \label{fig:1}
\end{figure*}

When Dirac points of graphene are energetically close to the band edge of the insulating substrate, charge carriers can be transferred between graphene and the substrate under the control of gate voltages due to quantum tunnelling effects. This may yield a long-wavelength electronic crystal (EC) at the surface of the substrate, given that the carrier density introduced to the substrate is below a threshold value, as schematically shown in Fig.~\ref{fig:1}(a,b). On the one hand, the long-wavelength EC at the surface of the substrate would impose an interlayer superlattice Coulomb potential to graphene, which would generate subbands with reduced non-interacting Fermi velocity of the Dirac cone, thus triggers gap opening at the Dirac points by $e$-$e$ interactions in graphene. Meanwhile, concomitant with the gap opening, the Fermi velocities around the charge neutrality point (CNP) are dramatically enhanced due to $e$-$e$ interactions effects. The subbands may also possess nontrivial topological properties with nonzero valley Chern numbers that can be controlled by superlattice constant and anisotropy. Especially, we find a number of ``magic lines" in the parameter space of superlattice's constant and anisotropy, at which the Fermi velocity along one direction vanishes exactly. The subbands would acquire non-zero Chern numbers when passing through these magic lines.
On the other hand, the gapped Dirac state at the CNP of graphene would further stabilize the long wavelength electronic-crystal state in the substrate by pinning the relative charge centers of the two layers in an anti-phase interlocked pattern, in order to optimize the interlayer Coulomb interactions.

\section{Coulomb interactions in graphene}
\label{sec:HF_gr_sl}
To describe the graphene-insulator heterostructure, we consider a model Hamiltonian consisted of a graphene part, an insulator substrate part, and the coupling between them (see Eqs.~\eqref{eq:full-ham} and Sec.~S6 of Supplementary Material \cite{supp}). As we are interested in the low-energy electronic properties, graphene's band structure is modelled by the low-energy Dirac cones around the $K$ and $K'$ valleys. The long-wavelength EC (charge ordered) state in the substrate is considered as a charge insulator, with the electrons being frozen to form a superlattice \cite{supp}. 
Thus, long-wavelength charge order of the substrate is coupled to the graphene layer via interlayer Coulomb interactions to exert a superlattice potential on the Dirac electrons. 
Neglecting the intervalley coupling thanks to the large superlattice constant $L_s$ ($\gtrapprox 50\,$\AA) \cite{comment_density}, we can construct an effective single-particle Hamiltonian for the continuum Dirac fermions in graphene that are coupled with a superlattice Coulomb potential \cite{supp}
\begin{equation}
    H^\mu_0(\mathbf{r}) = \hbar v_F \mathbf{k} \cdot \bm{\sigma}^{\mu} + U_d(\mathbf{r})
    \label{eq:ham}
\end{equation}
where $\bm{\sigma}^{\mu}$ are the Pauli matrices $(\mu \sigma_x, \sigma_y)$ with the valley index $\mu=\pm 1$, $v_F$ is the non-interacting Fermi velocity of graphene, and $U_d(\mathbf{r})$ is the background superlattice potential with the period $U_d(\mathbf{r})=U_d(\mathbf{r}+\mathbf{L_s})$. The superlattice of the EC is set to be rectangular, with anisotropy $r=L_y/L_x$ and $L_{x,y}$ being the superlattice constant in the $x,y$-direction, respectively. We denote $L_s=L_x$.
As a result, the superlattice potential $U_d(\mathbf{r})$ would fold Dirac cones into its small Brillouin zone (BZ), forming subbands and opening up a gap at the boundary of the supercell BZ, as shown in Fig.~\ref{fig:1}(c) for a rectangular superlattice with $r=1.2$ (same as that of CrOCl atomic lattice) in valley $K$ ($\mu=1$) with $L_s=600\,$\AA. 
The energy degeneracies from folding are all lifted by $U_d$, whose Fourier component reads \cite{supp}
\begin{equation}
    U_d(\mathbf{Q}) = \frac{e^2}{\epsilon_0 \epsilon_r \Omega_0} \, \frac{e^{-\vert\mathbf{Q}\vert d}}{\vert\mathbf{Q}\vert}\;,
    \label{eq:Ud}
\end{equation}
where $\mathbf{Q}\neq\mathbf{0}$ is the reciprocal lattice vector associated with $\mathbf{L_s}$, $\Omega_0\!=\!L_x L_y$ is the area of the primitive cell of the superlattice. The Coulomb potential $U_d$, screened by a dielectric constant $\epsilon_r$, decays exponentially in the reciprocal space $\sim \exp (-Q d)$, where $d$ is the distance between the substrate surface and graphene monolayer. 
Furthermore, the Fermi velocities near the Dirac points of the subbands are suppressed by $U_d$ \cite{park_natphys2008} as clearly shown in Fig.~\ref{fig:1}(c). Such a continuum-model description is adopted throughout the paper given that  $L_s \gg a$ ($a=2.46\,$\AA\ is graphene's lattice constant) is always fulfilled for low carrier density $\lessapprox 10^{13}\,$cm$^{-2}$, with $L_s\sim1/\sqrt{n}$ for the EC state.

While it is highly desirable to open a gap at the Dirac points in graphene for the purpose of field-effect device fabrication, the superlattice potential of Eq.~(\ref{eq:Ud}) alone cannot gap out Dirac points in graphene as the system still preserves $C_{2z}\mathcal{T}$ symmetry. However, the Dirac points can be unstable against $e$-$e$ Coulomb interactions (with the spontaneous breaking of $C_{2z}\mathcal{T}$ symmetry) once the Fermi velocity of the non-interacting band structure is suppressed below a threshold, which can be assisted by the superlattice potential from the long-wavelength charge order. One of the similar illustrations is twisted bilayer graphene (TBG) \cite{bistritzer_pnas2011}, where the Fermi velocity is strongly suppressed around the ``magic angle", leading to moir\'e flat bands exhibiting diverse correlated and topological phases \cite{cao-nature18-mott,cao-nature18-supercond,balents-review-tbg,andrei-review-tbg,Liu-OMmoire-NatRevPhys-2021,bernevig-nrp22}. 
Here we further calculate the Fermi velocity of the superlattice subbands around the Dirac point, denoted as $v_F(L_s,\epsilon_r)$, which depends on both the superlattice constant $L_s$ and the background dielectric constant $\epsilon_r$. Accordingly, the effective fine structure constant $\alpha(L_s,\epsilon_r)=e^2/(4\pi\epsilon_0\epsilon_r\hbar v_F(L_s,\epsilon_r))$ can also be tuned by $L_s$ and $\epsilon_r$, as shown in Fig.~\ref{fig:1}(d). We see that there is a substantial region in the $(L_s, \epsilon_r)$ phase space with $\alpha(L_s,\epsilon_r)>\alpha_c\approx 0.92$ \cite{alpha-qmc-prb10}, which indicates that the Dirac-semimetal phase of graphene may no longer be stable against $e$-$e$ interactions within this regime according to previous theoretical study \cite{alpha-qmc-prb10}. 

Such a picture is not unique to rectangular superlattice, but applies to various superlattice geometries. Treating the superlattice potential $U_d(\mathbf{Q})$ using second-order perturbation theory, the renormalized non-interacting effective Hamiltonian for arbitrary superlattice geometry can be expressed as
\begin{equation}
H^{0}_{\rm{eff}}(\k)=\hbar v_F\,\left( 1-\sum_{\vert\mathbf{Q}\vert\neq 0}\frac{\vert U_d(\mathbf{Q})\vert^2}{(\hbar v_F)^2\vert\mathbf{Q}\vert^2} \right)\,\left(\,\k-\sum_{\vert\mathbf{Q}\vert\neq 0}\frac{\vert U_d(\mathbf{Q})\vert^2}{(\hbar v_F)^2\vert\mathbf{Q}\vert^2}\,\left(\k-\frac{2\k\cdot\mathbf{Q}}{\vert\mathbf{Q}\vert^2}\,\mathbf{Q}\right)\right)\cdot\bm{\sigma}\;.
\end{equation}
We see that the effective non-interacting Hamiltonian as well as the Fermi velocity have similar dependence on $L_s$ and $\epsilon_r$ (through $U_d(\mathbf{Q})$) for all lattice geometries.  We have also calculated the effective fine-structure constants  $\alpha(L_s,\epsilon_r)=e^2/(4\pi\epsilon_0\epsilon_r\hbar v_F(L_s,\epsilon_r))$ for both triangular and square lattices \cite{supp}, and the results are very similar to that of rectangular lattice with $r=1.2$ shown in Fig.~\ref{fig:1}(d).

This motivates us to include $e$-$e$ interactions in the graphene layer in our model. 
Despite several theoretical predictions of gapped Dirac states in graphene \cite{kotov_rmp2012,drut-prl09,jung_prb2011,Tang_science2018,trushin_prl2011}, to the best of our knowledge no gap at the CNP has been experimentally observed in suspended graphene yet \cite{elias_natphys2011,faugeras_prl2015}. This can be attributed to interaction-enhanced Fermi velocity around the CNP, screening of $e$-$e$ interactions due to ripple-induced charge puddles, disorder effects, etc. \cite{dassarm_prb2007rc,borghi_sscom2009,novoselov-ripple-2011,stauber_prl2017,kotov_rmp2012}. Nevertheless, analogous to TBG, the subbands in our system with reduced non-interacting Fermi velocity would quench the kinetic energy and further promote the $e$-$e$ interaction effects in graphene. 

Our unrestricted Hartree-Fock calculations \cite{supp} confirm precisely the argument above. 
As interaction effects are most prominent around the CNP, we project the Coulomb interactions onto only a low-energy subspace including three valence and three conduction subbands ($n_\text{cut}\!=\!3$) that are closest to CNP for each valley and spin. 
To incorporate the influences of Coulomb interactions from the high-energy remote bands, the renormalized Fermi velocity within the low-energy subspace can be derived from the renormalization group (RG) approach \cite{gonzalez_nuclphysb1993,kotov_rmp2012,castroneto_rmp2009,nair_science2008} 
\begin{equation}
    v_F^* = v_F \left(1+\frac{\alpha_0}{4\epsilon_r }\log \frac{E_c}{E_c^{*}} \right)\;,
    \label{eq:RG_vF}
\end{equation}
where $\alpha_0\!=\!e^2/(4\pi\epsilon_0 \hbar v_F)$ is the ratio between the Coulomb interaction energy and kinetic energy, i.e., the effective fine-structure constant of free-standing graphene, 
$E_c^{*}$ delimits the low-energy window within which the unrestricted Hartree-Fock calculations are to be performed, and $E_c$ is an overall energy cut-off above which the Dirac-fermion description to graphene is no longer valid. 
Unlike TBG \cite{vafek_prl2020}, other parameters of the effective Hamiltonian (Eq.~\eqref{eq:ham}) such as $U_d$, are unchanged under the RG flow \cite{supp}.

\begin{figure*}[!htbh]
	\centering
	\includegraphics[width=0.85\textwidth]{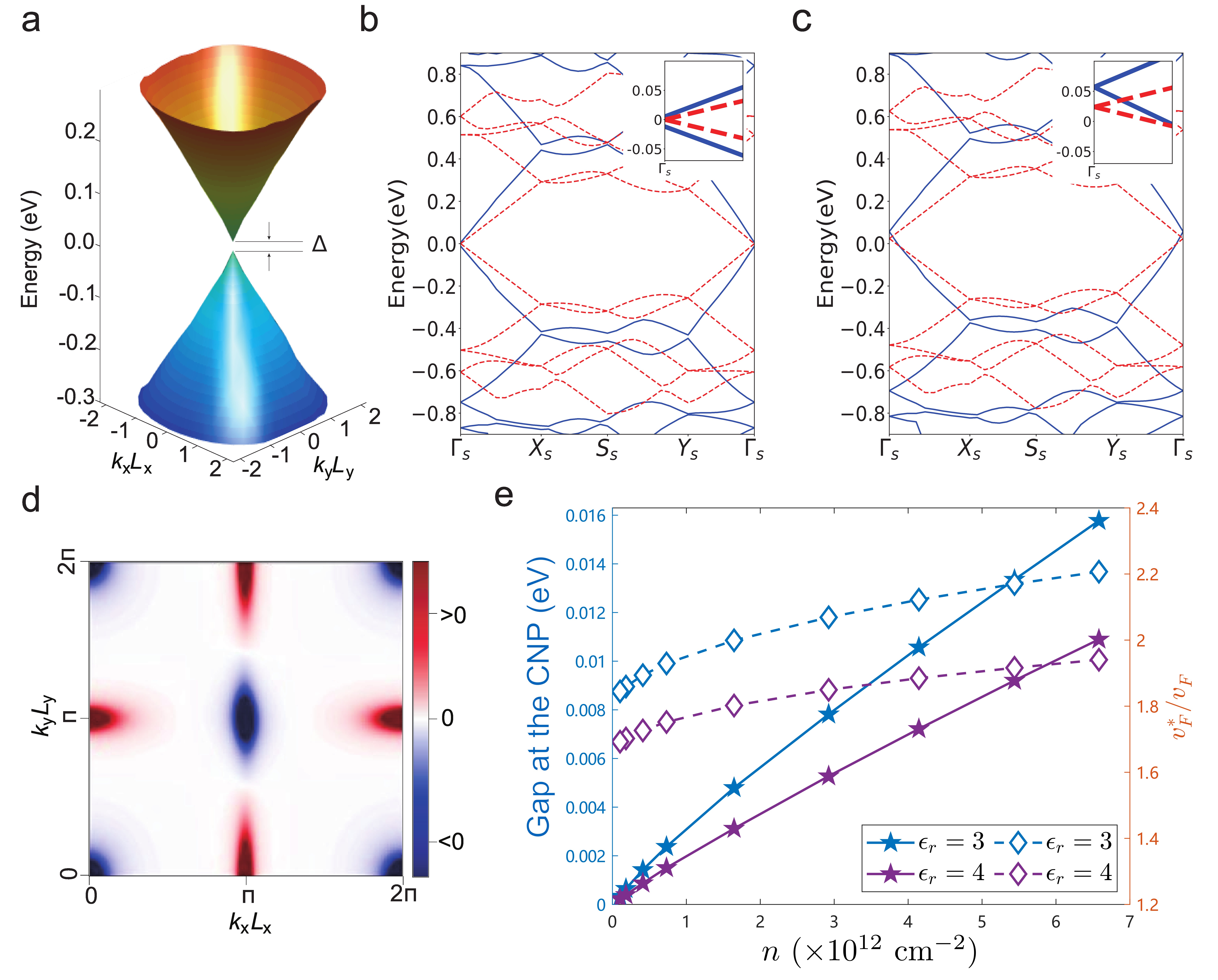}
	\caption{(a) Calculated Hartree-Fock single-particle excitation spectrum of graphene coupled to a long-wavelength Coulomb potential, with $\nu=0$. (b) and (c) show by blue solid lines the Hartree-Fock band structures of $L_s=50\,$\AA\ and $\epsilon_r\!=\! 3.0$, with the filling factor $\nu=0$ in (b) and $\nu=-0.003$ in (c). The red dashed lines represent the non-interacting Dirac cones. The insets zoom in energy close to the Dirac points. Zero energies in (b) and (c) are defined as the chemical potentials for $\nu=0$ and $\nu=-0.003$, respectively.
	(d) The calculated gaps at CNP (filled stars) and the interaction-enhanced Fermi velocities at slight hole dopings $\nu=-0.003$ (hollow diamonds) as a function of the substrate's carrier density $n$. (e) The thermal activation gap $\Delta$ measured on the devices in \cite{wang_arxiv2021} for different nominal dopings $n_\text{tot}$. (f) Distribution of Berry curvature of the
highest valence subband of $K$ valley for $r = 1.2$ and $L_s = 50$\,\AA, which gives zero valley Chern number. } 
	\label{fig:2}
\end{figure*}

We first study the interaction effects of graphene coupled to a rectangular superlattice potential with $r=1.2$ and 50\,\AA\,$\leq L_s\leq 400\,$\AA, corresponding to carrier density of the EC state at the surface of the substrate $0.1\!\times\!10^{12}\,\textrm{cm}^{-2}\!\leq\!n\!\leq\!6.58\!\times\!10^{12}\,\textrm{cm}^{-2}$ (with $n=2/(rL_s^2)$),
with $\epsilon_r=3, 4$, and $d=7\,$\AA\ (obtained from first principles density functional theory calculations for one particular commensurate CrOCl-graphene supercell \cite{supp}). Here, we consider two different filling factors:
exactly at the CNP ($\nu\!=\!0$) and a slight hole doping ($\nu\!\approx\!-0.003$). 
When $\nu\!=\!0$, a gap can be opened up due to interaction effects [see Fig.~\ref{fig:2}(a,b)], leading to two nearly degenerate insulating states, one is $\sigma_z$-sublattice polarized and the other is characterized by the order parameter $\tau_z\sigma_z$, where $\tau_z$ and $\sigma_z$ denote the third Pauli matrix in valley and sublattice space, respectively. Then, intervalley Coulomb interactions would split such degeneracy, and the sublattice polarized insulator with zero Chern number becomes the unique ground state \cite{supp}.
Notably, the gap decreases almost linearly with $n$ as  shown in Fig.~\ref{fig:2}(d), and eventually vanishes as $n\to 0$. This is because the superlattice Coulomb potential exerted on graphene is proportional to the carrier density of the long-wavelength order from the substrate. 
Consequently, the Fermi velocity of the bare Dirac dispersion of graphene would be less suppressed at smaller carrier density $n$, which disfavors gap opening. Eventually in the limit of $n\to 0$, with a charge ordered state of infinite lattice constant, graphene would recover its non-interacting behavior as a gapless Dirac semimetal.

\begin{figure*}[!htbh]
	\centering
	\includegraphics[width=0.9\textwidth]{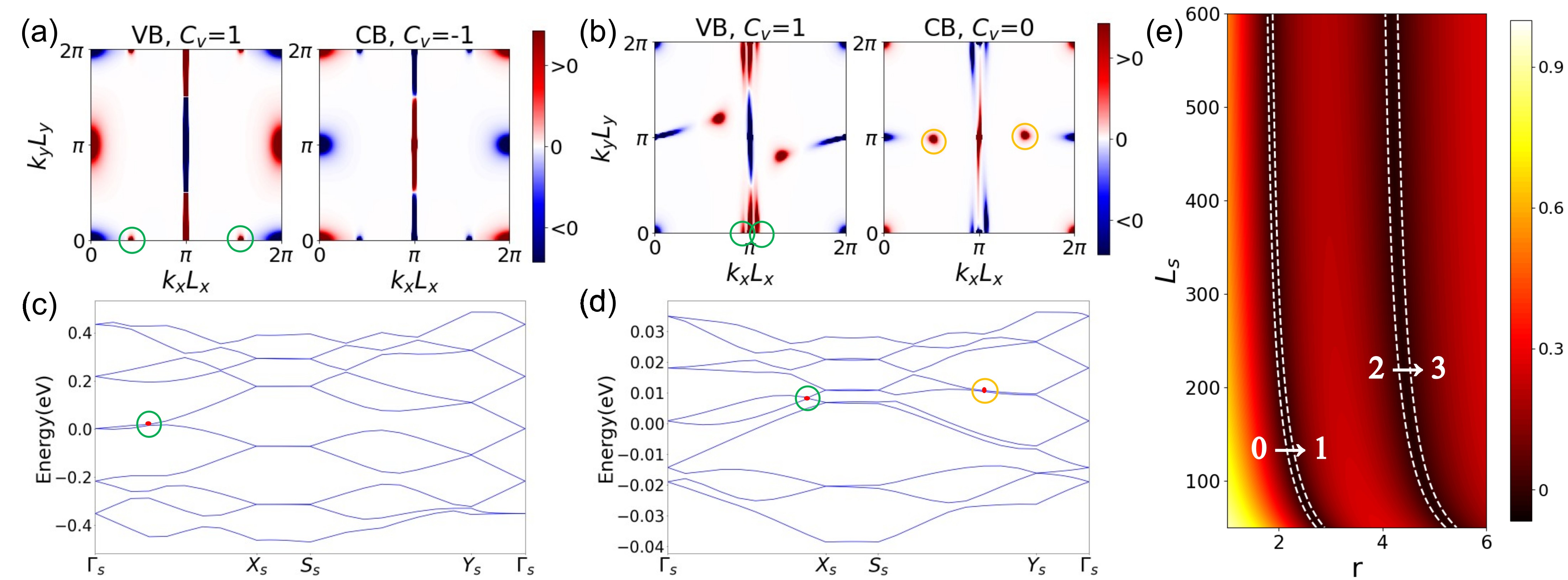}
	\caption{(a) and (b) shows the distribution of Berry curvature in the $r=3$ superlattice's BZ of the lowest valence and conduction band in valley $K$ for $L_s=50$ and $600\,$\AA, respectively. Their corresponding valley Chern number are also given on the top of each panel. (c) and (d) are the non-interacting band structure of the $r=3$ superlattice with $L_s=50$ and $600\,$\AA. (e) Colormap of Fermi velocity in the $x$-direction $v_x$ of the valence band for $\epsilon_r = 3$. The color coding indicates $v_x/v_F$. Here we vary $L_x$ from 50 to 600\,\AA\ and anisotropy parameter $r$ from 1 to 6. The white dashed line, i.e., the ``magic lines", mark the position in parameter space where $v_x$ vanishes.}
	\label{fig3}
\end{figure*}

To verify our theory, we have also experimentally measured the gaps at CNP in graphene-CrOCl heterostructure at different nominal carrier densities \cite{supp} using the same high-quality device reported in Ref.~\cite{wang_arxiv2021}. The details for the measurement set up and the device configuration are presented in Sec.~S8 of Supplementary Material.
The measured gaps also decrease linearly with $n_{\textrm{tot}}$, from 7.7\,meV with $n_{\rm{tot}}=3.4\times\,10^{12}\,\rm{cm}^{-2}$, to 5.8\,meV with $n_{\rm{tot}}=0.5\times 10^{12}\,$cm$^{-2}$ \cite{supp}, consistent with the trend from theoretical calculations, as shown in Fig.~\ref{fig:2}(e). Nevertheless, when $n_{\rm{tot}}\to 0$, such a linear dependence of the gap on $n_{\rm{tot}}$ may no longer be valid \cite{supp}. This is because in Eq.~\eqref{eq:Ud}, the interlayer Coulomb potential only applies to the situation of a single valley to accommodate  charge carriers in the substrate. In reality, there may be additional valley degeneracy in the substrate, which is crucial for the evolution of gap as $n_{\rm{tot}}\to 0$. Although the valley degeneracy of the substrate does not change our results qualitatively, the theoretically calculated gap \textit{vs.} $n_{\rm{tot}}$ fits to the experimental data of CrOCl-graphene heterostructure more precisely at low density once including the two-fold valley degeneracy of CrOCl (see Table~\ref{table:candidate}). The details are given in Supplementary Material (Fig.~S11) \cite{supp}.

We note that the electronic crystal at the surface of the substrate is expected to persist even if the carrier density exceeds the threshold value due to the extra energy gain from interlayer Coulomb coupling in such coupled system, which will be discussed in detail in the section ``Cooperative coupling between graphene and substrate'' below. 
Strain is also inevitable in such graphene-insulator heterostructures, which would give rise to pseudo-magnetic fields coupled to the Dirac electrons \cite{guinea-np10,guinea-gauge-2010,crommie-science10}, thus further enhance the $e$-$e$ interaction effects in graphene.

The single-particle excitation spectrum is also significantly altered by Coulomb interactions within the low-energy window, as shown in Fig.~\ref{fig:2}(b) and (c) with fillings $\nu\!=\!0$ and $\nu\!=\!-0.003$, respectively. 
We note that although the superlattice potential $U_d$ suppresses Fermi velocity in graphene [see Fig.~\ref{fig:1}(c)], $e$-$e$ interactions can compensate such effects. The Fermi velocity is not only enhanced by the Coulomb potentials from the remote energy bands [Eq.~(\ref{eq:RG_vF})], but also further boosted by $e$-$e$ interactions within the low energy window $E_c^{*}\sim n_{\rm{cut}}\hbar v_F 2\pi/L_s$.
Eventually, the Fermi velocity can be magnified up to more than twice of the non-interacting value of free-standing graphene ($v_F$) at slight hole doping $\nu=-0.003$, as shown in Fig.~\ref{fig:2}(d). This perfectly explains the recent experiment in gate-controlled graphene-CrOCl heterostructure, 
in which  the Fermi velocity around CNP is significantly enhanced compared to non-interacting value at slight carrier doping, such that robust quantum Hall effect can be observed under tiny vertical magnetic fields ($\sim 0.1\,$T) and at high temperatures \cite{wang_arxiv2021}. 
We note that the EC state may be stabilized by vertical magnetic fields even when the carrier density in the substrate exceeds the zero-field threshold value \cite{fukuyama-prb79,andrei-prl88}, which in turn boosts the low-field, high-temperature quantum Hall effect in the graphene layer due to the scenario discussed above.

Although it has been theoretically proposed that the magnetic proximity effect together with spin-orbit coupling could in principle give rise to topologically nontrivial states in graphene \cite{qiao-prl14}, it seems to be irrelevant to the graphene-insulator heterostructures considered in the present study. For example, in CrOCl-graphene device, no magnetic hysteresis has been observed in graphene, and the measured Landau level degeneracy is still compatible with that of spin-valley degenerate Dirac cones \cite{wang_arxiv2021}. Most saliently, the gap opening and the robust quantum Hall effect persist up to temperatures far above the N\'eel temperature of CrOCl ($\sim$14\,K) \cite{wang_arxiv2021}. Similarly, the magnetic proximity coupling was also reported to be negligible for CrI$_3$-graphene heterostructure \cite{Tseng-CrI3-arxiv22}. Therefore, compared to the power-law decaying interlayer Coulomb coupling, the exponentially decaying magnetic proximity coupling may not play an important role in such charge-transfer graphene-insulator heterostructures.

The essential results discussed above, i.e., the gap opening at CNP and the concomitant drastic enhancement of Fermi velocity, remain valid for different types of the background superlattices. Specifically, we have also performed calculations for the case of triangular superlattices, which lead to qualitatively the same conclusions, as presented in Sec.~S5 of Supplementary Material \cite{supp}.

\section{Topological properties}
\label{sec:topo}

Different from magic-angle TBG \cite{song-tbg-prl19, yang-tbg-prx19, origin-magic-angle-prl19, po-tbg-prb19, jpliu-prb19}, the low-energy subbands for graphene coupled to a rectangular superlattice potential $U_d(\mathbf{r})$ with small anisotropy ($r \sim 1$) turn out to be topologically trivial with a compensating Berry-curvature distribution, leading to zero Chern number. This remains true even in the gapped Dirac state after including $e$-$e$ interactions, as shown in  Fig.~\ref{fig:2}(f). The trivial band topology is somehow anticipated because the superlattice potential is non-chiral in the sense that it is coupled equally to the two sublattice of graphene, which does not have any pseudo-gauge-field structure such as that in TBG \cite{jpliu-prb19,guinea-prl12}.

Hence, it is unexpected that changing the anisotropy $r$ and the lattice size $L_s$ of the superlattice potential $U_d$ can make the subbands topological. For example, keeping $L_x = 50\,$\AA\  but with $r=3.0$, both the highest valence band and the lowest conduction band acquire nonzero valley Chern numbers $C= \pm 1$ (after adding an infinitesimal $C_{2z}$-breaking staggered sublattice potential). As shown in Fig.~\ref{fig3}(a), besides the four high symmetry points, it appears another two ``hot spots" (annotated by green circles) along the line connecting $\Gamma_s$ and $X_s$. This additional contribution breaks the balance between positive and negative contribution of Berry curvature to Chern number, leading to non-zero valley Chern number. Such contribution stems from another accidental crossing point between the low-energy valence and conduction bands along the $k_x$-direction through changing merely the anisotropy parameter $r$, as shown in Fig.~\ref{fig3}(c) by red dot within green circle. 

While increasing $r$ from unity (with fixed $L_s$), the Fermi velocity in the $x$-direction of the valence band around the Dirac point, $v_x$, is gradually reduced, as shown in Fig.~\ref{fig3}(e). As the same origin of Klein tunneling effects, the spinor structure of graphene's wavefunction forces the Fermi velocity in the $y$-direction to be intact \cite{park_natphys2008}. Further tuning $r$ at some point would totally flatten $v_x$. In Fig.~\ref{fig3}(e), we mark by white dashed lines ``the magic lines'' on which $v_x$ of the valence band closest to Dirac points vanishes exactly. The magic lines always come in pair as an effect of chiral (particle-hole) symmetry breaking induced by the superlattice potential. As particle-hole symmetry is broken in the energy spectrum, when $v_x$ vanishes in the valence band, the counterpart in the conduction band remains finite. The valence subband around the Dirac point has to curve upwards to create an accidental band crossing point, after that $v_x$ of the valence band becomes zero again. Therefore, a band crossing would be germinated at the Dirac point, and then move away along the $k_x$-direction with larger $r$. On the one hand, the band crossing moving away from $\Gamma_s$ is of accidental nature, which is generally avoided unless the lattice parameters are at some fine-tuned values. On the other hand, the Dirac point at $\Gamma_s$ remains stable as protected by $C_{2z}\mathcal{T}$ symmetry. If the Dirac point is gapped, say, by a tiny staggered sublattice potential, the low-energy subbands become topological with nonzero valley Chern numbers.
In particular, with the increase of $r$ at fixed $L_s$, the absolute value of valley Chern number of the valence subband (closest to Dirac points) increases by $1$  whenever one pair of the magic lines are passed through. The positions of these magic lines also depend on the background dielectric constant $\epsilon_r$ since  larger $\epsilon_r$ corresponds to weaker Fermi-velocity renormalization effect, which would shift the magic lines to larger $r$ values. In Supplementary Material, we provide animated figure (Fig.~S4) and videos  demonstrating the evolution of the band structures and Berry curvatures with increasing $r$ at fixed $L_s$.
Such topologically nontrivial subbands with highly anisotropic Fermi velocities may provide an alternative platform to realize  topological quantum matter.

We note that the anisotropic charge ordered superlattices may be realized in two ways. First, one can design a spatially modulated electrostatic potential, which has been realized in monolayer graphene by inserting a patterned dielectric superlattice between the gate and the sample \cite{dean-nn18}. Then, the anisotropy of the superlattice can be artificially tuned by the dielectric patterning in the substrate. Second, for some given carrier density, the Fermi surface of the conduction (or valence) band of the substrate may be (partially) nested, which may lead to a charge density wave (CDW) state with the nesting wavevector. For example, for CrOCl, the Fermi surfaces under different Fermi energies (above the conduction band minimum) are given in Fig.~S17(c) of Supplementary Material. Clearly, under some proper fillings, the Fermi surfaces are nested or partially nested, which may give rise to CDW states with anisotropic superlattices.  We note that topologically nontrivial flat bands have also been proposed to exist in Bernal bilayer graphene coupled with a background superlattice potential \cite{cano-bilayer-arxiv22}.

Furthermore, we find that changing $L_s$ is also able to control the valley Chern number of the subbands.  For example, with $r=3$ and $L_s=600\,$\AA, as shown in Fig.~\ref{fig3}(b), while the highest valence band remains topological with non-zero valley Chern number 1 for valley $K$ with the two aforementioned crossing points (green circles) merely moving to $X_s$, the lowest conduction band turns out to be topologically trivial. This is due to two additional band crossing points (orange circles) close to the $Y_s$-$S_s$ line between the lowest and the second lowest conduction bands, as annotated by red dots in an orange circle in Fig.~\ref{fig3}(d). 

The nontrivial topology must arise from the intrinsic Berry phases of the Dirac cones. Such topologically nontrivial bands are particularly surprising for our system, since the Dirac fermions are subjected to a ``trivial" superlattice potential, which couples identically with two sublattices of graphene. Nevertheless, the nontrivial subband topology is highly tunable by changing the superlattice's size and anisotropy \cite{supp}.

\section{Cooperative coupling between graphene and substrate}
\label{sec:HF_couple}
In the previous calculations, a charge ordered superlattice in the substrate is presumed, which exerts a classical superlattice Coulomb potential to graphene. However, this assumption should be re-examined. Moreover, besides the effects from the substrate to graphene, the feedback effects from graphene to the substrate should be discussed as well. Therefore, in this section, we study the coupled bilayer system as a whole, and treat the electrons in graphene layer and the substrate layer on equal footing. In particular, we model the carriers transferred to the substrate as 2D electron gas with long-range $e$-$e$ Coulomb interactions. Electrons in the substrate and in graphene interact with each other via long-range Coulomb potential, whose Fourier component of wavevector $\q$ reads $e^2\,\exp(-\vert\q\vert\,d)/(2\epsilon_0\epsilon_r\vert\q\vert)$. Thus, the total Hamiltonian for the Coulomb-coupled graphene-insulator heterostructure system includes \cite{supp}:

\begin{widetext}
	\begin{subequations}
		\begin{align}
			H^0_{\text{gr}}&=\sum_{\k,\mu,\alpha,\alpha',\sigma}\,\left(\hbar v_F\k\cdot\bm{\sigma}^{\mu} \right)_{\alpha,\alpha'}\,\hcd_{\sigma\mu\alpha}(\k)\,\hc_{\sigma\mu\alpha'}(\k)\;, \label{eq:full-ham-a}\\\
			H^0_{\text{sub}}&=\sum_{\k,\sigma}\,\left(\frac{\hbar^2\bk^2}{2m^*}+E_{\text{CBM}}\right)\,\hdd_{\sigma}(\k)\,\hd_{\sigma}(\k)\;, \label{eq:full-ham-b} \\
			H^{\text{intra}}_{\text{gr}}&=\frac{1}{2 S}\sum_{\substack{\sigma,\sigma' \\ \mu,\mu'}}\,\sum_{\substack{\alpha,\alpha'\\ \k,\k',\q}}\,V_\text{int}(\q)\,\hcd_{\sigma\mu\alpha}(\k+\q)\,\hcd_{\sigma'\mu'\alpha'}(\k'-\q)\,\hc_{\sigma'\mu'\alpha'}(\k')\,\hc_{\sigma\mu\alpha}(\k)\;, \label{eq:full-ham-c}\\
			H^{\text{intra}}_{\text{sub}}&=\frac{1}{2 S}\sum_{\k,\k',\q}\,\sum_{\sigma,\sigma'}\,V_\text{int}(\q)\,\hdd_{\sigma}(\k+\q)\,\hdd_{\sigma'}(\k'-\q)\,\hd_{\sigma'}(\k')\,\hd_{\sigma}(\k)\; , \label{eq:full-ham-d}\\
			H_{\text{gr-sub}}&=\frac{1}{ S}\sum_{\mu,\alpha,\sigma,\sigma'}\,\sum_{\k,\k',\q}\,\frac{e^2\,e^{-\vert\q\vert\,d}}{2\epsilon_0\epsilon_r\vert\q\vert} \hcd_{\sigma\mu\alpha}(\k)\,\hdd_{\sigma'}(\k')\,\hd_{\sigma'}(\k'-\q)\,\hc_{\sigma\mu\alpha}(\k+\q)\; . \label{eq:full-ham-e}
		\end{align}
		\label{eq:full-ham}
	\end{subequations}
\end{widetext}
On the graphene side, Eq.~\eqref{eq:full-ham-a} is the familiar Dirac Hamiltonian describing the non-interacting low-energy physics of graphene. The $e$-$e$ Coulomb interactions within graphene are  described by Eq.~\eqref{eq:full-ham-c}, where  the dominant intravalley long-range Coulomb interactions are considered and  $V_\text{int}(\q)$ is in the form of double-gate screened Coulomb potential (see Eq.~(\ref{eq:V_doublegate}). Here, $\hc_{\sigma\mu\alpha}(\k)$ and $\hcd_{\sigma\mu\alpha}(\k)$  denote annihilation and creation operators for the low-energy Dirac electrons with wavevector $\k$, valley $\mu$, spin $\sigma$, and sublattice $\alpha$. Note that $S$ refers to the total surface area of the coupled system, and the atomic wavevectors $\k,\k',\q$ are expanded around the Dirac points. On the substrate side, without loss of generality, we suppose that the chemical potential is close to the conduction band minimum (CBM) with its energy $E_{\text{CBM}}$, and the energy dispersion of the low-energy electrons around CBM can be modelled by a parabolic band as for 2D free electron gas with effective mass $m^*$. Other electrons in the deep valence bands are supposed to be integrated into the static dielectric screening constant thanks to a large gap of the substrate. Therefore, the non-interacting Hamiltonian Eq.~\eqref{eq:full-ham-b} for electrons in the substrate can be written in the plane wave basis with creation and annihilation operators $\{\hdd_{\sigma}(\k),\hd_{\sigma}(\k) \}$, where $\k$ is the plane wave wavevector expanded around the CBM, and $\sigma$ denotes spin. The $e$-$e$ Coulomb interactions within substrate [Eq.~(\ref{eq:full-ham-d})] is taken to be the long-range Coulomb interaction with the same double-gate screened form of $V_{\rm{int}}(\q)$. The coupling between graphene and substrate is only via the long-range Coulomb potential, which is captured by Eq.~\eqref{eq:full-ham-e}. The prefactor $e^2\,\exp(-\vert\q\vert\,d)/(2\epsilon_0\epsilon_r\vert\q\vert)$ in front of the field operators in Eq.~\eqref{eq:full-ham-e} is nothing but the 2D Fourier transform of 3D Coulomb potential. Interlayer hoppings can be neglected given that the interlayer distance $d\gtrapprox 5$\,\AA\ in such heterostructures (e.g., $d\approx 7$\,\AA\ in graphene-CrOCl heterostructure from first principles calculations), thus the exponentially decaying interlayer hopping amplitude is much weaker than the power-law-decaying interlayer Coulomb interaction. This is further confirmed by directly calculating the orbital projected band structures of a commensurate supercell of CrOCl-graphene heterostructure based on density functional theory. It turns out that the Dirac cone in such heterostructure supercell stems almost $100\%$ from carbon $p_z$ orbitals of graphene (see Sec.~S7 of Supplementary Material), which clearly indicates the absence of interlayer hybridization (hopping).

We use distinct letters to denote the ladder operators for electrons in graphene ($\hc,\hcd$) and substrate ($\hd,\hdd$). This implies in a notational manner the approximation of distinguishable electrons. In other words, the many-body wavefunction of the coupled bilayer system (denoted as $\vert\Psi\rangle$) can be written a separable fashion, namely a direct product of graphene's and substrate's part, i.e., 
\begin{equation}
	\vert\Psi\rangle=\vert\Psi\rangle_{c}\otimes\vert\Psi\rangle_{d}
	\label{eq:psi-product}
\end{equation}
In a mean-field treatment, the corresponding many-body wavefunction would thus be a direct product of two Slater determinants, $\vert\Psi\rangle_{c}$ and $\vert\Psi\rangle_{d}$ for the graphene layer and the substrate layer, respectively. This is reminiscent of the Born-Oppenheimer approximation for electrons and ions. Technically,  this means that order parameters $\sim\langle \hcd \hd \rangle \, (\langle \hdd \hc \rangle)$ are not allowed in our treatment.
A finite value of  $\langle \hcd \hd \rangle \, (\langle \hdd \hc \rangle)$ suggests the emergence of another phase, an interlayer excitonic condensate  in such coupled bilayer system. However, we note that such interlayer exciton has to be driven by intervalley Coulomb scattering between the $K/K'$ valley of graphene and (presumingly) $\Gamma$ valley of substrate's electrons, with the amplitude $\sim e^2 \exp(-\vert \mathbf{K}\vert d)/(2\epsilon_0\epsilon_r\vert\mathbf{K}\vert)$ being several orders of magnitudes smaller than the intravalley one in our problem. Thus, it is completely legitimate to neglect the interlayer particle-hole exchange in our problem, and the separable wavefunction ansatz Eq.~(\ref{eq:psi-product}) is very well justified.
Then, we solve the full interacting Hamiltonian Eqs.~\eqref{eq:full-ham} under the separable-wavefunction ansatz Eq.~\eqref{eq:psi-product}, and the workflow is presented in Methods section. Nevertheless, the interlayer excitonic insulator state consisted of Dirac electrons (holes) and quadratically dispersive holes (electrons) is possible in valley-matched graphene-insulator heterostructures, such as those consisted of graphene and transition metal dichalcogenides with the band extrema at $\mathbf{K}$ points. We leave this for future study.

\begin{widetext}
\begin{figure}
	\centering
	\includegraphics[width=0.7\textwidth]{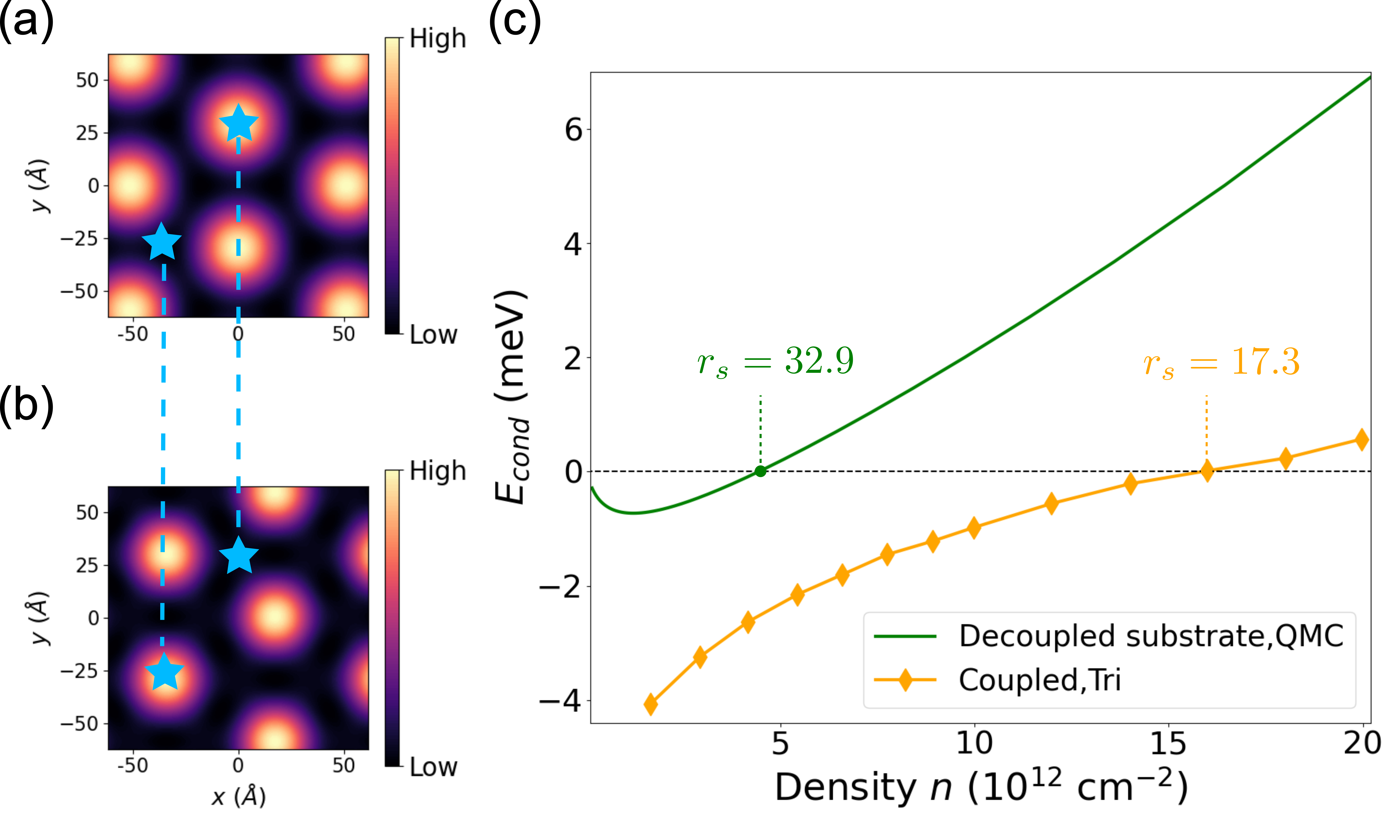}
	\caption{ Charge density modulations after minimizing interlayer Coulomb interactions for (a) gapped Dirac state in graphene, and (b) electronic-crystal state in the substrate, which forms triangular superlattice. (c) Condensation energy (per electron) of the electronic crystal state $E_{\rm{cond}}$ \textit{vs.} the carrier density $n$ in the substrate. The green line represents the condensation energy of the decoupled system using the quantum Monte Carlo data, and the orange lines shows that of the coupled bilayer system, which is significantly stabilized by interlayer Coulomb coupling. The critical $r_s$ for Wigner-crystal transitions are also indicated in the coupled and decoupled cases, respectively. } 
	\label{fig:4}
\end{figure}
\end{widetext}

To explore how the interlayer Coulomb coupling would affect the electronic crystal state of the substrate, we first consider the situation as a reference that the substrate is decoupled from graphene. The energy difference between the spin polarized EC state and Fermi-liquid (FL) state (condensation energy) as a function of the carrier density $n$ is given by quantum Monte Carlo calculations in \cite{drummond_prl2009,rapisarda-dqmc-1996}, as shown by the green line in Fig.~\ref{fig:4}(c), where  an  effective mass $m^{*}=1.3$, a background dielectric constant $\epsilon_r=4$, and valley degeneracy of 2  are considered in order to mimic the conduction band minimum of CrOCl.
  The condensation energy reaches zero when $n\!\approx\!4.5\times 10^{12}\,\rm{cm}^{-2}$ (corresponding to critical Wigner-Seitz radius $r_s^{*}\approx 32.9$),  suggesting the transition from the EC to the FL state. More details are given in Methods section. 

We further include the interlayer Coulomb coupling between the substrate and graphene (setting the chemical potential at the CNP of graphene), which can be treated using perturbation theory given that the interlayer Coulomb energy is always much smaller than the sum of the intralayer Coulomb energy and kinetic energy within the relevant parameter regime (see Fig.~\ref{fig:intra-vs-inter} in Methods). Specifically, with the separable wavefunction ansatz (Eq.~(\ref{eq:psi-product})), the ground-state charge densities for the graphene layer and the EC layer are separately obtained from unrestricted Hartree-Fock calculations, which are further used to estimate the interlayer Coulomb energy.
 More details about the perturbative treatment of interlayer Coulomb interactions are presented in Sec.~S6 of Supplementary Material \cite{supp}.

We find that the condensation energy (per electron) of the EC is substantially enhanced after including the interlayer interactions, as shown by the orange diamonds in Fig.~\ref{fig:4}. As a result, the EC-FL transition is postponed to a much higher density $n\approx 16\times 10^{12}\,\rm{cm}^{-2}$ (corresponding to  critical Wigner-Seitz radius $r_s^*\approx 17.3$). This is because the energy of the coupled bilayer can be further lowered by pinning the charge centers (marked as light blue stars in Fig.~\ref{fig:4}(a,b)) of the two layers in an anti-phase interlocked pattern, in order to optimize the repulsive interlayer Coulomb energy. The extra energy gain from such interlocking of charge centers compensates the energy cost of the EC state when $n\gtrapprox 4.5\times 10^{12}\,\rm{cm}^{-2}$, thus substantially stabilizes the EC state.

On the one hand, since the condensation energy of the free 2D electron gas in the decoupled substrate is estimated using the model that accurately fits to quantum Monte Carlo data \cite{drummond_prl2009}, the estimate of the critical density for the decoupled substrate is expected to be accurate. On the other hand, in the case of substrate coupled with graphene layer, although the interlayer Coulomb energy is estimated with Hartree-Fock approximation, the qualitative conclusion (that the EC state gets stabilized by a cooperative interlayer Coulomb coupling) is expected to be valid even in a beyond-mean-field treatment.
This is because under the separable-wavefunction ansatz, the interlayer Coulomb energy in the EC state is always negative (compared to that of FL state) under an optimal choice of relative charge centers, which thus always stabilizes the EC state even if the intralayer interactions are treated using beyond-mean-field approaches. 

We note that the stabilizing effect of EC is not unique to band-aligned graphene-insulator heterostructures considered in this work. In general, it only requires the presence of another (tunable) gapped state exhibiting non-uniform charge distribution atop of the EC. For example, remarkably robust EC state has been observed in a bilayer system consisting of two monolayer MoSe$_2$ separated by hexagonal boron nitride \cite{mose2-wc-nature21}, which was also argued to be stabilized by the interlocking of the EC states in the two layers.

\section{Materials realization}
\label{sec:materials} 
The scenario discussed above is not only closely related to CrOCl-graphene and CrI$_3$-graphene heterostructures \cite{wang_arxiv2021, Tseng-CrI3-arxiv22}, but can also be extended to various band-aligned graphene-insulator heterostructures. As along as the conduction band minimum (CBM) or valence band maximum (VBM) of the substrate is energetically close to the Dirac points of graphene, charges could be easily transferred between graphene and the substrate's surface by gate voltages. Furthermore, it is more likely to form long-wavelength ordered state at the surface of the substrate (with slight carrier doping) if the material has large effective masses at the CBM or VBM. Meanwhile, an insulator with relatively small dielectric constant would have weaker screening effects to $e$-$e$ interactions,  which also favours long-wavelength ordered state at small carrier doping.

Following these guiding principles, we have performed high-throughput first principles calculations based on density functional theory for various insulating van der Waals materials. Eventually we find twelve suitable candidate materials (including CrOCl and CrI$_3$), whose CBM and VBM energy positions, dielectric constants ($\epsilon_r$), effective masses at the band edges, and the corresponding Wigner-Seitz radii ($r_s$) are listed in Table~\ref{table:candidate}. Clearly, the Wigner-Seitz radii of these materials at the band edges (estimated under slight doping concentration $n\!=\!10^{12}$\,cm$^{-2}$) are all above the threshold of forming a Wigner-crystal state ($r_s\!\gtrapprox\! 31$) \cite{drummond_prl2009}. Additionally, the energy bands of these insulating substrate materials can be easily shifted using vertical displacement fields \cite{supp}, such that charge transfer between graphene and the substrate can be controlled by non-disruptive gate voltages.  
We have also considered heterostructures consisted of graphene and TMDs. Besides trilayer (or thicker) WS$_2$ as already listed in Table~\ref{table:candidate}, we further nominate WSe$_2$ (trilayer or thicker), MoSe$_2$ (bilayer or thicker), and MoTe$_2$ (bilayer or thicker) as possible candidate substrates to realize the effects discussed above.  More details are given in Sec.~S7 of Supplementary Material.

\begin{table*}
	\caption{Candidate substrate materials for the graphene-insulator heterostructure systems. The dielectric constants $\epsilon_r$ \cite{Fritz-dielectric-PhysRevB.93.115151-2016,Fritz-dielectric-Petousis2017,dielectric-choudhary2020joint}, conduction band minimum position ($E_{\rm{CBM}}$), valence band maximum position ($E_{\rm{VBM}}$), the corresponding effective mass $m^*$ at the band edge that is closer to the Dirac point (set to zero) in energy, and the dimensionless Wigner-Seitz radius $r_s=g_v m^*/\sqrt{\pi n}\epsilon_r m_0 a_{\rm{B}}^0$ ($a_{\rm{B}}^0$ is the Bohr radius and $m_0$ is the bare electron mass, $g_v$ is the valley degeneracy) estimated under a small doping concentration $n\!=\!10^{12}$\,cm$^{-2}$, are presented. Here ``bi" and ``tri" stand for bilayer and trilayer systems, respectively.}
	\label{table:candidate}
	\centering
	\begin{tabular}{ccccccc}
		\hline
		Materials & $\epsilon_r$ & $E_{\rm{CBM}}$  & $E_{\rm{VBM}}$ & $m^*/m_0$  & $g_v$  & $r_s$\\
		\hline
		AgScP$_2$S$_6$ (bi) & 3.67 & 0.07\,eV & -1.89\,eV & $3.94$ & 6 & 683.4 \\
		AgScP$_2$Se$_6$ (bi) & 4.06 & 0.15\,eV & -1.37\,eV & $2.63$ & 6 & 412.8 \\
		IrBr$_3$ (bi) & 6.53 & 0.23\,eV & -1.43\,eV & $8.08$ & 2 & 262.7 \\
		IrI$_3$ (bi) & 7.59 &  0.33\,eV & -0.95\,eV & $1.76$ & 2 &  49.1 \\
		YI$_3$ (tri) & 3.45 & 0.53\,eV & -2.1\,eV & $2.12$ & 1  & 65.3 \\
		YBr$_3$ (tri) & 6.78 & 0.68\,eV & -3.15\,eV & $2.76$ & 1 &  43.3 \\
		ReSe$_2$ (bi) & 6.38 & 0.32\,eV & -0.83\,eV & $1.82$ & 2 & 60.7 \\
		ScOCl (bi) & 5.27 & 0.21\,eV & -4.04\,eV & $3.29$ & 1  & 66.2 \\
		PbO (bi) & 8.47 & 2.02\,eV & -0.03\,eV &  $11.89$ & 4  & 595.8 \\
		CrI$_3$ (bi) & 3.00 & -0.32\,eV & -1.58\,eV & $2.02$ & 2 & 142.8 \\
		CrOCl (bi) &  3$\sim$\,4 &  -0.13\,eV & -3.26\,eV &  $1.31$ & 2 &  55.7-74.2  \\
		WS$_2$ (tri,quad) & 3.63 & 0$\sim$\,0.08\,eV &  -1.01$\sim$\,-0.97\,eV  &  $1.16$ & 6 &  201$\sim$203  \\
		WSe$_2$ (tri,quad) & 4.07 & 0.27$\sim$\,0.47\,eV &  -0.65$\sim$\,-0.52\,eV  &  $0.53$ & 6 &  87.4 \\
		MoSe$_2$ (bi, tri, quad) & 7.29 & -0.01$\sim$\,0.31\,eV & -0.97$\sim$\,-0.86\,eV & 0.73$\sim$\,0.77 & 6 & 66$\sim$70 \\
		MoTe$_2$ (bi, tri, quad) & 6.75 & 0.31$\sim$\,0.42\,eV & -0.54$\sim$\,-0.47\,eV & 0.7$\sim$\,0.75 & 6 & 68$\sim$73\\
		\hline
	\end{tabular}
\end{table*}

\section{Conclusions}
\label{sec:summary}
In conclusion, we have studied the synergistic correlated electronic states emerging from coupled graphene-insulator heterostructures with gate tunable band alignment.
Based on comprehensive theoretical studies, we have shown that the gate tunable carrier doping may yield a long-wavelength electronic crystal at the surface of the substrate driven by $e$-$e$ interactions within the substrate, which in turn exerts a superlattice Coulomb potential to the Dirac electrons in graphene layer. This would substantially change the low-energy spectrum of graphene, where a gapped Dirac state concomitant with drastically enhanced Fermi velocity would emerge as $e$-$e$ interaction effects. These theoretical results are quantitatively supported by our transport measurements in graphene-CrOCl heterostructure. Besides, the Dirac subbands in graphene can be endowed with nontrivial topological properties by virtue of the interlayer Coulomb coupling with the long-wavelength electronic crystal underneath. Reciprocally, the electronic crystal in the substrate can be substantially stabilized by virtue of a cooperative interlayer Coulomb coupling with the gapped Dirac state of graphene.
We have further performed high-throughput first principles calculations, and suggested a number of promising insulating materials as candidate substrates for graphene to realize such effects. 

However, the understanding of such coupled bilayer correlated electronic systems is still at a preliminary stage, and the study is far from being complete.
First,  the long-wavelength electronic crystal  cannot be the only possible candidate ground state. Other correlated states such as magnetic or even superconducting states may also occur  in the charge doped insulating substrate, e.g., in the case of high-temperature cuprate superconductor \cite{lnw-rmp06,Sigrist-uncSC-RMP-1991} and monolayer 1T'-WTe$_2$ \cite{wte2-supercond18}. This may give rise to diverse quantum states of matter in  graphene  due to interfacial proximity couplings with Dirac fermions. Moreover, so far we have only considered the ground state properties of such coupled bilayer correlated electronic systems. What is more intriguing is the collective excitations of the electronic crystal and their couplings with Dirac electrons in graphene. 
Around the quantum melting point of the electronic crystal, strong quantum fluctuations would be coupled with Dirac fermions with graphene via interlayer Coulomb interactions, which may give rise to unique quantum critical properties. Therefore, our work may stimulate further exploration of the intriguing physics in such a platform for correlated and topological electrons.

\section{Methods}
\subsection{Hartree-Fock approximations assisted by renormalization group approach}
\label{sec:hf}
When graphene is coupled to a superlattice potential, the Coulomb interactions are suitably expressed in the subband eigenfunction basis,  on which we have performed the Hartree-Fock calculations. Since interaction effects are most prominent around the CNP, we project the Coulomb interactions onto only a low-energy window including three valence and three conduction subbands that are closest to the Dirac point per valley per spin. We use a mesh of $18 \times 18$ $\k$-points to sample the mini Brillouin zone of the superlattice. 

To incorporate the influences of Coulomb interactions from the high-energy remote bands, we rescale the Fermi velocity within the low-energy window of the effective Hamiltonian using Eq.~\eqref{eq:RG_vF}. The other parameters of the non-interacting effective Hamiltonian are unchanged under the RG treatment since their corrections are of higher order, thus can be neglected. In other words, we find the following RG equations for Fermi velocity $v_F$ and leading superlattice potential $U_d$ with respect to energy cutoff $E_c$
\begin{align}
    &\frac{d\, v_F}{ d \log E_c} = - \frac{e^2}{16 \pi \epsilon_0 \epsilon_r}\;,\\
    &\frac{d \, U_d(\bm{Q})}{d \log E_c} = 0\;.
    \label{eq:rg}
\end{align}
The detailed derivations of the RG equations are presented in Sec.~S3 of Supplementary Material.

We also neglect on-site Hubbard interactions and intervalley coupling in $e$-$e$ Coulomb interactions, which turn out to be  one or two order(s) of magnitude weaker than the dominant intravalley long-range Coulomb interactions in such graphene-based superlattice systems \cite{zhang_prl2022}. To model the screening effects to the $e$-$e$ Coulomb interactions from the dielectric environment, we introduce the double gate screening form of $V_{int}$, whose Fourier transform is expressed as
\begin{equation}
	V_{\rm{int}}(\mathbf{q})=\frac{e^2 \tanh(q d_s)}{2 \Omega_0 \epsilon_r \epsilon_0 q}\;,
	\label{eq:V_doublegate}
\end{equation}
where $\Omega_0$ is the area of the  superlattice’s primitive cell, $\epsilon_r$ is a background dielectric constant and the thickness between two gates is $d_s = 400$\,\AA.
Then, we initialize the Hartree-Fock loop with the initial conditions in the form of various different order parameters and obtain the converged ground state self-consistently (see Sec.~S4 of Supplementary Material \cite{supp}).

When we consider electrons in graphene and substrate on equal footing in Eq.~\eqref{eq:full-ham}, the routine of Hartree-Fock calculations is exactly the same. However, we need to first consider solely the substrate side. After performing unrestricted Hartree-Fock calculations, we use the ground-state charge density of EC in the substrate as input for constructing the superlattice potential. Explicitly, we need to replace Eq.~\eqref{eq:Ud} by
\begin{equation}
	U_d(\Q)=\frac{e^2}{2 \epsilon_0\,\epsilon_r\,\Omega_0}\,\frac{e^{-\vert\Q\vert d}\,\rho_d(\Q)}{\vert\Q\vert}\;.
	\label{eq:Ud-Q}
\end{equation} 
where $\rho_d (\Q)$ is the Fourier component of the charge density in the substrate. More details can be found in  Sec.~S6 of Supplementary Material \cite{supp}.

\begin{figure}
	\center
	\includegraphics[width=0.45\textwidth]{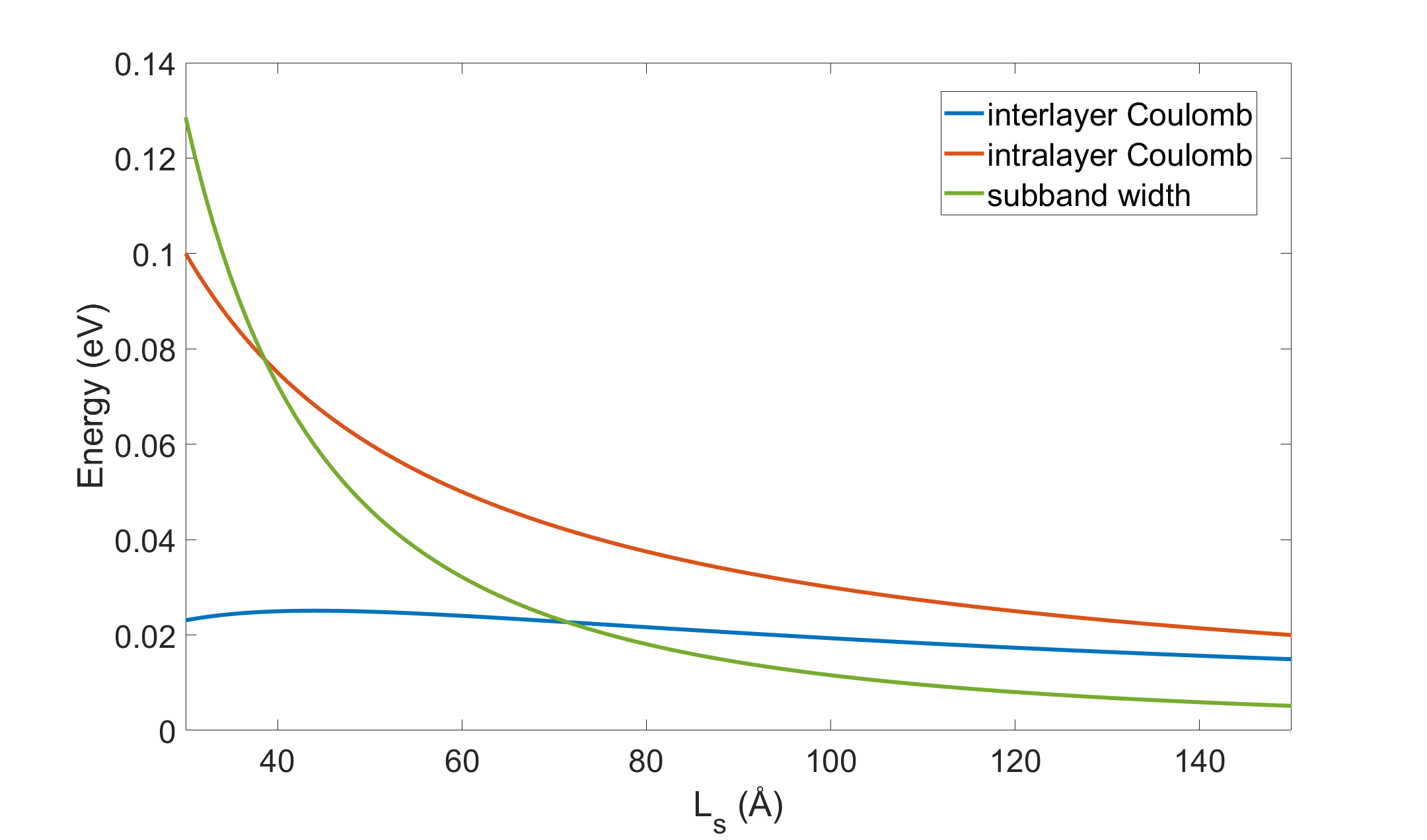}
	\caption{Order of magnitudes of subband width (green) and intra- (red) and inter-layer (blue) Coulomb potential strength for different $L_s$. The dielectric constant is selected to be 4.}
	\label{fig:intra-vs-inter}
\end{figure}

\subsection{Workflow to solve the coupled bilayer Hamiltonian}
\label{sec:workflow}
We solve the Hamiltonian of the coupled bilayer system described by Eqs.~\eqref{eq:full-ham} in the following workflow: 
\begin{itemize}
\item First, we start our calculations by considering solely the substrate Hamiltonian Eqs.~(\ref{eq:full-ham-b}) and (\ref{eq:full-ham-d}). 
We considered the case of triangular superlattice, which is the actual ground state for the EC of  2D electron gas. In particular, the total energy of the triangular EC can described by a fitting model given in Ref.~\cite{drummond_prl2009}:
\begin{equation} 
	E_{\mathrm{WC}}=\frac{c_1}{r_s}+\frac{c_{3/2}}{r_s^{3/2}}+\frac{c_2}{r_s^2}+\frac{c_{5/2}}{r_s^{5/2}}+\frac{c_3}{r_s^3}
\end{equation}
where $c_1=-1.106103$, $c_{3/2}=0.814$, $c_2=0.113743$, $c_{5/2}=-1.184994$ and $c_3=3.097610$.  These parameters are determined by fitting to quantum Monte Carlo data. The total energy for the Fermi-liquid state of 2D electron gas is given by the following model \cite{rapisarda-dqmc-1996}: 
\begin{subequations}
	\begin{align}
E_{\rm{FL}} &= E_{\rm{FL}}^{\rm{HF}} + E_{\rm{FL}}^{c} \\
E_{\rm{FL}}^{\rm{HF}} &= \frac{1}{2 r_s^2}-\frac{4\sqrt{2} }{3\pi r_s} \\
E_{\rm{FL}}^{c} &= a_0\left\{ 1 + A x^2  \left[ B \ln{\frac{x+a_1}{x}}+ C \ln{\frac{\sqrt{x^2+2a_2x+a_3}}{x}} + D \left( \arctan{\frac{x+a_2}{\sqrt{a_3-a_2^2}} }-\frac{\pi}{2}\right) \right]\right\}
	\end{align}
\end{subequations}
where $x=\sqrt{r_s}$ and 
\begin{subequations}
\begin{align}
A&=\frac{2\left(a_1+2a_2\right)}{2a_1a_2-a_3-a_1^2} \\
B&=\frac{1}{a_1}-\frac{1}{a_1+2a_2} \\
C&=\frac{a_1-2a_2}{a_3}+\frac{1}{a_1+2a_2} \\
D&=\frac{F-a_2C}{\sqrt{a_3-a_2^2}} \\
F&=1+\left(2a_2-a_1\right)\left(\frac{1}{a_1+2a_2}-\frac{2a_2}{a_3}\right)
\end{align}
\end{subequations}
with with $a_0=-0.1925$, $a_1=7.3218$, $a_2=0.16008$ and $a_3=3.1698$.  These parameters for the FL state are also determined by fitting to quantum Monte Carlo data \cite{rapisarda-dqmc-1996}.
The energies are given in Hartree atomic units. Then, one can extract the condensation energy for the isolated 2D electron gas in the substrate $E_{\rm{WC}}-E_{\rm{FL}}$, with the accuracy comparable to quantum Monte Carlo calculations.

\item Second, with the help of the separable wavefunction ansatz Eq.~(\ref{eq:psi-product}), we further calculate the ground-state charge density of the EC state in the substrate under Hartree-Fock approximations. Although the Wigner crystal condensation energy would be significantly overestimated with such mean-field approximation, the ground-state charge density can still be properly described by the unrestricted Hartree-Fock treatment \cite{wc-hf-prb03}.
Then, one can integrate out the charge degrees of freedom of the substrate so that the charge density modulation characterized by the Fourier components of the charge density $\{\rho_d (\Q)\}$ ($\Q$ denotes the reciprocal vector of the superlattice) can be used as an input for the superlattice potential $\widetilde{U}_d(\Q)$, as shown in Eq.~\eqref{eq:Ud-Q}.
Compared to Eq.~(\ref{eq:Ud}), this superlattice potential is more realistic and self-contained in our model. Eq.~(\ref{eq:Ud-Q}) would be recovered to Eq.~(\ref{eq:Ud}) by setting $\rho_d(\Q)=2$ for any reciprocal vector $\Q$, which is equivalent to say that two (spin degenerate) charges per primitive supercell are localized in real space in a Dirac-$\delta$-function form.

\item Third, we perform RG-assisted unrestricted HF calculations for the interacting Dirac electrons in graphene as explained in Methods~\ref{sec:HF_gr_sl}. If the chemical potential is at the CNP of graphene,  a gap opening will be triggered by $e$-$e$ interactions within the graphene layer as discussed previously. 

\item From the above procedures, we would separately obtain converged HF ground states, $\vert\Psi\rangle_d$ for the substrate, and $\vert\Psi\rangle_c$ for graphene, respectively. From the ground-state wavefunctions $\vert\Psi\rangle_d$ and $\vert\Psi\rangle_c$, one can extract the corresponding ground-state charge density modulations $\{\rho_d(\Q)\}$ and $\{\rho_c(\Q)\}$, based on which the interlayer Coulomb energy [the expectation value of Eq.~(\ref{eq:full-ham-e})] can be calculated. More details are given in Sec.~S6 of Supplementary Material.

\hspace{7pt} However, the ground states are obtained so far by minimizing (mostly) the intralayer parts of the full Hamiltonian, the interlayer Coulomb interaction Eq.~(\ref{eq:full-ham-e}) is not optimized yet. We note that the intralayer kinetic energy and intralayer Coulomb interaction energy for both graphene and the substrate are unchanged under constant lateral shifts of the charge centers, thus the ground state $\vert\Psi\rangle_d\otimes\vert\Psi\rangle_c$ obtained so far is massively degenerate up to global and relative shifts of the bilayer charge centers. Such degeneracy would be partially lifted by the interlayer Coulomb energy $\langle H_{\rm{gr-sub}} \rangle$. Obviously, $\langle H_{\rm{gr-sub}} \rangle$ is invariant under the global shift of the charge centers of the bilayer system, but it varies with respect to a relative charge-center shift. Therefore, by  virtue of perturbation theory, optimizing the interlayer Coulomb energy amounts to find the optimal relative shift vector between the charge centers of the two layers within the degenerate ground-state manifold obtained in the previous procedures. Such perturbative treatment of $H_{\rm{gr-sub}}$ is justified given that the interlayer Coulomb energy is always weaker than the sum of the kinetic energy and the intralayer Coulomb energy within relevant parameter regime, as shown in Fig.~\ref{fig:intra-vs-inter}. For example, the interlayer Coulomb energy $\sim\!20\,$meV for typical parameters $L_s=50$\,\AA\ and $\epsilon_r=4$, while the intralayer Coulomb energy $\sim\!60\,$meV. More details for the perturbative calculation of interlayer Coulomb energy can be found in Sec.~S6 of Supplementary Material \cite{supp}.

\item Finally, we gather all the contributions from Eq.~\eqref{eq:full-ham} to find out the total energy of the coupled bilayer system staying in a gapped Dirac state (at the CNP) for graphene and a long-wavelength EC state for the substrate. By comparing it with that of a non-interacting Dirac state for graphene and a 2D Fermi-liquid state for the substrate, we can then find out if the gapped graphene interplays with the long-wavelength charge-ordered substrate in a cooperative or competitive manner. 

\hspace{7pt} It turns out that the bilayer system tends to \textit{cooperate} with each other such that both the gapped Dirac state (at the CNP) of graphene and the long-wavelength charge ordered state in the substrate are \textit{substantially stabilized} by the interlayer Coulomb coupling. The results are presented in Fig.~\ref{fig:2} and \ref{fig:4} of the main text. 
\end{itemize}

\subsection{Density functional theory calculations}
\label{sec:dft}
The first principles calculations are performed with the projector augmented-wave method within the density functional theory~\cite{PAW}, as implemented in the Vienna ab initio simulation package software~\cite{VASP}. The crystal structure is fully optimized until the energy difference between two successive steps is smaller than 10$^{-6}$eV and the Hellmann-Feynman force on each atom is less than 0.01\,eV$\cdot \AA$. The generalized gradient approximation by Perdew, Burke, and Ernzerhof is taken as the exchange–correlation potential~\cite{PBE}. As Cr is a transition metal element with localized $3d$ orbitals, we use the on-site Hubbard parameter $U\!=\!5.48\,$eV for the Cr $3d$ orbitals in the CrOCl bilayer and $U\!=\!3\,$eV for Cr $3d$ orbitals in the CrI$_3$ bilayer. The so-called fully localized limit of the spin-polarized GGA+U functional is adopted as suggested by Liechtenstein and coworkers~\cite{UTYPE}, and the non-spherical contributions from the gradient corrections are taken into consideration. The “DFT+D2” type of vdW correction has been adopted for all multilayer calculations to properly describe the interlayer interactions~\cite{VDW-D2}. 

Our high-throughput filtering of the proper insulating substrate materials for graphene starts from the 2D materials computational database~\cite{Haastrup-C2DB-2dmater-2018}. We only focus at those with bulk van der Waals structures which have been previously synthesized in laboratory. This ensures that it is experimentally feasible to exfoliate few layers from their bulk sample and then stack them on graphene to form heterostructures.

\subsection{Experimental measurements of the gaps in graphene-CrOCl heterostructure}
\label{sec:exp-method}
By designing a dual-gated structure, we used few-layered CrOCl as an bottom dielectric while few-layered hexagonal boron nitride (h-BN) was served as top gate dielectric. The top and bottom gate voltages can then be converted into doping and displacement fields for further data analysis. Graphene, h-BN, and CrOCl flakes are mechanically exfoliated from high quality bulk crystals. The vertical assembly of few-layered hBN, monolayer graphene and few-layered CrOCl were made using the polymer-assisted dry-transfer method. Electron beam lithography was done using a Zeiss Sigma 300 SEM with a Raith Elphy Quantum graphic writer. Top and bottom gates as well as contacting electrodes were fabricated with an e-beam evaporator, with typical thicknesses of Ti/Au $\sim$ 5/50\,nm. Electrical transport measurements of the devices were performed using an Oxford TeslaTron 1.5\,K system. Gate voltages on the as-prepared multi-terminal devices were fed by a Keithley 2400 source meter. Channel resistances were recorded in 4-probe configurations using low frequency (13.33 Hz) lock-in technique with Stanford SR830 amplifiers. The gate dependencies of channel resistances were measured at various temperatures for the extraction of thermal gaps. More details about the device configuration, measurement set up, and sample quality can be found in Sec.~S8 of Supplementary Material.  


\section{Acknowledgement}
We would like to thank Jian Kang and Jinhai Mao for valuable discussions, and to thank Hanwen Wang for the help in making the plots. This work is supported by the National Natural Science Foundation of China (grant No. 12174257, No. 11974357, and No. U1932151), the National Key R \& D program of China (grant No. 2020YFA0309601 and No. 2019YFA0307800), and the start-up grant of ShanghaiTech University. 




\bibliography{references}

\begin{thebibliography}{89}
\bibitem [{com()}]{comment_density}%
  \BibitemOpen
  \href@noop {} {\bibinfo {title} {If one considers that the charge order in
  the {TMO} layer results from a {Wigner}-crystal-like instability, then the
  value of superlattice constant {$L_s=50$ \AA}\ would correspond to a carrier
  density $\sim 7\times 10^{12}$ cm$^{-2}$ transferred from graphene to the
  insulating substrate, which is close to the upper limit for a double-gated
  graphene device.}}\BibitemShut {Stop}%
\end{thebibliography}

\widetext
\clearpage

\begin{center}
\textbf{\large Supplementary Material for “Synergistic correlated states and nontrivial topology in graphene-insulator heterostructures"}
\end{center}
\maketitle

\setcounter{equation}{0}
\setcounter{figure}{0}
\setcounter{table}{0}
\setcounter{page}{1}
\makeatletter
\renewcommand{\theequation}{S\arabic{equation}}
\renewcommand{\thefigure}{S\arabic{figure}}
\renewcommand{\bibnumfmt}[1]{[S#1]}
\newcommand{\setlabel}[1]{\edef\@currentlabel{#1}\label}

\section{S1. Non-interacting Hamiltonian for a graphene-insulator heterostructure}
\setlabel{S1}{sec:noninter-gr-crocl}

The Hamiltonian for a graphene-insulator heterostructure can be always divided into three parts: graphene part $H_G$, the insulating substrate part $H_S$ and the coupling between them $H_{G\rm{-}S}$. The graphene part can be suitably described by a tight-binding model since we focus on the low-energy physics. As we have explained in the main text, with slight carrier doping band edge, the insulator substrate is supposed to form a long-wavelength charge order on the interface near graphene sheet thanks to Coulomb interactions between electrons occupying the band edge of the insulating substrate (transferred from graphene layer). The insulator substrate part is then modeled by a 2D Hamiltonian for electrons hopping on a 2D superlattice which forms a Wigner-crystal-like or long-wavelength charge ordered insulator state at some proper  filling. The geometry of the superlattice is determined by the long-wavelength order at the interface. Explicitly, the graphene part $H_G$ and the insulator substrate part $H_S$ can be generally written as
\begin{align}
    \hat{H}_G &= \sum_{\bk, \sigma, \alpha, \alpha'} \gamma_{\alpha, \alpha'} (\bk) \cop^\dagger_{\sigma \alpha}(\bk) \cop_{\sigma \alpha'}(\bk)
    \label{eq:HG}\\
    \hat{H}_S &= \sum_{\btk, \sigma} \eta (\btk) \dop^\dagger_{\sigma}(\btk) \dop_{\sigma}(\btk)
    \label{eq:HS}
\end{align}
where $\cop^{(\dagger)}_{\sigma\alpha}(\k)$ and $\dop^{(\dagger)}_{\sigma}(\kt)$ are fermionic annihilation (creation) operators for electrons in graphene and the insulator substrate, respectively. In the lower index of these operators, $\alpha$ is the sublattice index for the bipartite lattice of graphene and $\sigma$ is the spin degree of freedom of electrons. To emphasize the fact that graphene and the insulator substrate have different lattices and thus different Brillouin zone, we denote $\bk$ and $\btk$ as the wavevectors in the Brillouin zone of graphene and that of the long-wavelength superlattice in the substrate, respectively. In our calculations, the lattice for $H_S$ is set to be rectangular or triangular. The lattice symmetry turns out to make no qualitative differences.

Since electrons have negligible probability to hop between graphene and the insulator substrate due to rather large distance $d$  between two sheets in the $z$-direction ($d\sim7\,$\AA\ from DFT calculations in CrOCl-graphene heterostructure), we suppose that electrons from two sheets are coupled only via long-ranged Coulomb interactions. Unlike $H_G$ and $H_S$, such long-ranged Coulomb interactions are more easily written in real space. In terms of field operators $\hat{\psi}(\br)$, the inter-sheet coupling reads
\begin{equation}
    \hat{H}_{G\rm{-}S} = \int d^2 \bm{r} d^2 \bm{r'} \sum_{\sigma,\sigma '} \hat{\psi}^{\dagger}_{c, \sigma}(\bm{r}) \hat{\psi}^{\dagger}_{d, \sigma'}(\bm{r'}) V (| \bm{r}-\bm{r'} + d \bm{\hat{z}}|) \hat{\psi}_{d, \sigma'}(\bm{r'}) \hat{\psi}_{c, \sigma}(\bm{r})
    \label{eq:HG-S}
\end{equation}
where $V(| \bm{r}-\bm{r'}+d \bm{\hat{z}}|)$ is the 3D long-ranged Coulomb potential $e^2/4 \pi \epsilon_0 \epsilon_r r$ and electrons in graphene and the insulating substrate are described by the field operators with lower index $c$ and $d$, respectively. Here $\epsilon_0$ is the vacuum permittivity and $\epsilon_r$ is the dimensionless relative dielectric constant of the insulating substrate. In the spirit of tight-binding formalism, we write the field operators in terms of Wannier functions
\begin{align}
    \hat{\psi}^{\dagger}_{c, \sigma}(\bm{r}) &= \sum_{i,\alpha} \phi_\alpha ^* (\br-\bm{a}_i-\bm{\tau}_\alpha) \chi^{\dagger}_\sigma \cop^{\dagger}_{i,\sigma \alpha} \\
    \hat{\psi}^{\dagger}_{d, \sigma}(\bm{r}) &= \sum_{i,\alpha} \widetilde{\phi} ^* (\br-\bm{R}_i) \chi^{\dagger}_\sigma \dop^{\dagger}_{i,\sigma}
    \label{eq:cd-transform1}
\end{align}
where $\phi_\alpha$ and $\widetilde{\phi}$ are Wannier functions localized on the graphene and the insulator substrate Bravais lattice sites, which are described by $\bm{a}_{i}$ and $\bm{R}_{i}$, respectively. Here $\alpha$ refers to the sublattice index in graphene and $\bm{\tau}_{\alpha}$ is the vector denoting the position of the $\alpha$th sublattice inside the unit-cell. 
The spin degrees of freedom is included by the index $\sigma$ and also explicitly by spinor $\chi_\sigma$.  It is worthwhile to note that here the Bravais lattice sites and the corresponding Wannier functions for the substrate  refer to those of the spontaneously generated charge ordered superlattice, not the atomic lattices of the substrate. A more fundamental treatment based on the atomic lattice sites of the substrate lattice will be discussed in Sec.~\ref{sec:couple}.

After the transformations above, the Hamiltonian $H_{G\rm{-}S}$ in the Wannier basis reads
\begin{equation}
    \hat{H}_{G\rm{-}S} = \sum_{\substack{\sigma, \sigma' \\ \alpha, \alpha'}} \sum_{\substack{i,i' \\ j,j'}} U^{\sigma \sigma'}_{i \alpha j , i' \alpha' j'} \cop^\dagger_{i,\sigma \alpha} \dop^\dagger_{j,\sigma'} \dop_{j',\sigma'} \cop_{i',\sigma \alpha'} 
\end{equation}
with
\begin{equation}
    U^{\sigma \sigma'}_{i \alpha j, i' \alpha' j'} = \int d^2 \bm{r} d^2 \bm{r'} \phi_\alpha^* (\bm{r}-\bm{a}_i-\bm{\tau}_\alpha) \widetilde{\phi} ^* (\bm{r}-\bm{R}_j) V (| \bm{r}-\bm{r'} + d \bm{\hat{z}}|) \widetilde{\phi} (\bm{r'}-\bm{R}_{j'}) \phi_{\alpha'} (\bm{r}-\bm{a}_{i'}-\bm{\tau}_{\alpha'}) \chi^\dagger_\sigma \chi^\dagger_{\sigma'} \chi_{\sigma'} \chi_\sigma.
\end{equation}

If Wannier functions are so localized such that 
\begin{align}
    \phi_\alpha ^* (\bm{r}-\bm{a}_i-\bm{\tau}_\alpha) \phi_{\alpha'} (\bm{r}-\bm{a}_{i'}-\bm{\tau}_{\alpha'}) &\approx 0 \hspace{2cm} \text{if } (i,\alpha) \ne (i',\alpha') \\
    \widetilde{\phi} ^* (\bm{r}-\bm{R}_j) \widetilde{\phi} (\bm{r}-\bm{R}_{j'}) &\approx 0 \hspace{2cm} \text{if } j \ne j' \\
    |\phi_\alpha (\bm{r}-\bm{a}_i-\bm{\tau}_\alpha)|^2 &\approx \delta^{(2)}(\bm{r}-\bm{a}_i-\bm{\tau}_\alpha) \\
    |\widetilde{\phi} (\bm{r}-\bm{R}_j) |^2 &\approx \delta^{(2)}(\bm{r}-\bm{R}_j)\;,
    \label{eq:delta-loc}
\end{align}
with $\delta^{(2)}(\bm{r})$ is the 2D Dirac $\delta$-function distribution, we can simplify the previous expression to
\begin{equation}
    U^{\sigma \sigma'}_{i \alpha j, i' \alpha' j'} = U_{i \alpha j} \delta_{i,i'} \delta_{\alpha,\alpha'} \delta_{j,j'} 
\end{equation}
with $\delta_{\mu, \nu}$ is the Kronecker delta and 
\begin{equation}
    U_{i \alpha j} = V (|\bm{a}_i+\bm{\tau}_\alpha - \bm{R}_j + d \bm{\hat{z}}|).
\end{equation}
Then, we write $H_{G-S}$ in reciprocal space using the following Fourier transformation
\begin{align}
    \cop_{i,\sigma \alpha} &= \frac{1}{\sqrt{N_c}} \sum_{\bk} e^{i \bk \cdot \bm{a}_{i}} \cop_{\sigma \alpha}(\bk)\\
    \dop_{i,\sigma} &= \frac{1}{\sqrt{N_d}} \sum_{\btk} e^{i \btk \cdot \bm{R}_{i}} \dop_{\sigma}(\btk)
\end{align}
where $N_c$ and $N_d$ are the number of lattice sites for electron in graphene and the insulator substrate, respectively. The Hamiltonian $H_{G\rm{-}S}$ in the basis of $\cop_{\sigma \alpha}(\bk)$ and $\dop_{\sigma}(\btk)$ reads
\begin{equation}
    \hat{H}_{G-S} = \frac{1}{N_c N_d} \sum_{\substack{\sigma,\sigma' \\ i,\alpha,j}} \sum_{\substack{\bk,\bk ' \\ \btk, \btk'}} U_{i \alpha j} \  e^{i (\bk ' -\bk) \cdot (\bm{a}_{i}-\bm{R}_{j})} \  e^{i (\bk ' - \bk + \btk ' - \btk) \cdot \bm{R}_{j}} \  \cop^\dagger_{\sigma \alpha}(\bk) \dop^\dagger_{\sigma'}(\btk) \dop_{\sigma'} (\btk') \cop_{\sigma \alpha}(\bk').
\label{eq:hGS_k}
\end{equation}

Now we first define $\widetilde{\mathbf{R}}=\mathbf{a}_i-\mathbf{R}_j$, and let $\bk '-\bk=\bm{q}=\btq +\mathbf{G}$, where $\mathbf{G}$ is a reciprocal vector of the long-wavelength ordered superlattice and $\btq$ is the wavevector within the superlattice Brillouin zone.  Then we take use of the identity $\sum_j e^{i(\btk '-\btk +\bm{q})\cdot\mathbf{R}_j}=\sum_j e^{i(\btk '-\btk+\btq+\mathbf{G})\cdot\mathbf{R}_j}=N_d\delta_{\btk '-\btk,\btq}$, Eq.~(\ref{eq:hGS_k}) can be simplified as
\begin{equation}
    \hat{H}_{G-S} =  \sum_{\substack{\sigma,\sigma' \\ \alpha}} \sum_{\substack{\bk,\btk \\ \btq, \widetilde{\bm{G}}}} \widetilde{V}(\btq + \bm{G}) \   \cop^\dagger_{\sigma \alpha}(\bk) \dop^\dagger_{\sigma'}(\btk) \dop_{\sigma'} (\btk + \btq) \cop_{\sigma \alpha}(\bk - \btq - \bm{G})
\end{equation}
The coupling $\widetilde{V}(\btq+\mathbf{G})$ reads
\begin{align}
   \widetilde{V}(\btq+\bm{G}) &= \frac{1}{N_c} \sum_{i} V (|\bm{a}_{i}+\bm{\tau}_{\alpha} - \bm{R}_{j} + d \bm{\hat{z}}|) e^{-i (\btq+\bm{G}) \cdot (\bm{a}_{i} - \bm{R}_{j})} \nonumber \\
   &= \frac{1}{N_c} \sum_{\widetilde{\bm{R}}} V (|\widetilde{\bm{R}}+\bm{\tau}_{\alpha} + d \bm{\hat{z}}|) e^{-i (\btq+\bm{G}) \cdot \widetilde{\bm{R}}} \nonumber \\
   &= \frac{1}{N_d} \int \frac{d^2 r}{\Omega_d} V (|\bm{r}+\bm{\tau}_{\alpha} + d \bm{\hat{z}}|) e^{-i (\btq+\bm{G}) \cdot \bm{r}} \nonumber \\ 
   &= \frac{e^2}{2 \epsilon_0 \epsilon_r N_d \Omega_d} \frac{e^{-|\btq+\bm{G}|d}}{|\btq+\bm{G}|} 
\end{align}
where $\Omega_d$ is the area of the unit-cell of the surface superlattice of the substrate. 
In the third line of the above derivation, we smear the sum over $\widetilde{\bm{R}}=\bm{a}_i-\bm{R}_j$  by replacing it with an integral over the surface $S = N_d \Omega_d = N_c \Omega_c$ with $\Omega_c$ the area of graphene's unit-cell since we are interested in the physics in the length scale of the superlattice $\{ \bm{R}_{j} \}$, which is supposed to much larger than that of graphene. Finally, the last line is the 2D partial Fourier transformation of the 3D Coulomb potential.

Since we focus on the low-energy physics around the Dirac cones of graphene, we can attribute valley index $\mu$ to electrons in graphene and neglect intervalley coupling thanks to the exponential decay of $\widetilde{V}(\bm{q})$ so that 
\begin{equation}
    \hat{H}_{G-S} =  \sum_{\substack{\sigma,\sigma' \\ \alpha}} \sum_{\substack{\bk,\btk \\ \btq, \bm{G}}} \widetilde{V}(\btq + \bm{G}) \sum_{\mu}   \cop^\dagger_{\sigma \mu \alpha}(\bk) \dop^\dagger_{\sigma'}(\btk) \dop_{\sigma'} (\btk + \btq) \cop_{\sigma \mu \alpha}(\bk - \btq - \bm{G})  .  
\end{equation}
In the meantime, the Hamiltonian for graphene only $H_G$ [see Eq.~\eqref{eq:HG}] can be divided into two valley sectors
\begin{equation}
    \hat{H}_G = \sum_{\bk, \sigma, \alpha, \alpha', \mu} \left( \hbar v_F \bk \cdot \bm{\sigma}^\mu \right)_{\alpha, \alpha'} \cop^\dagger_{\sigma \mu \alpha}(\bk) \cop_{\sigma \mu \alpha'}(\bk)
    \label{eq:H_Gonly}
\end{equation}
where $\bm{\sigma}^\mu = (\mu \sigma_x, \sigma_y)$ with $\sigma_{x,y}$ are the Pauli matrices and the valley index $\mu =\pm 1$.

In the Hartree approximation by contracting $c$ and $d$ fermion operators separately, we have
\begin{equation}
    \hat{H}_{G-S} =  \sum_{\substack{\sigma, \alpha, \mu}} \sum_{\substack{\bk, \bm{G}}} \widetilde{V}(\bm{G}) \sum_{\btk, \sigma '}  \langle \dop^\dagger_{\sigma'}(\btk) \dop_{\sigma'} (\btk) \rangle \, \cop^\dagger_{\sigma \mu \alpha}(\bk) \cop_{\sigma \mu \alpha}(\bk - \bm{G}).  
\label{eq:HG-S-Hartree}  
\end{equation}
Since the long-wavelength charge order state is insulating presumingly with two spin degenerate electrons occupying each supercell, we have 
\begin{equation}
\sum_{\btk, \sigma '}  \langle \dop^\dagger_{\sigma'}(\btk) \dop_{\sigma'} (\btk) \rangle = 2 N_d.
\end{equation}
Writing $\bk = \btk + \bm{G}$ with $\bm{G}$ in the superlattice reciprocal lattice, the final form of the coupling between graphene and insulating substrate used in our calculations reads
\begin{equation}
    \hat{H}_{G-S} =  \sum_{\substack{\sigma, \alpha, \mu}} \sum_{\substack{\bm{G}, \bm{Q} \\ \in \{ \bm{G_i} \} } } \widetilde{U}_d (\bm{Q})  \  \cop^\dagger_{\sigma \mu \alpha, \bm{G}+\bm{Q}}(\btk) \cop_{\sigma \mu \alpha, \bm{G}}(\btk).   
    \label{eq:hamGS_final}
\end{equation}
where 
\begin{equation}
    \widetilde{U}_d (\bm{Q}) = \frac{e^2}{\epsilon_0 \epsilon_r \Omega_d} \frac{e^{-|\bm{Q}|d}}{|\bm{Q}|}
    \label{eq:Udfourier}
\end{equation}

In the meantime, we integrate out the Hamiltonian for insulating substrate $H_S$ [see Eq.~\eqref{eq:HS}] so that it becomes a constant charge density, which is omitted in our calculations. To wrap up, we get the effective non-interacting Hamiltonian in continuum in the valley $\mu$ 
\begin{equation}
    H^\mu_0(\bm{r}) = \hbar v_F \bm{k} \cdot \bm{\sigma}^{\mu} + U_d(\bm{r})
    \label{eq:ham_0}
\end{equation}
where the Fourier component of $U_d(\mathbf{r})$ is precisely $\widetilde{U}_d (\bm{G})$ [see Eq.~\eqref{eq:Udfourier}] with $\bm{G}$ in the reciprocal lattice of the underlying insulating substrate's surface superlattice.

\begin{figure}[htb]
    \centering
    \includegraphics[width=0.85\textwidth]{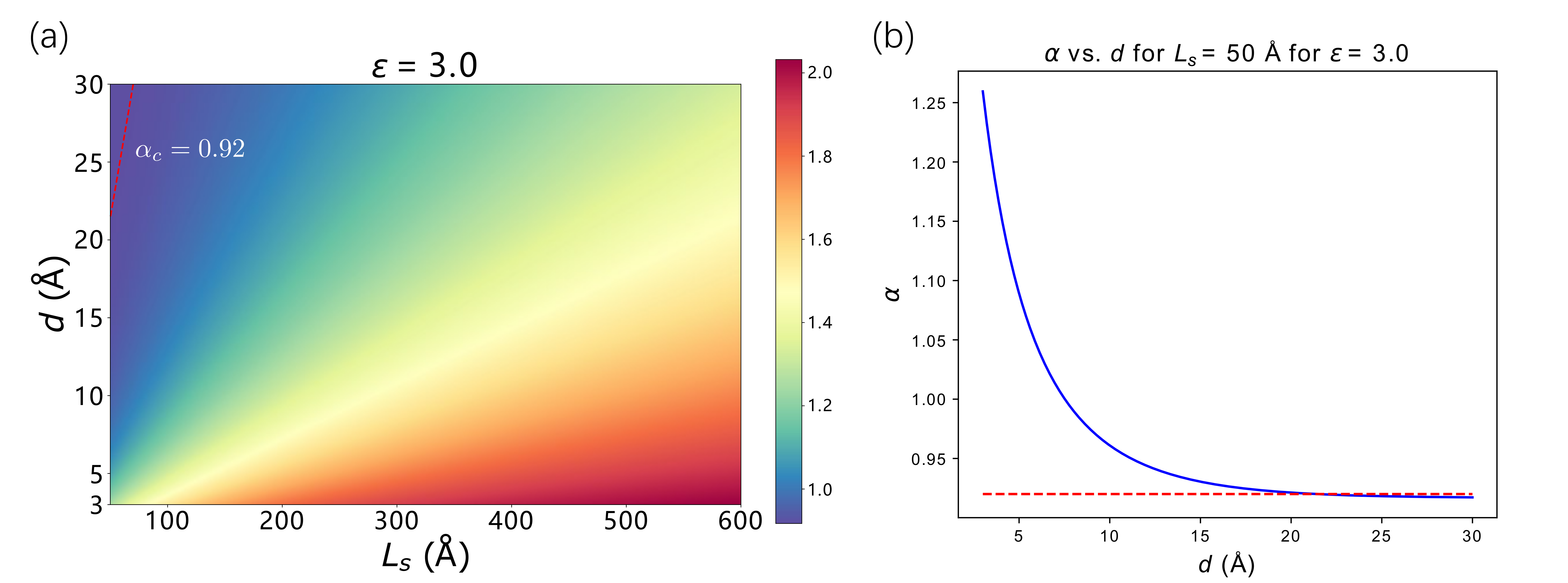}
    \caption{(a) Calculated effective fine-structure constant $\alpha$ of graphene in the parameter space of the background Wigner-crystal superlattice $L_s$, and the interlayer distance between graphene and the substrate $d$, with the background dielectric constant fixed as $\epsilon_r=3$. (b) Line plot of $\alpha$ vs. $d$, with $L_s=50$ \AA, and $\epsilon_r=3$. The red lines mark the critical value of $\alpha_c=0.92$ above which the Dirac point of graphene would be gapped by e-e interactions.}
    \label{alpha_SI}
\end{figure}

\begin{figure}[htb]
	\centering
	\includegraphics[width=0.85\textwidth]{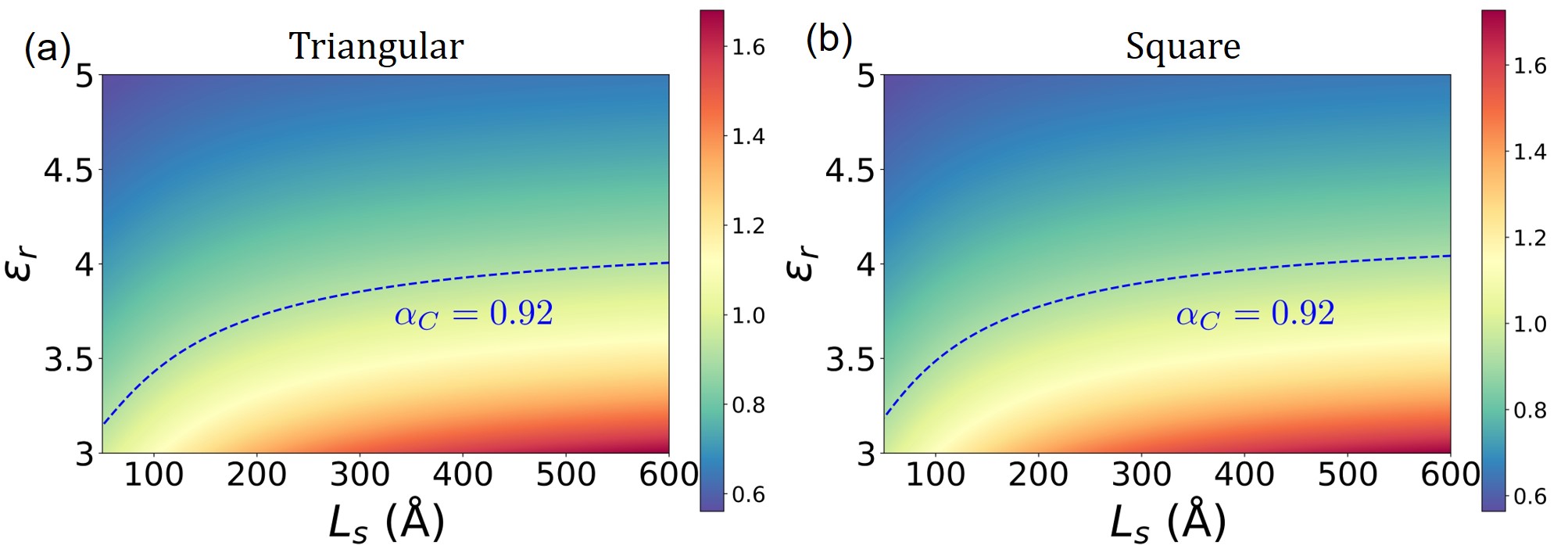}
	\caption{Calculated effective fine-structure constant $\alpha$ of graphene in the parameter space of the background Wigner-crystal superlattice $L_s$, and the interlayer distance between graphene and the substrate $d$, with the background dielectric constant fixed as $\epsilon_r=3$, for (a) triangular and (b) square lattice.}
	\label{alpha_sqr_tri}
\end{figure}

As revealed by Eq.~\eqref{eq:Udfourier}, increasing the interlayer distance $d$ would diminish the interlayer Coulomb interaction between the charges of the Wigner crystal at the surface of the substrate and the Dirac electrons in the graphene layer [see Eq. (2) in the main text]. To estimate the e-e interaction effects in graphene with different interlayer distance $d$, we have calculated the effective fine structure constant of graphene: $\alpha=e^2/4 \pi \epsilon_0 \epsilon_r  \hbar v_F^{\text{SL}}$, where the Fermi velocity $v_F^{\text{SL}}$ is the non-interacting Fermi velocity of the graphene being coupled to the Coulomb superlattice potential. Note that $v_F^{\text{SL}}$ is reduced compared to the free-standing non-interacting Fermi velocity $v_F$ due to the effects of Coulomb superlattice potential, whose strength depends on the interlayer distance $d$, the background dielectric constant $\epsilon_r$, and the superlattice constant $L_s$ of the Wigner crystal formed at the surface of the substrate. In Fig.~\ref{alpha_SI}(a), we plot the effective fine structure constant $\alpha$ as a function of the interlayer distance $d$, and the superlattice constant $L_s$, with a fixed background dielectric constant $\epsilon_r=3$. We note that there is small region in the upper left corner where $\alpha \le \alpha_c=0.92$, which means that graphene would remain as a semimetal in this region. We further plot $\alpha$ vs. $d$ in Fig.~\ref{alpha_SI}(b), and find that $\alpha$ decreases with the increase of $d$, and $\alpha$ becomes smaller than the critical value 0.92 when $d>20$  \AA, the moment that is expected to undergo a transition from a gapped to a gapless Dirac state. It turns out that the colormaps of $\alpha$ remains qualitatively the same for different lattice geometries, comparing Fig.~\ref{alpha_SI}(a) for rectangular lattice with Fig.~\ref{alpha_sqr_tri} for triangular and square lattice. This also justifies why we can consider only two particular cases, rectangular and triangular lattice, in our work without loss of generality.

In our numerical implementations, the lattice of insulating substrate is set to be rectangular or triangular, from which we obtain qualitatively the same correlated states in the graphene layer. The range of $\{ \bm{G_i} \}$ is limited to $|n_x|,|n_y| \leq 4$ with $\bm{G} = n_x \bm{g_x} + n_y \bm{g_y}$. $\bm{g_{x,y}}$ are the two reciprocal lattice vectors for the rectangular lattice of insulating substrate. The sum over $\bm{Q}$ in Eq.~\eqref{eq:hamGS_final} stops at the limit $|n_x|+|n_y|\leq 2$.

\section{S2. Topological Properties of the non-interacting Hamiltonian}
\setlabel{S2}{sec:topoS2}

We further study the topological properties of our model Hamiltonian derived in the previous section. 
Different from magic-angle TBG \cite{song-tbg-prl19, yang-tbg-prx19,po-tbg-prb19, origin-magic-angle-prl19, jpliu-prb19}, the low-energy subbands for graphene coupled to a rectangular superlattice potential $U_d(\mathbf{r})$ with small anisotropy ($r \sim 1$) turns out to be topologically trivial. To be specific, in Fig.~2d of the main text, we have shown the Berry curvature distribution of the highest valence band of valley $K$ in the mini BZ of the superlattice with $L_x=50$\,\AA\  and $r=L_y/L_x=1.2$. 
We see that the Berry curvature is mostly concentrated at the band crossing points, i.e., the four high symmetry points $\Gamma_s$, $S_s$, $X_s$, and $Y_s$. The contributions from the  $\Gamma_s$  and the $S_s$ points are exactly compensated by those from the $X_s$ and $Y_s$ points, resulting in a band with zero Chern number. This is anticipated because the superlattice potential is non-chiral in the sense that it is coupled equally to the two sublattice of graphene. 
This remains true even including $e$-$e$ interactions.

\begin{figure}[htb]
    \centering
    \includegraphics[width=0.7\textwidth]{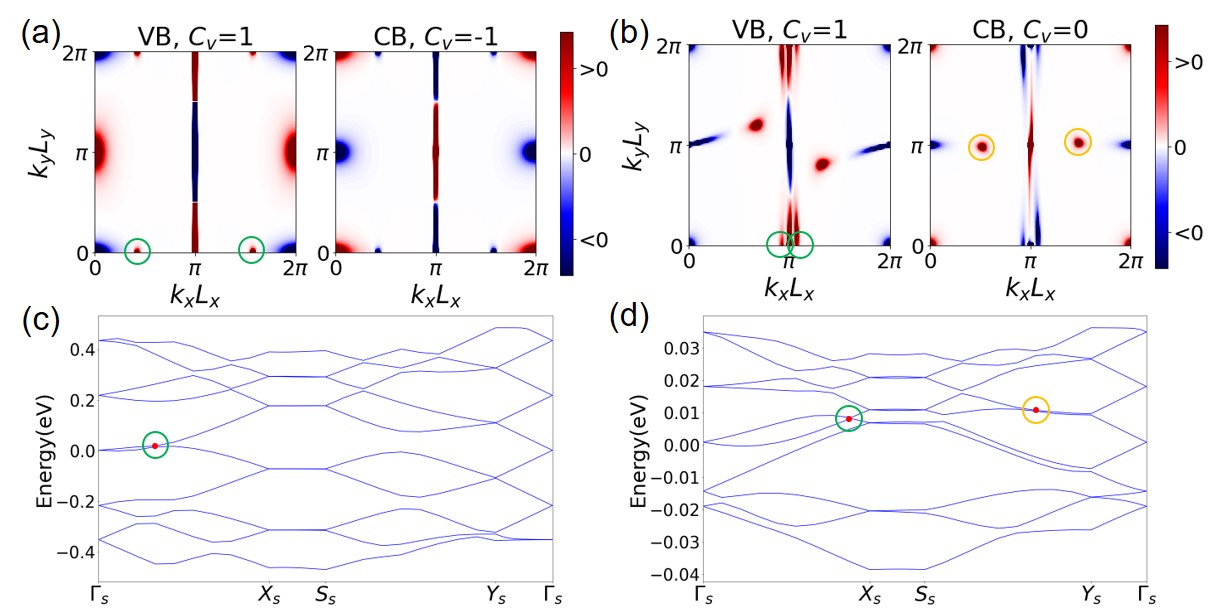}
    \caption{(a) and (b) shows the distribution of Berry curvature in the $r=3$ superlattice's BZ of the lowest valence and conduction band in valley $K$ for $L_s=50$ and $600$ \AA, respectively. Their corresponding valley Chern number are also given on the top of each panel. (c) and (d) are the non-interacting band structure of the $r=3$ superlattice with $L_s=50$ and $600$ \AA.}
    \label{fig:3}
\end{figure}

Hence, it is unexpected that changing the anisotropy $r$ and the lattice size $L_s$ of the superlattice potential $U_d$ can make the subbands topological. For example, keeping $L_x = 50$ \AA \  but with $r=3.0$, the valley Chern number of the low-energy subbands become nonzero. For valley $K$, the highest valence band and lowest conduction band now have a Chern number $C= \pm 1$, respectively. As shown in Fig.~\ref{fig:3}(a), besides the four high symmetry points, it appears another two ``hot spots" (annotated by green circles) along the line connecting $\Gamma_s$ and $X_s$.  This new contribution breaks the balance between positive and negative contribution of Berry curvature to Chern number,
leading to non-zero valley Chern number. Such contribution stems from a new crossing point between the low-energy valence and conduction bands along the $k_x$-direction through changing merely the anisotropy parameter $r$, as shown in Fig.~\ref{fig:3}(c) with red dot in green circle. From the animated Fig.~\ref{gif}, we see that while increasing $r$, the Fermi velocity is gradually reduced, and at $r=2.7$, becomes vanishingly small along one direction due to Klein tunneling effects. Then, further tuning $r$ germinates a band crossing originated from the Dirac point then gradually moving away from it.
Alternatively, we find that changing $L_s$ can also control the valley Chern number of the subbands, since $L_s$ is encoded in the superlattice potential (see Eq.~\eqref{eq:Udfourier}). 
For example, with $r=3$ and $L_s=600$ \AA, as shown in Fig.~\ref{fig:3}(b), while the highest valence band remains topological with non-zero valley Chern number 1 for valley $K$ with the two aforementioned crossing points (green circles) merely moving to $X_s$, the lowest conduction band turns out to be topologically trivial. This is due to two new band crossing points (orange circles) close to the $Y_s$-$S_s$ line between the lowest and the second lowest conduction bands, as annotated by red dots in an orange circle in Fig.~\ref{fig:3}(d). Such topologically nontrivial bands are particularly surprising for our system, since the Dirac fermions are coupled to a ``trivial" superlattice potential coupled identically on two sublattices, different from the case of magic-angle TBG. Thus, the nontrivial topology must arise by virtue of the intrinsic Berry phases of the Dirac cones.

\begin{figure}[htb]
    \centering
    \animategraphics[width=0.35\textwidth]{2}{fig/band_Ls50_ratio_}{1}{41}
    \animategraphics[width=0.5\textwidth]{2}{fig/Chern_Ls50_}{1}{41}
    \caption{Animation (click the figure in, for example, Adobe Acrobat PDF reader to activate) for the non-interacting band structures of graphene on a superlattice with $L_s=50$ \AA \  and the anisotropy parameter $r$ varying from 1 to 5. }
    \label{gif}
\end{figure}

\begin{figure}[h]
    \centering
    \includegraphics[width=0.8\textwidth]{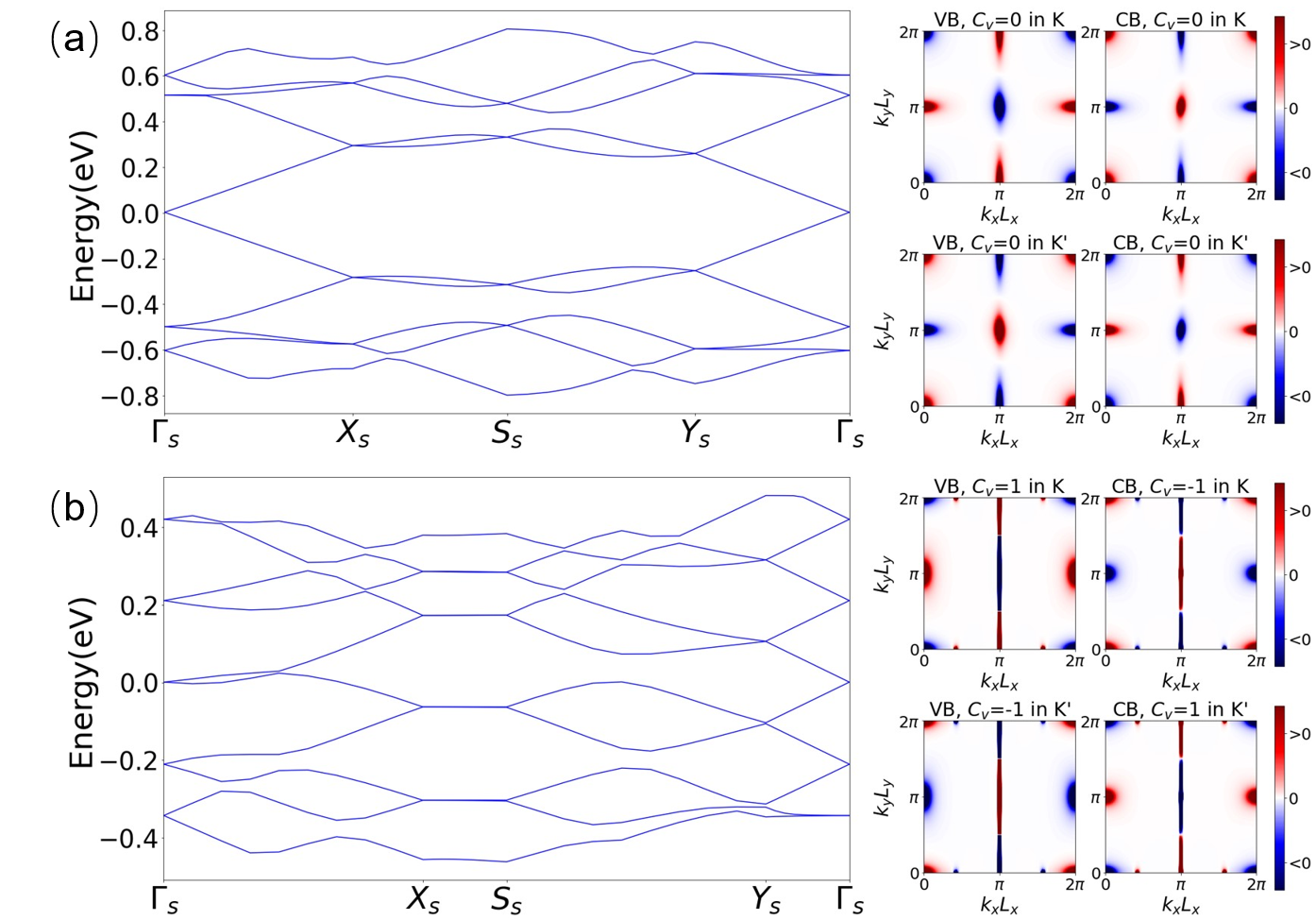}
    \caption{Non-interacting spectrum and distribution of Berry curvature in the first Brillouin zone for $L_s=50$ \AA \ with (a) $r=1.2$ and (b) $r=3$. Please refer to Sec.~\ref{sec:topoS2}.}
    \label{noninter_50}
\end{figure}

\begin{figure}[h]
    \centering
    \includegraphics[width=0.8\textwidth]{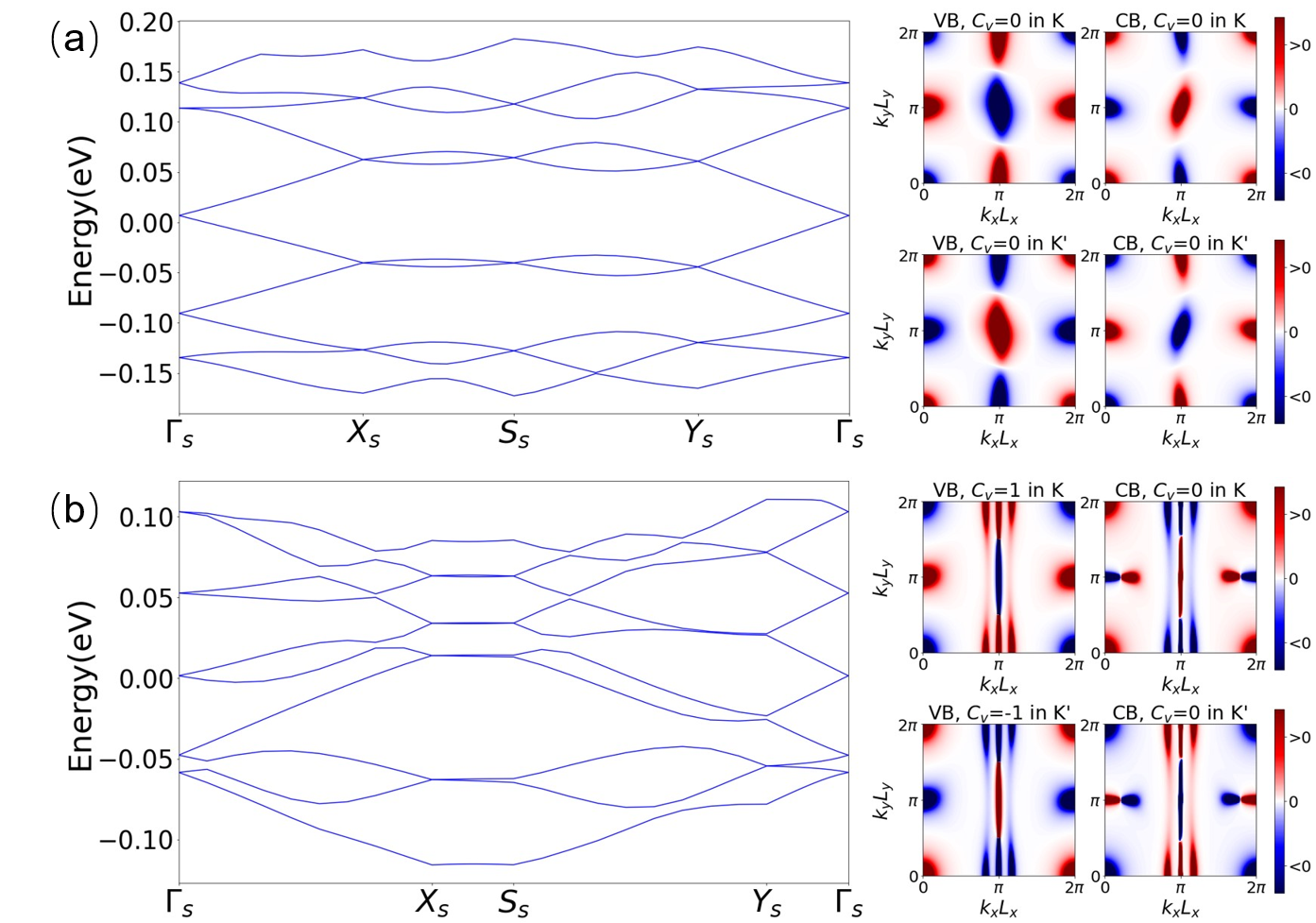}
    \caption{Non-interacting spectrum and distribution of Berry curvature in the first Brillouin zone for $L_s=200$ \AA \ with (a) $r=1.2$ and (b) $r=3$. Please refer to Sec.~\ref{sec:topoS2}.}
    \label{noninter_200}
\end{figure}

\begin{figure}[htb]
    \centering
    \includegraphics[width=0.8\textwidth]{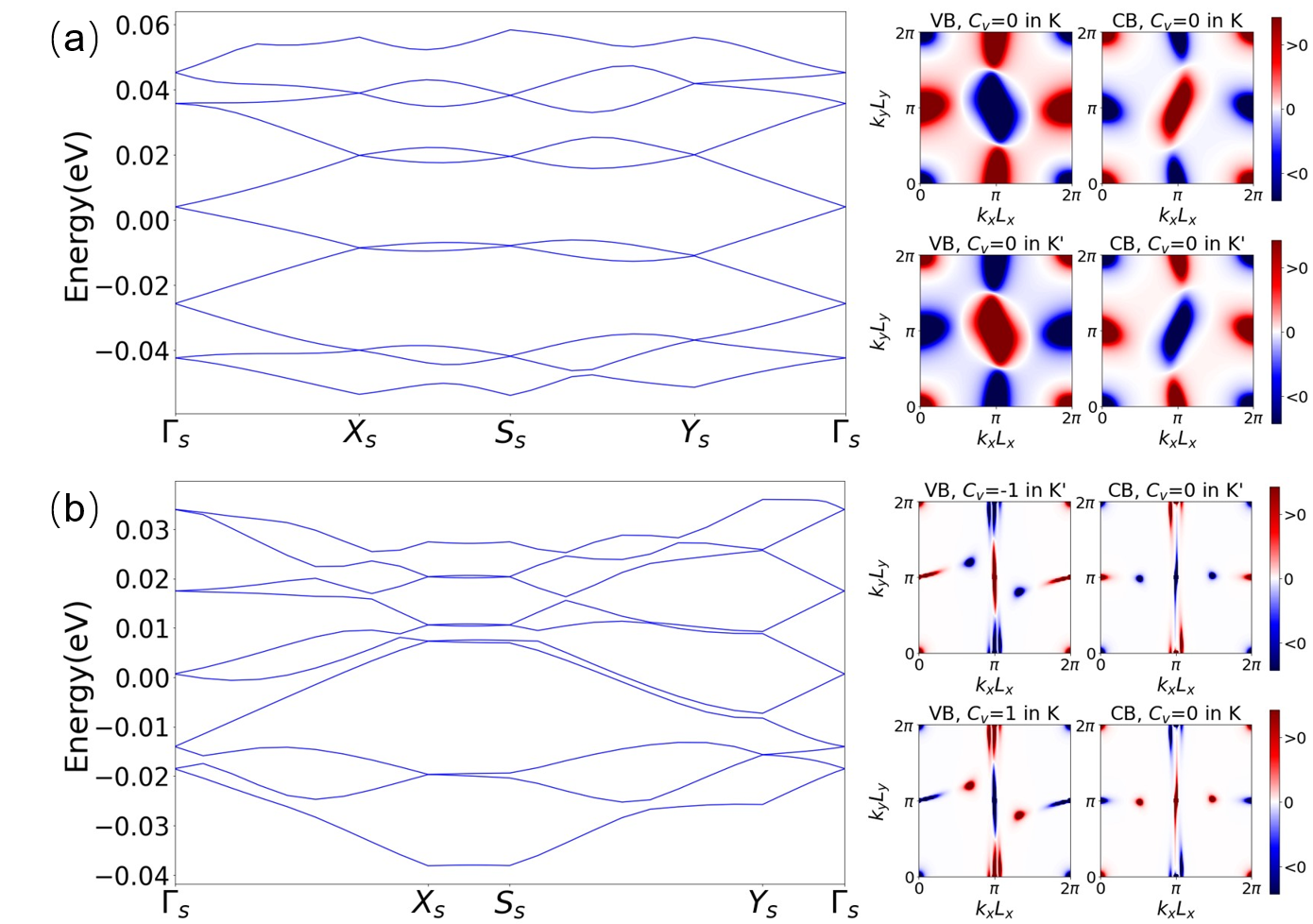}
    \caption{Non-interacting spectrum and distribution of Berry curvature in the first Brillouin zone for $L_s=600$ \AA \ with (a) $r=1.2$ and (b) $r=3$. Please refer to Sec.~\ref{sec:topoS2}.}
    \label{noninter_600}
\end{figure}

We further show the non-interacting energy spectra and distributions of Berry curvature in the first Brillouin zone for various superlattice constants ($L_s$) and anisotropy parameters ($r$), i.e., $L_s = 50$, 200, 600 \AA \  with $r=1.2$ and 3. The plots are given in Fig.~\ref{noninter_50}, \ref{noninter_200}, \ref{noninter_600} for $L_s = 50$, 200, 600 \AA, respectively. Since the system preserves time-reversal symmetry and the superlattice potential $U_d$ is diagonal in the sublattice subspace, the non-interacting energy spectrum in valley $K'$ is exactly the same as that in valley $K$ so that we only plot the spectrum for valley $K$ here. As shown below, the distribution of Berry curvature of the highest valence and the lowest conduction band in valley $K$ is exactly opposite to that in valley $K'$ as another consequence of time-reversal symmetry of the system. 

We also provide separately six videos in other Supplementary Data, which show the non-interacting energy spectra and distributions of Berry curvature in the first Brillouin zone for $L_s = 50$, 200, 600 \AA \  with $r=1$-10, respectively. We note that the anisotropic charge ordered superlattices may be realized in two ways. First, one can design a spatially modulated electrostatic potential, which has been realized in monolayer graphene by inserting a patterned dielectric superlattice between the gate and the sample \cite{dean-nn18}. Then, the anisotropy of the superlattice can be artificially tuned by the dielectric patterning in the substrate. Second, for some given carrier density, the Fermi surface of the conduction (or valence) band of the substrate may be (partially) nested, which may lead to a charge density wave (CDW) state with the nesting wavevector. For example, for CrOCl, the Fermi surfaces under different Fermi energies (above the conduction band minimum) are given in Fig.~\ref{type3}(c). Clearly, under some proper fillings, the Fermi surfaces are nested or partially nested, which may give rise to CDW states with anisotropic superlattices.  We note that topologically nontrivial flat bands have also been proposed to exist in Bernal bilayer graphene coupled with a background superlattice potential \cite{cano-bilayer-arxiv22}.

\section{S3. Renormalization group derivations}

\setlabel{S3}{sec:RG}
The derivation shown in this section is inspired from Ref.~\onlinecite{vafek_prl2020}. The e-e Coulomb interaction operator in our derivations is written as
\begin{equation}
    \hat{V}_{\text{int}} = \frac{1}{2} \int d^2 \br d^2 \br ' V_c(\br - \br ') \hat{\rho} (\br) \hat{\rho} (\br ') 
\end{equation}
where $V_c(\br) = e^2/4 \pi \epsilon_0 \epsilon_r r$ and $\hat{\rho}(\br)$ is the density operator of electrons at $\br$. The Hamiltonian Eq.~\eqref{eq:ham_0} is defined at some high energy cut-off $\pm E_c$. We focus in the valley $\mu=+1$ by the virtue of which the derivation for the valley $\mu=-1$ is immediate and the results are identical. Remember that the parameters $v_F$ and $\widetilde{U}_d(G)$ should be thought of as being fixed by a measurement at $E_c$ without e-e interactions. This also amounts to $\hat{\rho}(\br)= \hat{\psi}^\dagger (\br) \hat{\psi} (\br)$ with the non-interacting field operator $\hat{\psi} (\br)$
\begin{equation}
    \hat{\psi} (\br) = \sum_{\substack{\sigma, n, 
    \bk;\\ |\epsilon_{n,\bk}| \leq E_c }}  \phi_{\sigma n \bk} (\br) \cop_{\sigma n}(\bk) 
\end{equation}
where $\phi_{\sigma n \bk} (\br)$ is the wavefunction of an eigenstate of the non-interacting Hamiltonian $H_0$ [see Eq.~\eqref{eq:ham_0}] with energy $\epsilon_{n,\bk}$ and its associated annihilation operator is $\cop_{\sigma n}(\bk)$.

\subsection{Electron-electron interaction in a lower energy window}
Now we change the cut-off $E_c$ to a smaller one $E_c '$ and see how these parameters are modified by $\hat{V}_{\text{int}}$. $\hat{V}_{\text{int}}$ can be treated perturbatively when $E_c '$ is much larger than any other energy scale in the system. To do so, we split the field operator $\hat{\psi} (\br) = \hat{\psi}^{<} (\br) + \hat{\psi}^{>} (\br)$ where 
\begin{align}
    \hat{\psi}^{<} (\br) &= \sum_{\substack{\sigma, n, 
    \bk;\\ |\epsilon_{n,\bk}| \leq E_c ' }} \phi_{\sigma n \bk} (\br) \cop_{\sigma n}(\bk) \\
    \hat{\psi}^{>} (\br) &= \sum_{\substack{\sigma, n, 
    \bk;\\ E_c ' < |\epsilon_{n,\bk}| \leq E_c }} \phi_{\sigma n \bk} (\br) \cop_{\sigma n}(\bk). \\
\end{align}
Then, we integrate out the fast modes $\hat{\psi}^{>} (\br)$ in the expansion of $\hat{\rho}(\br) \hat{\rho}(\br')$. Note that $\hat{\psi}^{>} (\br)$ and $\hat{\psi}^{> \dagger} (\br)$ must appear equal times in each terms of the expansion otherwise it would vanish by taking the non-interacting mean value $\langle \dots \rangle_0$. Explicitly, these terms are retained up to a constant: 
\begin{align*}
    \hat{\rho}(\br) \hat{\rho}(\br') &= \hat{\rho}^{<} (\br) \hat{\rho}^{<} (\br') \\ 
    & \quad + \bar{\rho}^{>} (\br) \hat{\psi}^{< \dagger} (\br ') \hat{\psi}^{<} (\br ') + \bar{\rho}^{>} (\br ') \hat{\psi}^{< \dagger} (\br) \hat{\psi}^{<} (\br) \\
    & \quad +  \underbrace{\hat{\psi}^{< \dagger} (\br) \  \langle \hat{\psi}^{>} (\br) \hat{\psi}^{> \dagger} (\br ') \rangle_0 \ \hat{\psi}^{<} (\br ') + \hat{\psi}^{<} (\br) \  \langle \hat{\psi}^{> \dagger} (\br) \hat{\psi}^{>} (\br ') \rangle_0 \ \hat{\psi}^{< \dagger} (\br ')}_{(*)}
\end{align*}
with
\begin{align}
    \hat{\rho}^{<} (\br) &= \hat{\psi}^{< \dagger} (\br) \hat{\psi}^{<} (\br)\\
    \bar{\rho}^{>} (\br) &= \sum_{\substack{\sigma, n, 
    \bk;\\ E_c ' < |\epsilon_{n,\bk}| \leq E_c}} \phi_{\sigma n \bk}^* (\br) \phi_{\sigma n \bk} (\br).
\end{align}
The first term gives the Coulomb e-e interaction between electrons of the slow modes $\hat{\psi}^{<} (\br)$ below the new cut-off $E_c '$. The second and third term could be omitted if the system has particle-hole (p-h) symmetry as in twisted bilayer graphene \cite{vafek_prl2020}. In our system described by Eq.~\eqref{eq:ham_0}, the first nearest-neighbor coupling in $\widetilde{U}_d (\bm{G})$ preserves particle-hole (p-h) symmetry. The p-h symmetry is broken if further-neighbor coupling is included, which is exponentially smaller [see Eq.~\eqref{eq:Udfourier}]. So, it is legitimate in our RG derivation to neglect such weak p-h asymmetry in order to omit the second and the third term in the expansion. 

Then, we evaluate the rest of the terms in the expansion, which represents precisely the correction to $H_0$ from the fast modes $\hat{\psi}^{>} (\br)$ via Coulomb e-e interactions. Let us write
\begin{align*}
    (*) &= \hat{\psi}^{< \dagger} (\br) \left(  \sum_{\substack{\sigma, n, 
    \bk;\\ E_c ' < \epsilon_{n,\bk} \leq E_c}} \phi_{\sigma n \bk} (\br) \phi_{\sigma n \bk}^* (\br ')  \right) \hat{\psi}^{<} (\br ') + \hat{\psi}^{<} (\br) \left(  \sum_{\substack{\sigma, n, 
    \bk;\\ -E_c ' > \epsilon_{n,\bk} \geq -E_c}} \phi_{\sigma n \bk}^* (\br) \phi_{\sigma n \bk} (\br ') \right) \hat{\psi}^{< \dagger} (\br ') \\
    &= \hat{\psi}^{< \dagger} (\br) \left(  \sum_{\substack{\sigma, n, 
    \bk;\\ E_c ' < \epsilon_{n,\bk} \leq E_c}} \phi_{\sigma n \bk} (\br) \phi_{\sigma n \bk}^* (\br ')  \right) \hat{\psi}^{<} (\br ') + \hat{\psi}^{< \dagger} (\br ') \left(  \sum_{\substack{\sigma, n, 
    \bk;\\ -E_c ' > \epsilon_{n,\bk} \geq -E_c}} -\phi_{\sigma n \bk}^* (\br) \phi_{\sigma n \bk} (\br ') \right) \hat{\psi}^{<} (\br) 
\end{align*}
where the minus sign in the second line comes from the exchange the two fermionic operators and the constant arising from the exchange is omitted. Then, the e-e interaction $\hat{V}_{\text{int}}$ in the lower energy window delimited by $E_c '$ is
\begin{equation}
    \hat{V}_{\text{int}} = \frac{1}{2} \int d^2 \br d^2 \br ' V_c(\br - \br ') \hat{\rho}^{<} (\br) \hat{\rho}^{<} (\br ') + \frac{1}{2} \int d^2 \br d^2 \br ' V_c(\br - \br ') \hat{\psi}^{< \dagger} (\br) \mathcal{F}(\br, \br ') \hat{\psi}^{<} (\br ') 
    \label{eq:vint_RG}
\end{equation}
with 
\begin{equation}
   \mathcal{F}(\br, \br ') = \sum_{\substack{\sigma, n, 
    \bk;\\ E_c ' < |\epsilon_{n,\bk}| \leq E_c}} \text{sign} (\epsilon_{n,\bk}) \phi_{\sigma n \bk} (\br) \phi_{\sigma n \bk}^* (\br ').
    \label{eq:Frrp}
\end{equation}

\subsection{Evaluation of the correction to the non-interacting Hamiltonian from the fast modes}
In the following, we set $\hbar=1$ for the simplicity in mathematical expressions. Note that $\mathcal{F}(\br, \br ')$ has the structure of the residue of the Green's function $\hat{G}(z) = (z - \hat{H}_0)^{-1}$ taking only the valley $\mu=+1$ part in $\hat{H}_0=\hat{H}_G + \hat{H}_{G-S}$ [see Eqs.~\eqref{eq:H_Gonly} and \eqref{eq:hamGS_final}], namely
\begin{equation}
    \mathcal{F}(\br, \br ') = \oint_\mathcal{C} \frac{dz}{2 \pi i} \bra{\br} \hat{G}(z) \ket{\br'}
\end{equation}
where the contour $\mathcal{C}$ encloses the $z$-plane real line segment $[-E_c, -E_c']$ in the clockwise, and segment $[E_c ', E_c ]$ in the counterclockwise, sense.
As long as $E_c '$ dominates over all other energy scales such as $\widetilde{U}_d (\bm{G_0})$ and $v_F G_0$ with $\bm{G_0}$ denoting the primitive reciprocal vector of the underlying superlattice, the dominant contribution to the contour integral can be evaluated perturbatively using $\hat{G}(z) \approx \hat{G}_0(z) + \hat{G}_0(z) \hat{H}_{G-S} \hat{G}_0(z) + \mathcal{O}\left( \widetilde{U}_d^2 (\bm{G_0})/E_c'^2,  v_F^2 G_0^2/E_c'^2 \right)$ with $\hat{G}_0 (z) = (z - \hat{H}_G)^{-1}$.

It is easier to calculate the Green's function in the plane wave basis $\ket{\bk}$ \begin{equation}
   \mathcal{F}(\br, \br ') = \int \frac{d^2 k d^2 k'}{(2 \pi)^4} e^{i (\bk \cdot \br -  \bk ' \cdot \br ' )} \oint_\mathcal{C} \frac{dz}{2 \pi i} \bra{\bk} \hat{G}(z) \ket{\bk'} 
\end{equation}
with
\begin{align}
    \ket{\br} &= \int \frac{d^2 k}{(2 \pi)^2} e^{-i \bk \cdot \br} \ket{\bk} \\
    \bracket{\br}{\br '} &= \delta^{(2)} (\br - \br') \\
    \bracket{\bk}{\bk '} &= (2 \pi)^2 \delta^{(2)} (\bk - \bk')
\end{align}
where $\delta^{(2)}(\bm{x})$ is the 2D Dirac distribution. In the plane wave basis, the evaluation of Green's functions is straightforward
\begin{align}
    \bra{\bk} \hat{G}_0(z) \ket{\bk'} &= (2 \pi)^2 \delta^{(2)} (\bk - \bk') \frac{1}{2} \sum_{\lambda=\pm} \frac{1-\lambda \frac{\bk}{k} \cdot \bm{\sigma}}{z - \lambda v_F k} \\
    \bra{\bk} \hat{G}_0(z) \hat{H}_{G-S} \hat{G}_0(z) \ket{\bk'} &= (2 \pi)^2 \delta^{(2)} (\bk - \bk' + \bm{G}) \frac{1}{4} \sum_{\bm{G}} \widetilde{U}_d (\bm{G}) \sum_{\lambda,\lambda' = \pm} \frac{\left(1-\lambda \frac{\bk}{k} \cdot \bm{\sigma} \right) \left(1-\lambda' \frac{\bk+\bm{G}}{|\bk + \bm{G}|} \cdot \bm{\sigma} \right)}{ \left( z - \lambda v_F k \right) \left( z - \lambda ' v_F |\bk+\bm{G}| \right)}.
\end{align}
Then, the contour integral can be easily done:
\begin{align}
     \oint_\mathcal{C} \frac{dz}{2 \pi i} \bra{\br} \hat{G}_0(z) \ket{\br'} &= \int_{ E_c ' < v_F k \leq E_c} \frac{d^2 k}{(2 \pi)^2} e^{i \bk \cdot (\br - \br ')} \frac{\bk}{k} \cdot \bm{\sigma} \\
    \oint_\mathcal{C} \frac{dz}{2 \pi i} \bra{\br} \hat{G}_0(z) \hat{H}_{G-S} \hat{G}_0(z) \ket{\br'} &= \int_{ E_c ' < v_F k \leq E_c} \frac{d^2 k}{(2 \pi)^2} e^{i \bk \cdot (\br - \br ') - i \bm{G} \cdot \br '} \frac{1}{4} \sum_{\bm{G}} \widetilde{U}_d (\bm{G}) \mathcal{I}(\bk,\bm{G}) \\
    \mathcal{I}(\bk,\bm{G}) &= \frac{2}{v_F k + v_F |\bk + \bm{G}|} \left( 1 - \frac{\bk \cdot (\bk + \bm{G})}{k|\bk + \bm{G}|} + \frac{i \sigma_z (\bk \times \bm{G}) \cdot \bm{\hat{z}}}{k|\bk + \bm{G}|} \right).
\end{align}

\subsection{Renormalization group flow equations}

Now we only have to insert the previous results into the second term in Eq.~\eqref{eq:vint_RG} to derive the RG equations for $v_F$ and $\widetilde{U}_d(\bm{G})$. Let us compute first the integral for $\bra{\br} \hat{G}_0(z) \ket{\br'}$. After writing the 2D Coulomb potential in Fourier space $\widetilde{V}_{\text{2D}}(\bm{q}) = e^2 / 2 \epsilon_0 \epsilon_r q$, we have
\begin{align*}
      &\quad \frac{1}{2} \int d^2 \br d^2 \br ' V_c(\br - \br ') \oint_\mathcal{C} \hat{\psi}^{< \dagger} (\br) \frac{dz}{2 \pi i} \bra{\br} \hat{G}_0(z) \ket{\br'} \hat{\psi}^{<} (\br ') \\
      &= \int \frac{d^2 q}{(2 \pi)^2} \hat{\widetilde{\psi}}^{< \dagger} (\bm{q}) \underbrace{\left( \int_{ E_c ' < v_F k \leq E_c} \frac{d^2 k}{(2 \pi)^2} \frac{e^2}{4 \epsilon_0 \epsilon_r |\bm{q}-\bk|} \frac{\bk}{k} \cdot \bm{\sigma} \right)}_{\text{(A)}} \hat{\widetilde{\psi}}^{<} (\bm{q}) 
\end{align*}
with $\hat{\widetilde{\psi}}^{<} (\bm{q})$ is the Fourier transform of $\hat{\psi}^{<} (\br)$. Since $v_F q \ll E_c'$, we can Taylor expand (A) in terms of $q/k$. The leading order reads
\begin{equation}
    \text{(A)} =  \frac{e^2}{16 \pi \epsilon_0 \epsilon_r} \log \left( \frac{E_c}{E_c'} \right) \,\bm{q} \cdot \bm{\sigma}.
\end{equation}
Therefore, the RG equation reads
\begin{equation}
    \frac{d v_F}{ d \log E_c} = - \frac{e^2}{16 \pi \epsilon_0 \epsilon_r}.
\end{equation}
Actually, we find the famous result of the Fermi velocity renormalization in graphene due to the e-e interactions. 

In the same way, we calculate the integral for $\bra{\br} \hat{G}_0(z) \hat{H}_{G-S} \hat{G}_0(z) \ket{\br'}$:
\begin{align*}
      &\quad \frac{1}{2} \int d^2 \br d^2 \br ' V_c(\br - \br ') \oint_\mathcal{C} \hat{\psi}^{< \dagger} (\br) \frac{dz}{2 \pi i} \bra{\br} \hat{G}_0(z) \ket{\br'} \hat{\psi}^{<} (\br ') \\
      &= \int \frac{d^2 q}{(2 \pi)^2} \hat{\widetilde{\psi}}^{< \dagger} (\bm{q}-\bm{G}) \underbrace{\left( \int_{ E_c ' < v_F k \leq E_c} \frac{d^2 k}{(2 \pi)^2} \widetilde{V}_{\text{2D}}(\bm{q}-\bk-\bm{G}) \frac{1}{8} \sum_{\bm{G}} \widetilde{U}_d (\bm{G}) \mathcal{I}(\bk,\bm{G})  \right)}_{\text{(B)}} \hat{\widetilde{\psi}}^{<} (\bm{q}).
\end{align*}
Since $v_F q, v_F G \ll E_c'$, we can Taylor expand (B) in terms of $q/k$ and $G/k$ (considered as if they have the same order of magnitude). The leading order reads
\begin{equation}
    \text{(B)} = \frac{e^2}{16 \epsilon_0 \epsilon_r} G^2 \left( \frac{1}{E_c'} - \frac{1}{E_c }\right) + \mathcal{O} \left(\frac{v_F^3 q^3}{E_c'^3},\frac{v_F^3 G^3}{E_c'^3}\right),
\end{equation}
which can be neglected under the first-order RG procedure, namely
\begin{equation}
    \frac{d \widetilde{U}_d(\bm{G})}{d \log E_c} = 0.
\end{equation}

In summary, we have shown that the Fermi velocity in graphene $v_F$ is renormalized by the e-e Coulomb interaction in the standard way while the superlattice potential $U_d(\br)$ keep its value unchanged. In our numerical study of e-e interactions, we use the renormalized Fermi velocity $v_F^*$ in the Hartree-Fock calculations, where we have to take a cut-off $n_{\text{cut}}$ to the number of bands, to include the contributions from the higher energy bands outside the cut-off. Technically, we use
\begin{equation}
v_F^* = v_F \left(1 + \frac{e^2}{16 \pi \epsilon_0 \epsilon_r v_F} \log \left( \frac{L_s}{n_{\text{cut}} a_0} \right) \right)  
\end{equation}
where $L_s$ and $a_0$ are the lattice constant of the superlattice of $U_d(\br)$ and the carbon-carbon bond length in graphene, respectively. Here, the ratio $L_s/n_{\text{cut}} a_0$ plays the role of $E_c/E_c'$.

\section{S4. Hartree-Fock approximations to electron-electron interactions}

\setlabel{S4}{sec:HF_gr_crocl}
The derivation shown in this section is inspired from Ref.~\onlinecite{zhang_prl2022}. We consider the Coulomb interactions in graphene
\begin{equation}
\hat{V}_\text{int}=\frac{1}{2}\int d^2 r  d^2 r' \sum _{\sigma, \sigma '} \hat{\psi}_\sigma ^{\dagger}(\br)\hat{\psi}_{\sigma '}^{\dagger}(\br ') V_\text{int} (|\br -\br '|) \hat{\psi}_{\sigma '}(\br ') \hat{\psi}_{\sigma}(\br)
\label{eq:coulomb}
\end{equation}
where $\hat{\psi}_{\sigma}(\br)$ is real-space electron annihilation operator at $\br$ with spin $\sigma$. This interaction can be written as 
\begin{equation}
\hat{V}_\text{int}=\frac{1}{2}\sum _{i i' j j'}\sum _{\alpha \alpha '\beta \beta '}\sum _{\sigma \sigma '} \hat{c}^{\dagger}_{i, \sigma \alpha}\hat{c}^{\dagger}_{i', \sigma ' \alpha '} V^{\alpha \beta \sigma , \alpha ' \beta ' \sigma '} _{ij,i'j'}\hat{c}_{j', \sigma ' \beta '} \hat{c}_{j, \sigma \beta}\;,
\end{equation}
where
\begin{align}
V^{\alpha \beta \sigma , \alpha ' \beta ' \sigma '} _{ij,i'j'}=\int d^2 r d^2 r'  & V_\text{int} (|\br -\br '|) \,\phi ^*_\alpha (\mathbf{r}-\mathbf{R}_i-\tau _\alpha)\,\phi _\beta (\mathbf{r}-\mathbf{R}_j-\tau _\beta) \phi^*_{\alpha  '}(\br-\mathbf{R}_i'-\bm{\tau}_{\alpha '})\phi _{\beta  '}(\br-\mathbf{R}_j'-\bm{\tau} _{\beta '}) \nonumber \\
&\times \chi ^{\dagger}_\sigma \chi ^{\dagger}_{\sigma '}\chi _{\sigma '}\chi _{\sigma} .
\end{align}
Here $i$, $\alpha$, and $\sigma$ refer to Bravis lattice vectors, layer/sublattice index, and spin index. $\phi$ is Wannier function and $\chi$ is the two-component spinor wave function. We further assume that the "density-density" like interaction is dominant in the system, i.e., $V^{\alpha \beta \sigma , \alpha ' \beta ' \sigma '} _{ij,i'j'}\approx V^{\alpha \alpha \sigma , \alpha ' \alpha ' \sigma '} _{ii,i'i'}\equiv V_{i \sigma \alpha  ,i' \sigma ' \alpha '}$,  then the Coulomb interaction is simplified to
\begin{align}
\hat{V}_\text{int}=&\frac{1}{2}\sum _{i i'}\sum _{\alpha \alpha '}\sum _{\sigma \sigma '}\hat{c}^{\dagger}_{i, \sigma \alpha}\hat{c}^{\dagger}_{i', \sigma' \alpha} V_{i\sigma \alpha, i' \sigma ' \alpha '}\hat{c}_{i', \sigma ' \alpha '}\hat{c}_{i, \sigma \alpha} \nonumber \\
=&\frac{1}{2}\sum _{i\alpha \neq i'\alpha '}\sum _{\sigma \sigma '}\hat{c}^{\dagger}_{i, \sigma \alpha} \hat{c}^{\dagger}_{i', \sigma ' \alpha '} V_{i\alpha,i'\alpha '}\hat{c}_{i', \sigma ' \alpha '}\hat{c}_{i, \sigma \alpha} \nonumber \\
&+\sum _{i\alpha} U_0 \hat{c}^{\dagger}_{i,\uparrow \alpha} \hat{c}^{\dagger}_{i,\downarrow\alpha}\hat{c}_{i,\downarrow\alpha}\hat{c}_{i,\uparrow \alpha}
\end{align}
Here we can see that the Coulomb interaction can be divided into intersite Coulomb interaction and on-site Coulomb interaction. Given that the electron density is low ($10^{11}$ cm$^{-2}$), i.e., a few electrons per supercell, the chance that two electrons meet at the same atomic site is very low. The Coulomb correlations between two electron are mostly contributed by the inter-site Coulomb interactions. Therefore, the on-site Hubbard interaction has been neglected in our calculations. 

In order to model the screening effects to the e-e Coulomb interactions from the dielectric environment, we introduce the double-gate screening form of $V_\text{int}$, whose Fourier transform is expressed in Eq.~\eqref{eq:V_doublegate}.
Since we are interested in the low-energy bands, the intersite Coulomb interactions can be divided into the intra-valley term and the inter-valley term. The intra-valley term $\hat{V}^{\text{intra}}$ can be expressed as
\begin{equation}
\hat{V}^{\rm{intra}}=\frac{1}{2N_s}\sum_{\alpha\alpha '}\sum_{\mu\mu ',\sigma\sigma '}\sum_{\bk \bk ' \bq} V_{\rm{int}}(\bq)\,
\hat{c}^{\dagger}_{\sigma \mu \alpha}(\bk+\bq) \hat{c}^{\dagger}_{\sigma' \mu ' \alpha '}(\bk ' - \bq) \hat{c}_{\sigma ' \mu ' \alpha '}(\bk ')\hat{c}_{\sigma \mu \alpha}(\bk)\;,
\label{eq:h-intra}
\end{equation}
with $N_s$ is the total number of the superlattice's sites. The inter-valley term $\hat{V}^{\rm{inter}}$ is expressed as
\begin{equation}
\hat{V}^{\rm{inter}}=\frac{1}{2N_s}\sum_{\alpha\alpha '}\sum_{\mu ,\sigma\sigma '}\sum_{\mathbf{k} \mathbf{k} '\mathbf{q}} V_{\rm{int}}(\vert\mathbf{K}-\mathbf{K}'\vert)\, \hat{c}^{\dagger}_{\sigma \mu \alpha}(\bk+\bq) \hat{c}^{\dagger}_{\sigma'  -\mu \alpha '}(\bk' - \bq) \hat{c}_{\sigma' \mu \alpha '}(\bk') \hat{c}_{\sigma  -\mu \alpha}(\bk)\;.
\label{eq:h-inter}
\end{equation}
$\hat{V}^{\rm{intra}}$ includes the Coulomb scattering processes of two electrons created and annihilated in the same valley, and $\hat{V}^{\rm{inter}}$ includes the processes that two electrons are created in $\mu$ and $-\mu$ and get annihilated in $-\mu$ and $\mu$ valleys. Here the atomic wavevector $\mathbf{k}$ is expanded around the valley $K^{\mu}$ in the big Brillouin zone of graphene, which can be decomposed as $\mathbf{k}=\btk+\mathbf{G}$, where $\btk$ is the superlattice wavevector in the superlattice Brillouin zone, and $\mathbf{G}$ denotes a superlattice reciprocal lattice vector. 

The electron annihilation operator can be transformed from the original basis to the band basis:
\begin{equation}
\hat{c}_{\sigma\mu\alpha}(\bk)=\sum_n C_{\sigma \mu \alpha \mathbf{G},n}(\btk)\,\hat{c}_{\sigma \mu,n}(\btk)\;,
\label{eq:transform}
\end{equation}
where $C_{\sigma \mu \alpha \mathbf{G},n}(\btk)$ is the expansion coefficient in the $n$-th Bloch eigenstate at $\btk$ of valley $\mu$: 
\begin{equation}
\ket{\mu, n; \btk}=\sum_{\alpha \mathbf{G}}C_{\sigma\mu \alpha \mathbf{G},n}(\btk)\,\ket{ \sigma, \mu, \alpha, \mathbf{G}; \btk }\;.
\end{equation}
We note that the non-interacting Bloch functions are spin degenerate due to the separate spin rotational symmetry ($SU(2)\otimes SU(2)$ symmetry) of each valley. Using the transformation given in Eq.~(\ref{eq:transform}), the intra- and inter-valley Coulomb interaction can be written in the band basis
\begin{align}
\hat{V}^{\rm{intra}}=&\frac{1}{2N_s}\sum _{\btk \btk'\btq}\sum_{\substack{\mu\mu' \\ \sigma\sigma'}}\sum_{\substack{nm\\ n'm'}} \left(\sum _{\mathbf{Q}}\,V_{\rm{int}} (\mathbf{Q}+\btq)\,\Omega^{\mu \sigma,\mu'\sigma'}_{nm,n'm'}(\btk,\btk',\btq,\mathbf{Q})\right) \nonumber \\
&\times \hat{c}^{\dagger}_{\sigma\mu,n}(\btk+\btq) \hat{c}^{\dagger}_{\sigma'\mu',n'}(\btk'-\btq) \hat{c}_{\sigma'\mu',m'}(\btk') \hat{c}_{\sigma\mu,m}(\btk)
\label{eq:Hintra-band}
\end{align}
and
\begin{align}
\hat{V}^{\rm{inter}}=&\frac{1}{2N_s}\sum _{\btk \btk'\btq}\sum_{\substack{\sigma\sigma' \\ \mu}}\sum_{\substack{nm\\ n'm'}}\left(\sum _{\mathbf{Q}}\,V_{\rm{int}}(|\mathbf{K}-\mathbf{K}'|)\,\widetilde{\Omega}^{\mu, \sigma\sigma'}_{nm,n'm'}(\btk,\btk',\btq,\mathbf{Q})\right) \nonumber\\ 
&\times \hat{c}^{\dagger}_{\sigma\mu,n}(\btk+\btq) \hat{c}^{\dagger}_{\sigma'-\mu,n'}(\btk'-\btq) \hat{c}_{\sigma'\mu,m'}(\btk') \hat{c}_{\sigma-\mu,m}(\btk)
\label{eq:Hinter-band}
\end{align}
where the form factors $\Omega ^{\mu \sigma,\mu'\sigma'}_{nm,n'm'}$ and $\widetilde{\Omega}^{\mu, \sigma\sigma'}_{nm,n'm'}$ are written respectively as
\begin{equation}
\Omega ^{\mu \sigma,\mu'\sigma'}_{nm,n'm'}(\btk,\btk',\btq,\mathbf{Q})
=\sum _{\alpha\alpha'\mathbf{G}\mathbf{G}'}C^*_{\sigma\mu\alpha\mathbf{G}+\mathbf{Q},n}(\btk+\btq) C^*_{\sigma'\mu'\alpha'\mathbf{G}'-\mathbf{Q},n'}(\btk'-\btq)C_{\sigma'\mu'\alpha'\mathbf{G}',m'}(\btk')C_{\sigma\mu\alpha\mathbf{G},m}(\btk)
\end{equation}
and
\begin{equation}
\widetilde{\Omega}^{\mu, \sigma\sigma'}_{nm,n'm'}(\btk,\btk',\btq,\mathbf{Q})
=\sum _{\alpha\alpha'\mathbf{G}\mathbf{G}'}C^*_{\sigma\mu\alpha\mathbf{G}+\mathbf{Q},n}(\btk+\btq) C^*_{\sigma'-\mu\alpha'\mathbf{G}'-\mathbf{Q},n'}(\btk'-\btq)C_{\sigma'\mu \alpha'\mathbf{G}',m'}(\btk')C_{\sigma-\mu\alpha\mathbf{G},m}(\btk) \;.
\end{equation}

We make Hartree-Fock approximation to Eq.~\eqref{eq:Hintra-band} and Eq.~\eqref{eq:Hinter-band} such that the two-particle Hamiltonian is decomposed into a superposition of the Hartree and Fock single-particle Hamiltonians, where the Hartree term is expressed as
\begin{equation}
\begin{split}
\hat{V}_H^{\rm{intra}}=&\frac{1}{2N_s}\sum _{\btk \btk'}\sum _{\substack{\mu\mu'\\ \sigma\sigma'}}\sum_{\substack{nm\\ n'm'}}\left(\sum _{\mathbf{Q}} V_{\rm{int}} (\mathbf{Q}) \, \Omega^{\mu \sigma,\mu'\sigma'}_{nm,n'm'}(\btk,\btk',0,\mathbf{Q})\right)\\
&\times \left(\langle \hat{c}^{\dagger}_{\sigma\mu,n}(\btk)\hat{c}_{\sigma\mu,m}(\btk)\rangle \hat{c}^{\dagger}_{\sigma'\mu',n'}(\btk')\hat{c}_{\sigma'\mu',m'}(\btk') + \langle \hat{c}^{\dagger}_{\sigma'\mu',n'}(\btk')\hat{c}_{\sigma'\mu',m'}(\btk')\rangle \hat{c}^{\dagger}_{\sigma\mu,n}(\btk)\hat{c}_{\sigma\mu,m}(\btk)\right)
\end{split}
\end{equation}
and
\begin{equation}
\begin{split}
\hat{V}_H^{\rm{inter}}=&\frac{1}{2N_s}\sum _{\btk \btk'}\sum _{\substack{\sigma\sigma'\\ \mu}}\sum_{\substack{nm\\ n'm'}}\left(\sum _{\mathbf{Q}} V_{\rm{int}}(|\mathbf{K}-\mathbf{K}'|) \, \widetilde{\Omega}^{\mu, \sigma\sigma'}_{nm,n'm'}(\btk,\btk',0,\mathbf{Q})\right)\\
&\times \left(\langle \hat{c}^{\dagger}_{\sigma\mu,n}(\btk)\hat{c}_{\sigma-\mu,m}(\btk)\rangle \hat{c}^{\dagger}_{\sigma'-\mu,n'}(\btk')\hat{c}_{\sigma'\mu,m'}(\btk') + \langle \hat{c}^{\dagger}_{\sigma'-\mu,n'}(\btk')\hat{c}_{\sigma'\mu,m'}(\btk')\rangle \hat{c}^{\dagger}_{\sigma\mu,n}(\btk)\hat{c}_{\sigma-\mu,m}(\btk)\right) \;.
\end{split}
\end{equation}
The Fock term is expressed as:
\begin{align*}
\hat{V}_F^{\rm{intra}}=&-\frac{1}{2N_s}\sum _{\btk \btk'}\sum _{\substack{\mu\mu'\\ \sigma\sigma'}}\sum_{\substack{nm\\ n'm'}}\left(\sum _{\mathbf{Q}} V_{\rm{int}} (\btk’-\btk+\mathbf{Q}) \, \Omega^{\mu \sigma,\mu'\sigma'}_{nm,n'm'}(\btk,\btk',\btk’-\btk,\mathbf{Q})\right)\\
&\times \left(\langle \hat{c}^{\dagger}_{\sigma\mu,n}(\btk')\hat{c}_{\sigma'\mu',m'}(\btk')\rangle \hat{c}^{\dagger}_{\sigma'\mu',n'}(\btk)\hat{c}_{\sigma\mu,m}(\btk) + \langle \hat{c}^{\dagger}_{\sigma'\mu',n'}(\btk)\hat{c}_{\sigma\mu,m}(\btk)\rangle \hat{c}^{\dagger}_{\sigma\mu,n}(\btk')\hat{c}_{\sigma'\mu',m'}(\btk')\right)\;.   
\end{align*}
and
\begin{align*}
\hat{V}_F^{\rm{inter}}=&-\frac{1}{2N_s}\sum _{\btk \btk'}\sum _{\substack{\sigma \sigma ' \\ \mu}}\sum_{\substack{nm\\ n' m'}}\left(\sum _{\mathbf{Q}} V_{\rm{int}}(|\mathbf{K}-\mathbf{K}'|) \, \widetilde{\Omega}^{\mu, \sigma\sigma'}_{nm,n' m'}(\btk,\btk',\btk’-\btk,\mathbf{Q})\right)\\
&\times \left(\langle \hat{c}^{\dagger}_{\sigma\mu,n}(\btk')\hat{c}_{\sigma' \mu, m'}(\btk')\rangle \hat{c}^{\dagger}_{\sigma' -\mu,n'}(\btk)\hat{c}_{\sigma -\mu,m}(\btk) + \langle \hat{c}^{\dagger}_{\sigma' -\mu,n'}(\btk)\hat{c}_{\sigma-\mu,m}(\btk)\rangle \hat{c}^{\dagger}_{\sigma\mu,n}(\btk')\hat{c}_{\sigma' \mu,m'}(\btk')\right)\;.   
\end{align*}

We note that the typical intravalley interaction energy $\sim 240$ meV for $L_s = 50$~\AA \, and $\epsilon_r = 3$; while the intervalley interaction $\sim 30\,$meV, which is one order of magnitudes smaller than the intravalley interaction, thus we neglect the intervalley term [see Eq.~(\ref{eq:h-inter})] in most of our calculations. We also check \textit{a posteriori} that the intervalley Hartree and Fock energies are at least two orders of magnitude smaller than their intravalley counterpart. However, the intervalley interaction is crucial to lift the degeneracy between many-body ground state, as shown in the following section.


\newpage

\section{S5. Results of Hartree-Fock calculations}
\setlabel{S5}{sec:HF_results}

In this section, we gather the results of Hartree-Fock calculations including Hartree-Fock single-particle spectra and distributions of Berry curvature in the first Brillouin zone for $L_s = 50$, 200, 600 \AA. 

\begin{figure}[htb]
    \centering
    \includegraphics[width=0.8\textwidth]{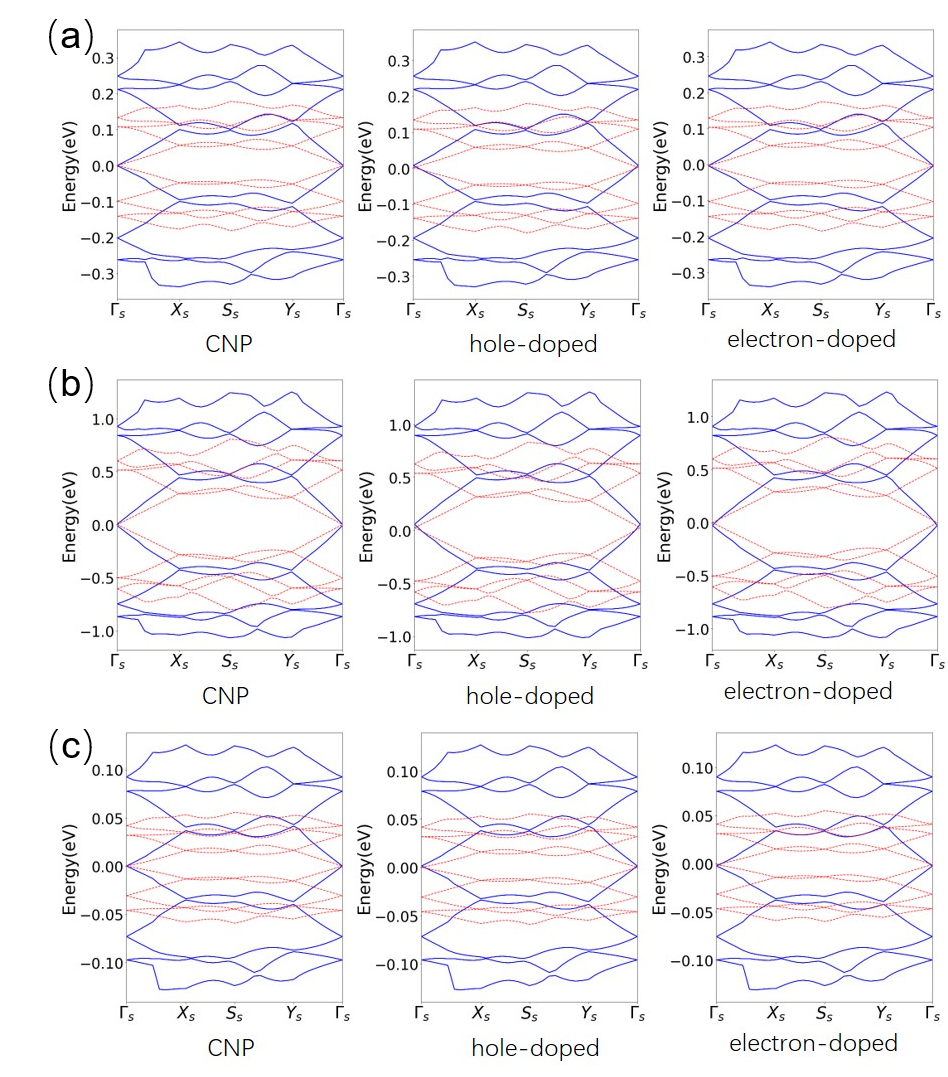}
\caption{{Hartree-Fock single-particle spectra for three different dopings with $r=1.2$ for (a) $L_s=50$ \AA, (b) $L_s=200$ \AA\,and (c) $L_s=600$ \AA.}}
\label{fig:HF_spec_Ls_r}
\end{figure}

First, we show the Hartree-Fock single-particle spectrum with a superlattice potential with $r=1.2$ of $L_s = 50$, 200, 600 \AA\  in Fig.~\ref{fig:HF_spec_Ls_r}. Here, we use $n_\text{cut}=5$ and study three types of doping: CNP ($\nu=0$), slight hole doping ($\nu=-0.003$) and slight electron doping ($\nu=+0.003$). As you can see from Table \ref{tab:gap_vf_supp} and the Hartree-Fock single-particle spectra, the results of a slightly electron-doped system is similar to those for a slightly hole-doped one. Note that we include only intravalley Coulomb interactions in these calculations. As shown in the following, the role of intervalley Coulomb interactions is merely to lift the ground state degeneracy and favor the $\sigma_z$-state.

\begin{table}[h]
    \caption{Parameters extracted from the Hartree-Fock single-particle spectra: gap opened at the CNP ($\nu=0$) and the ratio between interaction-renormalized Fermi velocity $v_F^*$ and the non-interacting one $v_F$ for different $L_s = 50$, 200, 600 \AA \  with fixed $r=1.2$.}
        \label{tab:gap_vf_supp}
\renewcommand{\arraystretch}{1.2}
    \centering
    \begin{tabular}{c|c|c|c}
        \hline 
         $L_s$(\AA) & 50 & 200 & 600 \\ 
         \hline \hline
         Gap at $\nu=0.0$ (meV) & 17 & 1.7 & 0.15 \\
         \hline
         $v_F^*/v_F$ at $\nu=-0.003$  & 2.1 & 1.8 & 1.7\\
         \hline
         $v_F^*/v_F$ at $\nu=+0.003$  & 2.1 & 1.8 & 1.7\\
         \hline
    \end{tabular} 

\end{table}

\begin{figure}[htb]
    \centering
    \includegraphics[width=0.95\textwidth]{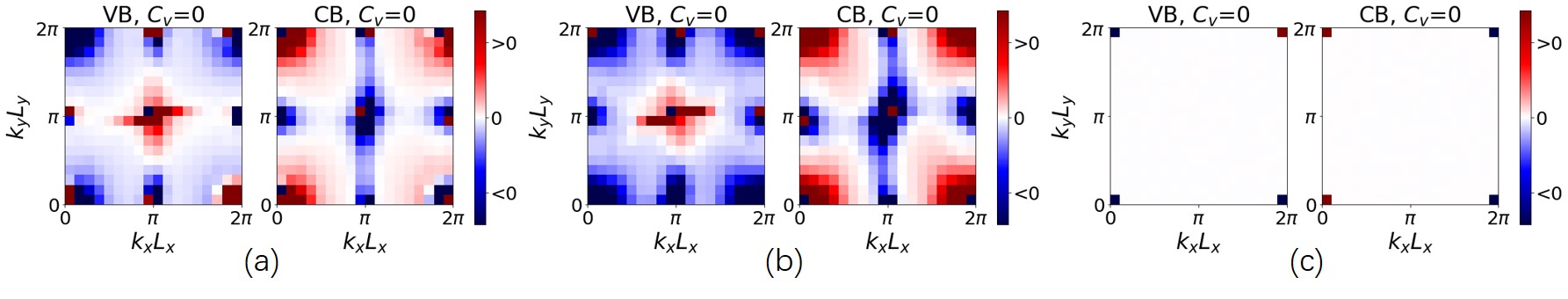}
    \caption{Distributions of Berry curvature in the first Brillouin zone of $r=1.2$ for $L_s =$ (a) 50 \AA, (b) 200 \AA, (c) 600 \AA}
    \label{fig:berry_HF_ncut5}
\end{figure}

Then, we show in Fig.~\ref{fig:berry_HF_ncut5} the distributions of Berry curvature in the first Brillouin zone of $r=1.2$ for $L_s = 50$, 200, 600 \AA. Here, $n_\text{cut}=5$. 

\begin{figure}[htb]
    \centering
    \includegraphics[width=0.6\textwidth]{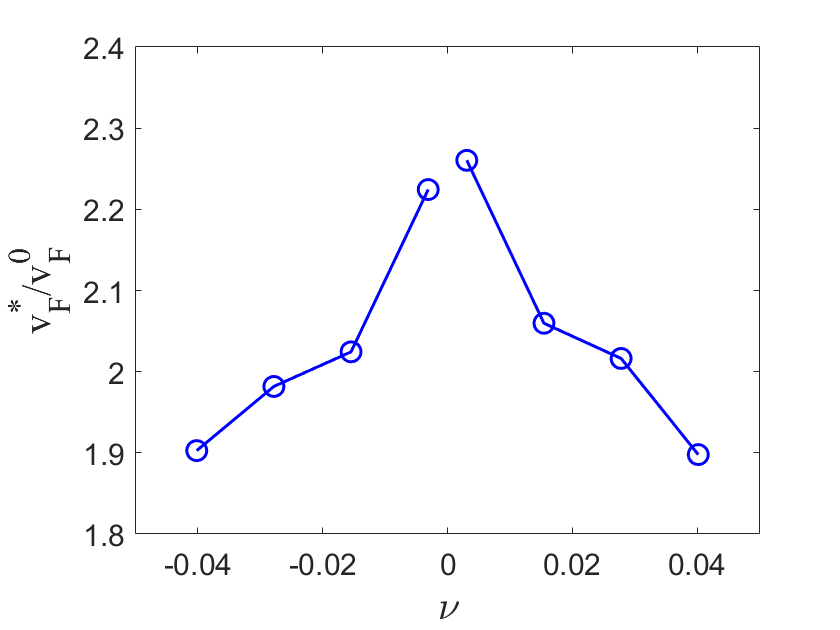}
    \caption{The ratio between the interacting Fermi velocity ($v_F^{*}$) and the non-interacting one of free-standing graphene ($v_F^{0}$) \textit{vs.} the filling factor ($\nu$) with the superlattice constant $L_s=50\,$\AA, and a background dielectric constant $\epsilon_r=3$.}
    \label{fig:vFvsnu}
\end{figure}

We further calculate the interacting electronic structure of graphene at different filling factors (denoted by $\nu$) away from the CNP, as presented in Fig.~\ref{fig:vFvsnu}. We find that the ground state is generally gapless at nonzero fillings, and the Fermi velocity enhances as the chemical potential approaches the charge neutrality point ($\vert\nu\vert\to 0$). In particular, the Fermi velocity increases from $v_F^*=1.9v_F$ at $\nu=-0.04$, to $v_F^*=2.2v_F$ at $\nu=-0.003$.

Now we show the effect of intervalley Coulomb interactions by comparing the total energy of $\sigma_z$-state with $\tau_z \sigma_z$-state for different $L_s = 50$, 200, 600\,\AA \  with fixed $r=1.2$. We calculate the difference (always negative) between them and see how it changes when we include the intervalley Coulomb interactions. Here, we use $n_\text{cut}=3$. 

As you can see from Table~\ref{tab:intervalley_lift}, the energy difference between the total energy of $\sigma_z$-state and $\tau_z \sigma_z$-state is enhanced by two orders of magnitude for $L_s=50$ and 200 \AA. However, the energy difference for $L_s=600$ \AA \  does not benefit anything from intervalley interactions. 

\begin{table}[h]
\renewcommand{\arraystretch}{1.2}
    \centering
            \caption{Difference between the total energy of the $\sigma_z$-state and $\tau_z \sigma_z$-state, with or without intervalley interactions, for $L_s = 50$, 200, 600 \AA \  with fixed $r=1.2$.}
    \begin{tabular}{c|c|c|c}
        \hline 
         $L_s$(\AA) & 50 & 200 & 600 \\ 
         \hline \hline
         $\Delta E$ with only intravalley ($\mu$eV) & -0.008 & -0.005 & -0.05 \\
         \hline
         $\Delta E$ with intra- and inter-valley ($\mu$eV) & -1.6 & -0.1 & -0.03 \\
         \hline
    \end{tabular} 
        \label{tab:intervalley_lift}
\end{table}

We also have performed Hartree-Fock calculations on a triangular lattice including three valence and three conduction bands ($n_{\rm{cut}}=3$) for $L_s = 50$, 200, 600 \AA \  using $18 \times 18$ $k$-mesh in the BZ. As shown in Table \ref{tab:gap_vf_tri_supp}, the results on a triangular lattice are qualitatively the same as those on a rectangular lattice. This ensures that our conclusions are lattice-independent.

\begin{table}[!h]
    \caption{Parameters extracted from the Hartree-Fock single-particle spectra on a triangular lattice: gap opened at the CNP ($\nu=0$) and the ratio between interaction-renormalized Fermi velocity $v_F^*$ and the non-interacting one $v_F$ for different $L_s = 50$, 200, 600 \AA.}
    \label{tab:gap_vf_tri_supp}
\renewcommand{\arraystretch}{1.2}
    \centering
    \begin{tabular}{c|c|c|c}
        \hline 
         $L_s$(\AA) & 50 & 200 & 600 \\ 
         \hline \hline
         Gap at $\nu=0.0$ (meV) & 21 & 1.9 & 0.24 \\
         \hline
         $v_F^*/v_F$ at $\nu=-0.003$  & 2.2 & 1.7 & 1.7\\
         \hline
    \end{tabular} 

\end{table}

\begin{figure}
	\includegraphics[width=0.5\textwidth]{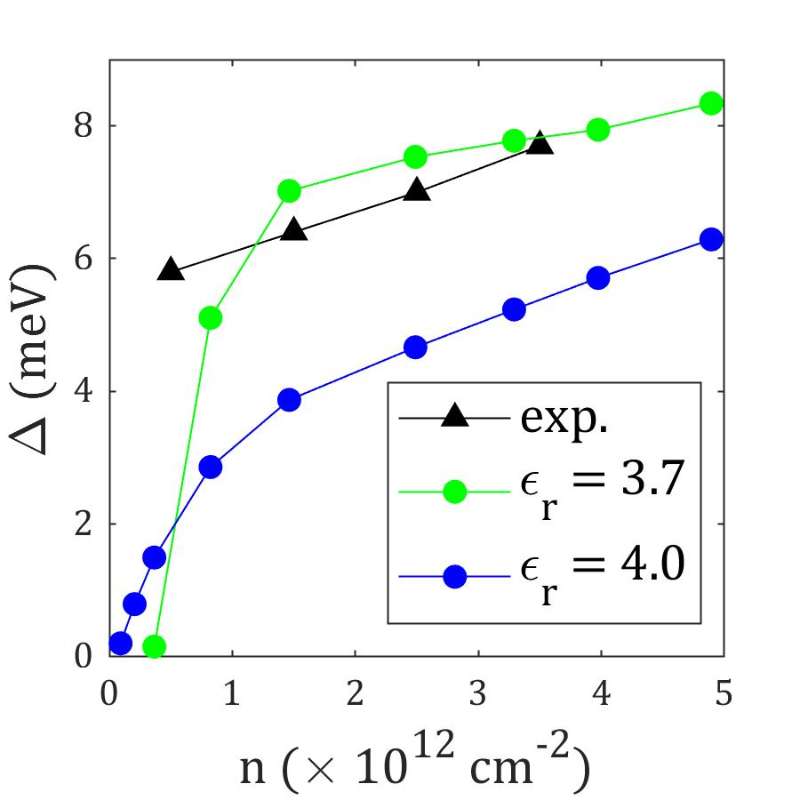}
	\caption{Comparison between the experimentally measured gap and the theoretically calculated gaps in graphene-CrOCl heterostructure.}
	\label{fig:compare_theo_exp}
\end{figure}

In some material, there are several valleys to accommodate the charges transferred from graphene so that we need to consider this valley degeneracy in the superlattice potential as follows:
\begin{equation}
	U_d (\Q) = e^2 g_v\frac{n (\Q) e^{-|\Q|d}} {2 \epsilon_0 \epsilon_r \Omega_0 |\Q|}
	\label{eq:UdQ_gv}
\end{equation}
where $n(\Q)$ is the Fourier transformed charge density per valley at superlattice’s reciprocal vector $\Q$ and the valley degeneracy $g_v=2$ for CrOCl. After including this valley degeneracy, our theoretical results are quantitatively consistent with the experimental data, as shown in Fig. \ref{fig:compare_theo_exp}. More details on experimental measurements are given in the last section Sec.~\ref{sec:exp}.

\section{S6. General coupled bilayer system in graphene-insulator heterostructures}
\setlabel{S6}{sec:couple}

Previous parts have already proved that the presence of a superlattice potential underneath monolayer graphene sheet helps the latter to open a gap at the CNP, and concomitantly enhance the Fermi velocity around the Dirac point. Now we would like to reconsider the assumptions made for the purpose of writing a simplistic Hamiltonian of our coupled bilayer system given by Eqs. \eqref{eq:HG},\eqref{eq:HS}, and \eqref{eq:HG-S}. 

\subsection{Hamiltonians}

First, when the Fermi level in graphene aligns closely to the band edge (say, the conduction band minimum) of the insulating substrate, the density of electrons in the conduction band has been assumed to be so low that electrons spontaneously break translational symmetry and form a charge-ordered superlattice. Second, we have assumed that the quantum degrees of freedom on the substrate are completely frozen so that the only effect (other than dielectric screening) acting by substrate to electrons in graphene is to apply a superlattice potential via the long-range Coulomb interactions, stemming from a non-uniform charge density of the charge-ordered state. Lastly, we have drastically supposed that the Wannier functions and the distribution of charge density at the surface of the substrate is as localized as a Dirac-$\delta$-function like distribution in real space [see Eq.~(\ref{eq:delta-loc})]. These assumptions were good enough if we focus on the physics on the graphene side. In particular, it turns out that the assumption on the extreme localization of the charges and/or Wannier functions [see Eq.~(\ref{eq:delta-loc})] in the substrate is a good starting point. This is because more extended charge distributions only make the Fourier components of the superlattice potential weaker (to be discussed in the following), which would not change our conclusion qualitatively. However, if we are interested in especially the synergistic interplay between graphene and the insulating substrate, we have to consider integrally the graphene-insulator heterostructure as a coupled bilayer system, and treat both layers on equal footing.

In this section, we aim to study, to the best of our efforts, the coupled bilayer system as a whole, and to treat quantum mechanically the electrons both in the graphene layer and the substrate layer. Most of the above assumptions have to be discarded or replaced by a less harsh one. To this end, in addition to the kinetic energy Eq.~(\ref{eq:HG}) and the long-range intralayer Coulomb interactions of graphene Eq.~(\ref{eq:coulomb}), we also need to consider the interacting many-body Hamiltonian for electrons on the surface of the substrate. Without loss of generality, we consider the situation that the conduction band minimum (CBM) of the substrate is charge doped.

Specifically, the non-interacting kinetic Hamiltonian of the low-energy electrons around the CBM of the substrate can be modeled by the following one
\begin{equation}
H^{0}_s=\sum_{\k_s,\sigma}\,\left(\frac{\hbar^2\,k_s^2}{2m^{*}}+E_{\text{CBM}} \right) \, \hdd_{\sigma}(\k_s) \, \hd_{\sigma}(\k_s)\;,
\label{eq:h0-sub}
\end{equation}
where $E_{\text{CBM}}$ is the energy position of CBM with respect to the Fermi level of graphene, and $m^{*}$ is the effective mass. For CrOCl, $m^{*}\approx 1.3 m_e$. $\hd^{(\dagger)}_{\sigma}(\k_s)$ is the annihilation (creation) operator of the low-energy electrons doped to the surface of the substrate, with the wavevector $\k_s$ (expanded around the CBM) in the atomic Brillouin zone of the substrate and spin index $\sigma$. 
Then we consider the long-range Coulomb interactions between electrons at the surface of the substrate:
\begin{equation}
H^{\text{int}}_{\text{sub}}=\frac{1}{2N_s\Omega_s}\sum_{\k_s,\k_s',\q_s} \sum_{\sigma,\sigma'} \,V_{\text{int}}(\q_s)\,\hdd_{\sigma}(\k_s+\q_s) \, \hdd_{\sigma'}(\k_s'-\q_s) \, \hd_{\sigma'}(\k_s')\, \hd_{\sigma}(\k_s)\;,
\label{eq:int-sub}
\end{equation}
where $N_s$ is the total number of \textit{atomic lattice sites} in the substrate layer, $\Omega_s$ is the area of the atomic primitive cell in the substrate layer, and $\k_s,\k_s',\q_s,$ denote the atomic wavevectors in the Brillouin zone of the atomic lattice of the substrate. The Coulomb potential $V_{\text{int}}(\q)$ is given by Eq.~\eqref{eq:V_doublegate}.

We continue to address the interlayer Coulomb interaction, i.e., Eq.~(\ref{eq:HG-S}). We first transform Eq.~(\ref{eq:HG-S}) to \textit{the basis of Wannier functions of the atomic lattices of graphene and the substrate} using the following transformations of the field operators:
\begin{align}
    \hat{\psi}^{\dagger}_{c, \sigma}(\bm{r}) &= \sum_{i,\alpha} \phi_\alpha ^* (\br-\bm{a}_i-\bm{\tau}_\alpha) \chi^{\dagger}_\sigma \cop^{\dagger}_{i,\sigma \alpha} \\
    \hat{\psi}^{\dagger}_{d, \sigma}(\bm{r}) &= \sum_{i,\alpha} \widetilde{\phi} ^* (\br-\mathbf{t}_i) \chi^{\dagger}_\sigma \dop^{\dagger}_{i,\sigma}
    \label{eq:cd-transform2}
\end{align}
where $\mathbf{t}_i$ denotes the \textit{atomic lattice sites} of the substrate, which is different from the presumed superlattice sites $\mathbf{R}_i$ in Eq.~(\ref{eq:cd-transform1}). $\widetilde{\phi}(\br-\mathbf{t}_i)$ and $\phi_\alpha ^* (\br-\bm{a}_i-\bm{\tau}_\alpha)$ denote the atomic-scale Wannier functions for the electrons in the substrate and in the graphene layer, respectively. As already mentioned around Eq.~(\ref{eq:cd-transform1}), $\bm{a}_i$ denotes the atomic lattice sites of graphene, and $\bm{\tau}_{\alpha}$ is the position of sublattice $\alpha$. Using Eq.~(\ref{eq:cd-transform2}), Eq.~(\ref{eq:HG-S}) can be expressed in the atomic Wannier function basis as follows
\begin{equation}
H_{\text{gr-sub}}=\sum_{ij,\alpha}\sum_{\sigma,\sigma'}\,V(\vert\mathbf{a}_i+\bm{\tau}_{\alpha}-\mathbf{t}_j +d\hat{\mathbf{z}}\vert)\,\hcd_{i,\alpha\sigma}\,\hdd_{j,\sigma'}\,\hd_{j,\sigma'}\,\hc_{i,\alpha\sigma}\;
\end{equation}
where $V(\br)$ is the 3D Coulomb potential. Here, we still use approximations similar to Eq.~\eqref{eq:delta-loc} to get this expression except that we now work on the atomic lattice sites for substrate.

One continues to perform the Fourier transform
\begin{align}
&\hc_{i,\alpha\sigma}=\frac{1}{\sqrt{N_c}}\sum_{\k}\,e^{i\k\cdot(\mathbf{a}_i+\mathbf{\tau}_{\alpha})}\,\hc_{\sigma\alpha}(\k)\;\nn
&\hd_{j,\sigma}=\frac{1}{\sqrt{N_s}}\sum_{\k_s}\,e^{i\k_s\cdot\mathbf{t}_j}\,\hd_{\sigma}(\k_s)\;,
\end{align}
 Then the interlayer Coulomb interaction becomes
\begin{align}
&H_{\text{gr-sub}}\;\nn
=&\frac{1}{N_c N_s}\sum_{ij\alpha}\sum_{\sigma\sigma'}\sum_{\k,\k',\k_s,\k_s'}\,V(\vert\mathbf{a}_i+\bm{\tau}_{\alpha}-\mathbf{t}_j +d\hat{\mathbf{z}}\vert)\,e^{i(\k-\k')\cdot(\mathbf{a}_i+\mathbf{\tau}_{\alpha})}\,e^{i(\k_s-\k_s')\cdot\mathbf{t}_j}\,\hcd_{\sigma\alpha}(\k)\,\hdd_{\sigma'}(\k_s)\,\hd_{\sigma'}(\k_s')\,\hc_{\sigma\alpha}(\k')\;\nn
=&\frac{1}{N_c N_s}\sum_{ij\alpha}\sum_{\sigma\sigma'}\sum_{\k,\k',\k_s,\k_s'}\,V(\vert\mathbf{a}_i+\bm{\tau}_{\alpha}-\mathbf{t}_j +d\hat{\mathbf{z}}\vert)\,e^{i(\k-\k')\cdot(\mathbf{a}_i+\mathbf{\tau}_{\alpha}-\mathbf{t}_j)}\,e^{i(\k_s-\k_s'+\k-\k')\cdot\mathbf{t}_j}\,\hcd_{\sigma\alpha}(\k)\,\hdd_{\sigma'}(\k_s)\,\hd_{\sigma'}(\k_s')\,\hc_{\sigma\alpha}(\k')\;.
\end{align}
Let $\k'-\k=\q_s+\mathbf{g}_s$, where $\q_s$ is a wavevector within the substrate's Brillouin zone, and $\mathbf{g}_s$ is the corresponding reciprocal vector, then
$\sum_j\,e^{i(\k_s'-\k_s+\q_s+\g_s)\cdot\mathbf{t}_j}=N_s\delta_{\k_s-\k_s',\q_s+\g_s}$.
%
Thus the interlayer Coulomb interaction becomes
\begin{equation}
H_{\text{gr-sub}}=\frac{1}{N_c}\sum_{\alpha,\sigma,\sigma'}\sum_{\k,\k_s,\q_s,\g_s}\sum_{\widetilde{\mathbf{R}}}
\,V(\vert\widetilde{\mathbf{R}}+d\hat{\mathbf{z}}\vert)\,e^{i(\q_s+\g_s)\cdot\widetilde{\mathbf{R}}}\,\hcd_{\sigma\alpha}(\k)\,\hdd_{\sigma'}(\k_s)\,\hd_{\sigma'}(\k_s-\q_s-\g_s)\,\hc_{\sigma\alpha}(\k+\q_s+\g_s)\;,
\label{eq:inter-coulomb2}
\end{equation}
where $\widetilde{\mathbf{R}}=\mathbf{a}_i+\bm{\tau}_{\alpha}-\mathbf{t}_j$. Then, 
\begin{align}
&\frac{1}{N_c}\sum_{\widetilde{\mathbf{R}}}\,V(\vert\widetilde{\mathbf{R}}+d\hat{\mathbf{z}}\vert)\,e^{i(\q_s+\g_s)\cdot\widetilde{\mathbf{R}}}\;\nn
=&\frac{1}{N_c\Omega_0}\sum_{\widetilde{\mathbf{R}}}\Omega_0\,V(\vert\widetilde{\mathbf{R}}+d\hat{\mathbf{z}}\vert)\,e^{i(\q_s+\g_s)\cdot\widetilde{\mathbf{R}}}\;\nn
=&\frac{1}{S}\int\,d^2\mathbf{r}\,V(\vert\mathbf{r}+d\hat{\mathbf{z}}\vert)\,e^{i(\q_s+\g_s)\cdot\mathbf{r}}\;\nn
=&\frac{e^2\,e^{-\vert\q_s+\g_s\vert\,d}}{2\epsilon_0\epsilon_r\vert\q_s+\g_s\vert\,S}
\label{eq:vq}
\end{align}
where $S=N_c\Omega_c$ is the total area of the system. Plugging Eq.~(\ref{eq:vq}) into Eq.~(\ref{eq:inter-coulomb2}), one obtains
\begin{align}
H_{\text{gr-sub}}&=\frac{1}{S}\sum_{\alpha,\sigma,\sigma'}\sum_{\k,\k_s,\q_s,\g_s}\,\frac{e^2\,e^{-\vert\q_s+\g_s\vert\,d}}{2\epsilon_0\epsilon_r\vert\q_s+\g_s\vert}\,\hcd_{\sigma\alpha}(\k)\,\hdd_{\sigma'}(\k_s)\,\hd_{\sigma'}(\k_s-\q_s-\g_s)\,\hc_{\sigma\alpha}(\k+\q_s+\g_s)\;\nn
&\approx \frac{1}{S}\sum_{\alpha,\sigma,\sigma'}\sum_{\k,\k_s,\q_s}\,\frac{e^2\,e^{-\vert\q_s\vert\,d}}{2\epsilon_0\epsilon_r\vert\q_s\vert}\,\hcd_{\sigma\alpha}(\k)\,\hdd_{\sigma'}(\k_s)\,\hd_{\sigma'}(\k_s-\q_s)\,\hc_{\sigma\alpha}(\k+\q_s)\;\nn
&\approx \frac{1}{S}\sum_{\mu,\alpha,\sigma,\sigma'}\sum_{\k,\k',\q}\,\frac{e^2\,e^{-\vert\q\vert\,d}}{2\epsilon_0\epsilon_r\vert\q\vert}\,\hcd_{\sigma\mu\alpha}(\k)\,\hdd_{\sigma'}(\k')\,\hd_{\sigma'}(\k'-\q)\,\hc_{\sigma\mu\alpha}(\k+\q)
\end{align}
where in the second line of the above equation we only keep the $\mathbf{g}_s=\mathbf{0}$ scattering channels due to the exponential decaying form of the interlayer Coulomb interactions in reciprocal space. In the last line of the above equation, we expand the wavevectors of electrons in the graphene layer around the Dirac point, i.e., let $\k\to\mathbf{K}_{\mu}+\k$, then assign valley index $\mu$ to the annihilation (creation) operators, $\hc^{(\dagger)}_{\sigma\alpha}(\k)\to\hc^{(\dagger)}_{\sigma\mu\alpha}(\k)$. We have also neglected the intervalley Coulomb scattering for the electrons in the graphene layer arising from the interlayer Coulomb interactions, which is an excellent approximation given that the superlattice constant $L_s$ is much larger than graphene's atomic lattice constant $a$. In the last line of the above equation, we have also let $\k_s\to\k'$, $\q_s\to\q$, in the sense that if we are interested in the low-energy states in both the graphene layer and the substrate layer, we do not have to distinguish whether the wavevector is defined in graphene's or substrate's Brillouin zone as those wavevectors are far from reaching the Brillouin zone boundary. Note that the mathematical derivations are for now practically the same as in Sec.~\ref{sec:noninter-gr-crocl}. However, The idea of using atomic lattice $\{ \bm{t}_j\}$ in substrate would make a difference. In short, a charge-ordered state would emerge spontaneously while treating together the intralayer $e$-$e$ Coulomb interactions Eq.~\eqref{eq:int-sub} and the kinetic part Eq.~\eqref{eq:h0-sub}. The spatial distribution of charges is localized, but it should also be smooth on the superlattice scale. We will show the technical details in the next subsection. After collecting all the contributions, we obtain the Hamiltonians Eqs.~(5a-e) in the main text. Comparing with what we have studied in Sec.~S1-S5, the really new flavors we add into the menu are the kinetic energy and Coulomb interactions in the substrate. In the following, we provide technical details on: first, Hartree-Fock calculations on the substrate's side, which offers us charge modulation of the EC state; second, perturbative approach to estimate the stabilizing effect via interlayer Coulomb potential.

\subsection{Hartree-Fock calculations on the coupled bilayer system}

We first start working on the sum of the Hamiltonians Eq.~(5b) and (5d). If electronic-crystal state is formed, electrons will be spontaneously arranged in a  superlattice.
This amounts to fold the plane wave wavevector $\k$ in the mini Brillouin zone of the presumed lattice so that $\k = \kt + \G$, where $\kt$ is within the first mini Brillouin zone and $\G$ is a reciprocal lattice vector of the superlattice. In Fig.~4 of the main text, we have considered the case of a triangular superlattice, which is the actual ground state of Wigner crystal for free 2D electron gas.
This leads to
\begin{equation}
	\hd_{\sigma,\G} (\kt) \equiv \hd_\sigma (\k) 
\end{equation}
so that the kinetic part becomes
\begin{equation}
	H^0_{\text{sub}}=\sum_{\kt,\G,\sigma}\,\left(\frac{\hbar^2 (\kt+\G)^2}{2m^*}+E_{\text{CBM}}\right)\,\hdd_{\sigma,
	\G}(\kt)\,\hd_{\sigma,\G}(\kt)\;.
\end{equation}
Similarly, we also write momentum transfer $\q$ in terms of $\qt + \Q$ so that Eq.~(5d) becomes
\begin{equation}
	H^{\text{intra}}_{\text{sub}}=\frac{1}{2 S}\sum_{\kt,\kt',\qt}\,\sum_{\substack{\sigma,\sigma' \\ \G,\G',\Q}}\,V_\text{int}(\qt+\Q)\,\hdd_{\sigma,\G+\Q}(\kt+\qt)\,\hdd_{\sigma',\G'-\Q}(\kt'-\qt)\,\hd_{\sigma',\G'}(\kt')\,\hd_{\sigma,\G}(\kt)\;.
\end{equation}
where we always use a dielectric constant $\epsilon_r=4$ in this section.

Then, we treat the interacting part Eq.~(5b) with standard HF approximations, whose formalism exactly parallel with what we have done for graphene except simpler with only spin index. By momentum and spin conservation,
where 
\begin{align}
	\left\langle\hdd_{\sigma,\G+\Q}(\kt+\qt)\,\hd_{\sigma',\G'}(\kt') \right\rangle_d=\left\langle\hdd_{\sigma,\G+\Q}(\kt')\,\hd_{\sigma,\G'}(\kt') \right\rangle_d\,\delta_{\kt+\qt,\kt'}\delta_{\sigma,\sigma'}
	\label{eq:dm}
\end{align}
So, the Hartree term reads
\begin{equation}
	V^{\text{H}}_{\text{sub}}=\frac{1}{S}\sum_{\kt,\kt'}\,\sum_{\substack{\sigma,\sigma' \\ \G,\G',\Q}}\,V_\text{int}(\Q)\,\left\langle\hdd_{\sigma',\G'-\Q}(\kt') \hd_{\sigma',\G'}(\kt') \right\rangle_d\,\hdd_{\sigma',\G+\Q}(\kt)\,\hd_{\sigma,\G}(\kt)\,\;.
\end{equation}
and the Fock term reads
\begin{align}
	V^{\text{F}}_{\text{sub}}&=-\frac{1}{S}\sum_{\kt,\kt'}\,\sum_{\substack{\sigma,\sigma' \\ \G,\G',\Q}}\,V_\text{int}(\kt'-\kt+\Q)\,\delta_{\sigma,\sigma'}\,\left\langle\hdd_{\sigma,\G+\Q}(\kt') \hd_{\sigma,\G'}(\kt') \right\rangle_d\,\hdd_{\sigma,\G'-\Q}(\kt)\,\hd_{\sigma,\G}(\kt)\,\nn
	&=-\frac{1}{S}\sum_{\kt,\kt'}\,\sum_{\substack{\sigma,\sigma' \\ \G,\G',\Q}}\,V_\text{int}(\kt'-\kt+\Q+\G-\G')\,\delta_{\sigma,\sigma'}\,\left\langle\hdd_{\sigma,\G'-\Q}(\kt') \hd_{\sigma,\G'}(\kt') \right\rangle_d\,\hdd_{\sigma,\G+\Q}(\kt)\,\hd_{\sigma,\G}(\kt)\,
\end{align}
where we recenter $\Q$ to get the last line. Here, $\langle \dots \rangle_d$ means the expectation value of observable after integrating only the substrate part of the many-body ground state $\ket{\Psi}^{0}_d$. Spin polarization is allowed in the Hartree-Fock treatment of the 2D electron gas in the substrate, and indeed a spin polarized Wigner crystal state always has a lower energy than a spin degenerate one.

The HF calculations have been done for a series of different superlattice constant $L_s$. 
Once $L_s$ is given, the charge density in substrate is then fixed by $n_d=g_v/(\sqrt{3}L_s^2/2$) for a spin polarized Wigner crystal state with triangular superlattice, where $g_v=2$ comes from the valley degeneracy of the conduction band minimum of CrOCl. Here we set up our parameters including the effective mass $m^*=1.3 m_0$, the background dielectric constant $\epsilon_r=4$, and valley degeneracy $g_v=2$, to mimic the properties of conduction band minimum of CrOCl.
Such mapping amounts to always let only the lowest  subband in the folded Brillouin zone be filled. During the iterations, we keep track all the subbands. To initialize the HF self-consistent loop, the terms like $\left\langle\hdd_{\sigma,\G'-\Q}(\kt') \hd_{\sigma,\G'}(\kt') \right\rangle_d $ are set to be non-zero, corresponding to a spontaneous charge order with Fourier component $\Q$, where $\Q$ is one of the primitive reciprocal vectors for the triangular superlattice. 
In other words, we start the HF loop from a charge-ordered state, which facilitates the convergence to the same phase. If the final converged state is gapped, we can extract its charge modulation as input for the next step. In our calculations, a  $9\times 9$ mesh of reciprocal lattice  (centered at $\Gamma$ point) has been adopted, with the mini Brillouin zone sampled by a $18\times 18$ $\k$ mesh. The former corresponds to the real-space mesh within a primitive cell, while the latter corresponds to the system size. We have also performed calculations adopting a $13\times 13$ mesh of reciprocal lattice points (corresponding to finer real-space mesh), and find completely consistent results with those calculated using a $9\times 9$ reciprocal lattice points.

With the HF results on the substrate side in hands, we are ready to study the rest of three Hamiltonians, namely Eqs.~(5a), (5c) and (5e) of the main text. The mean-field treatment is exactly the same as in Sec.~\ref{sec:RG} and Sec.~\ref{sec:HF_gr_crocl}. However, the interlayer coupling Hamiltonian Eq.~(5e) worth special attention. Under the separable wavefunction ansatz, we can still integrate out all the $d$-operators using the wavefunctions resulted from the previous HF calculations solely on the substrate side:
\begin{align}
	\langle H_{\text{gr-sub}}\rangle_{d} =&\frac{1}{S}\sum_{\mu,\alpha,\sigma,\sigma'}\sum_{\k,\k',\q}\,\frac{e^2\,e^{-\vert\q\vert\,d}}{2\epsilon_0\epsilon_r\vert\q\vert}\,\hcd_{\sigma\mu\alpha}(\k)\,\left\langle\hdd_{\sigma'}(\k')\,\hd_{\sigma'}(\k'-\q)\right\rangle_d\,\hc_{\sigma\mu\alpha}(\k+\q)\;\nn
	=&\frac{1}{S}\sum_{\mu,\alpha,\sigma,\sigma'}\sum_{\kt,\kt',\qt}\sum_{\G,\G',\Q}\,\frac{e^2\,e^{-\vert\qt+\Q\vert\,d}}{2\epsilon_0\epsilon_r\vert\qt+\Q\vert}\,\hcd_{\sigma\mu\alpha}(\kt+\G)\,\left\langle\hdd_{\sigma',\G'}(\kt')\,\hd_{\sigma',\G'-\Q}(\kt'\qt)\right\rangle_d\,\hc_{\sigma\mu\alpha}(\kt+\G+\qt+\Q)\;\nn
	=& \frac{1}{S}\sum_{\mu,\alpha,\sigma,\sigma'}\sum_{\kt,\kt'}\sum_{\G,\G',\Q}\,\frac{e^2\,e^{-\vert\Q\vert\,d}}{2\epsilon_0\epsilon_r\vert\Q\vert}\,\hcd_{\sigma\mu\alpha,\G}(\kt)\,\left\langle\hdd_{\sigma',\G'}(\kt')\,\hd_{\sigma',\G'-\Q}(\kt')\right\rangle_{d}\,\hc_{\sigma\mu\alpha,\G+\Q}(\kt)\;\nn
	=&\sum_{\sigma,\mu,\alpha} \, \sum_{\kt,\G,\Q}\,\frac{e^2\,e^{-\vert\Q\vert\,d}}{2\epsilon_0\epsilon_r\,\Omega_d\,\vert\Q\vert}\,\rho_d(\Q)\,\hcd_{\sigma\mu\alpha,\G}(\kt)\,\hc_{\sigma\mu\alpha,\G+\Q}(\kt)\;.
	\label{eq:HG-S_d_final}
\end{align}
Here, we define the charge modulation $\rho_d(\Q)$ of the charge-ordered state as
\begin{align}
	\rho_d(\Q)&=\frac{1}{N_d} \sum_{\kt',\G',\sigma'} \left\langle\hdd_{\sigma',\G'}(\kt')\,\hd_{\sigma',\G'-\Q}(\kt')\right\rangle_{d}\,\nn
	&=\frac{1}{N_d}\sum_{\kt',\G',\sigma'}\sum_n\,D_{\sigma\G'+\Q,n}^{*}(\kt')\,D_{\sigma\G',n}(\kt')\,\theta(E_F-E_{n\kt'}^{d})\;,
\end{align}
where we write the expectation values in the substrate's subband basis in the presence of HF potentials 
\begin{equation}
	\hd_{\sigma,\G}(\kt)=\sum_{n}\,D_{\sigma\G,n}(\kt)\,\hd_{\sigma,n}(\kt) \;.
\end{equation}
$\{D_{\sigma \G,n}(\kt)\}$ relates precisely the $d$-operators in the original plane-wave basis to that in the Bloch-function basis, and $E_{n\kt'}^{d}$ denotes the subband energy dispersion. 
Remarkably, the two conceptually different treatments give rise to an interlayer Coulomb potential with the same analytical properties. It justifies our previous drastic assumptions on the localization of charge distribution. Here we only do better since we can self-consistently solve the charge-ordered ground state of the substrate and its charge density distributions. 

Exactly parallel with what we have done earlier, we now solve the problem of graphene on the superlattice defined $\rho_d(\Q)$ while only keeping the dominant terms with small $\Q$. The $e$-$e$ interactions are still treated in the band basis by HF approximations in a low-energy window given by the band index cut-off. We keep track only three valence and three conduction bands per valley per spin. The contributions from high-energy bands are taken into account using the RG approach, replacing the Fermi velocity by a renormalized one. 
In the continuum model treatment, a  $9\times 9$ mesh of reciprocal lattice points has been adopted, and the mini Brillouin zone is sampled in a $18 \times 18$  $\k$ mesh. The filling of graphene is set to be at the CNP so that a sublattice gap would be opened in the presence of superlattice potential. Once found the final converged ground state using all the types of initial order parameters, we calculate the charge modulation of gapped graphene $\rho_c (\Q)$
\begin{align}
	\rho_c(\Q)=\frac{1}{N_c} \, \sum_{\sigma,\mu,\alpha}\,\sum_{\kt,\G} \left\langle\hcd_{\sigma\mu\alpha,\G}(\kt)\,\hc_{\sigma\mu\alpha,\G-\Q}(\kt)\right\rangle_{c}\,
\end{align}
where $\langle \dots \rangle_c$ is the expectation value of integrating only the graphene part of the many-body ground state $\ket{\Psi}^{0}_c$. 

So far, we have separately found the HF ground state for substrate only and that for graphene coupled to the superlattice potential arising from the charge-ordered state in substrate. We note $\langle \dots \rangle_{\text{HF}}$ as the expectation value of observable using the HF ground state, i.e., charge-ordered state for substrate and gapped state for graphene. As a reference, we also note $\langle \dots \rangle_{\text{FL}}$ as the expectation value of observable using the non-interacting plane-wave state, i.e., 2D electron gas for substrate and non-interacting Dirac fermions for graphene. If the charge-ordered state cooperates with the gapped state in graphene, the coupled bilayer system must be energetically more stable than the substrate alone. Explicitly, this means that the condensation energy $E_\text{cond, sub} > E_\text{cond, coupled}$,  where 
\begin{align}
	E_\text{cond, sub} &= \langle H^0_{\text{sub}} + H^{\text{intra}}_{\text{sub}} \rangle_{\text{HF}} - \langle H^0_{\text{sub}} + H^{\text{intra}}_{\text{sub}} \rangle_{\text{FL}} \\
	E_\text{cond, coupled} &= \langle H^0_{\text{sub}} + H^{\text{intra}}_{\text{sub}} + H^0_{\text{gr}} + H^{\text{intra}}_{\text{gr}} \rangle_{\text{HF}} - \langle H^0_{\text{sub}} + H^{\text{intra}}_{\text{sub}} + H^0_{\text{gr}} + H^{\text{intra}}_{\text{gr}} \rangle_{\text{FL}} + E_\text{gr-sub,opt} \;.
	\label{eq:energy}
\end{align} 
$E_\text{gr-sub,opt}$ is the interlayer Coulomb energy resulted from non-uniform charge modulations of the two layers, after optimization with respect to relative charge-center shift. It will be treated  using perturbation theory shortly.

Besides, there are two subtleties in this comparison. First one is related to the cut-off. When we study graphene in the presence of superlattice potential, only the bands in low-energy window are considered. In principle, we need to add the energy contribution of all the electrons in the occupied bands. However, this would not make a significant difference since electrons deep in the valence bands are little affected by all the interaction effects discussed here. Particularly, the relevant interaction energy scale considered in this work is always far below our choice of low-energy cutoff for graphene ($\sim 3\times\hbar v_F 2\pi/L_s$). 
On the other hand, the contribution from the remote energy bands outside the low-energy window has already been included in our RG approach. Our second concern comes from the feedback effect from graphene to substrate. In principle, the non-uniform charge modulation of graphene would also affect the profile of the charge density distribution of the electronic-crystal state in the substrate by minimizing the interlayer Coulomb interaction. However, such interlayer Coulomb potential is weak compared to the intralayer counterpart for the substrate, as shown in Fig.~5 in the main text. This allows us to treat the feedback effect perturbatively, i.e., include $\langle H_\text{gr-sub} \rangle_c$ as a perturbation to $H^0_\text{sub}+H^\text{intra}_\text{sub}$. 

The first order effect is precisely the charge interlocking between graphene and substrate. There exists an optimal relative shift vector $\bm{t}_m$ between the two layers that minimize the interlayer Coulomb energy cost
 \begin{align}
 	E_\text{gr-sub}^{(1)} = N_c \sum_{\Q\neq\mathbf{0}}\,\frac{e^2\,e^{-\vert\Q\vert\,d}}{2\epsilon_0\epsilon_r\,\Omega_d\,\vert\Q\vert}\,e^{-i \Q \cdot \bm{t}_m}\,\rho_d(\Q)\,\rho_c(-\Q)\;.
	\label{eq:EG-S_1} 
\end{align}  

However, to the second order, one has to consider how the HF wavefunction of the electronic-crystal state in substrate is changed by the charge modulation of graphene. In the HF subband basis, 
\begin{align}
	\langle H_{\text{gr-sub}}\rangle_{c}
	=&\sum_{\Q\neq\mathbf{0}}\sum_{\kt,\G, \sigma}\,\frac{e^2\,e^{-\vert\Q\vert\,d}}{2\epsilon_0\epsilon_r\,\Omega_d\,\vert\Q\vert}\,\rho_c(-\Q)\,\hdd_{\sigma,\G}(\kt)\,\hc_{\sigma,\G-\Q}(\kt)\,\nn
	=&\sum_{\Q\neq\mathbf{0}}\sum_{\kt,\G, \sigma}\,\frac{e^2\,e^{-\vert\Q\vert\,d}}{2\epsilon_0\epsilon_r\,\Omega_d\,\vert\Q\vert}\,\rho_c(-\Q)\,\sum_{n,m}\, D_{\sigma \G'+\Q,m}^{*}(\kt)\,D_{\sigma\G',n}(\kt)\hdd_{\sigma,m}(\kt)\,\hd_{\sigma,n}(\kt)\,.
	\label{eq:HG-S_c}
\end{align}
The energy shift due to such modification is 
\begin{align}
	E_\text{gr-sub}^{(2)}=\sum_{\kt,\sigma}\,\sum_{\substack{n\in\text{occ.} \\ m \in \text{emp.}}}\, \frac{\left\vert \sum_{\G',\Q}\frac{e^2\,e^{-\vert\Q\vert\,d}}{2\epsilon_0\epsilon_r\,\Omega_d\,\vert\Q\vert}\,e^{-i\Q\cdot\mathbf{t}_m}\rho_c(-\Q)\, D_{\sigma \G'+\Q,m}^{*}(\kt)\,D_{\sigma\G',n}(\kt) \right\vert^2}{E_{n\kt}^{d}-E_{m\kt}^{d}} < 0\,,
	\label{eq:EG-S_2}
\end{align}
where $\mathbf{Q}\neq\mathbf{0}$ in the above equation.
Note that $E_\text{gr-sub}^{(2)}$ is also sensitive to the relative shift between the two layers so one need to find the optimal $\bm{t}_\text{opt}$ such that it minimizes the sum of $E_\text{gr-sub}^{(1)}$ and $E_\text{gr-sub}^{(2)}$, namely $E_\text{gr-sub,opt}$ in Eq.~(\ref{eq:energy}). This is the value we need to use in the comparison between two condensation energy $E_\text{cond, sub}$ and $ E_\text{cond, coupled}$. We also note that all the  kinetic energies and intralayer Coulomb interaction energies are unchanged under the relative shift between the two layers.
The results will answer the question on how synergistic correlated states occur in such coupled bilayer heterostructure.

\section{S7. Details of DFT calculations for the substrate materials}
\setlabel{S7}{sec:DFT}
\subsection{Lattice structures, deformation potentials, and band structures of candidate substrate materials}
In this section and the following one, we present the details for the density function theory (DFT) calculations of the 14 candidate substrate materials presented in Table~\uppercase\expandafter{\romannumeral 1} of the main text. In this subsection, we show the results for all but CrOCl and transition dichalcogenides. These materials, which worth special attention, are given in the next subsections. The first principles calculations are performed with the projector augmented-wave method within the density functional theory~\cite{PAW}, as implemented in the Vienna ab initio simulation package software~\cite{VASP}. The crystal structure is fully optimized until the energy difference between two successive steps is smaller than 10$^{-6}$\,eV and the Hellmann-Feynman force on each atom is less than 0.01\,eV/\AA. The generalized gradient approximation by Perdew, Burke, and Ernzerhof is taken as the exchange–correlation potential~\cite{PBE}. As Cr is a transition metal element with localized $3d$ orbitals, we use the on-site Hubbard parameter $U\!=\!5.48\,$eV for the Cr $3d$ orbitals in the CrOCl bilayer and $U\!=\!3\,$eV for Cr $3d$ orbitals in the CrI$_3$ bilayer. The so-called fully localized limit of the spin-polarized GGA+U functional is adopted as suggested by Liechtenstein and coworkers~\cite{UTYPE}, and the non-spherical contributions from the gradient corrections are taken into consideration. The “DFT+D2” type of vdW correction has been adopted for all multilayer calculations to properly describe the interlayer interactions~\cite{VDW-D2}. 

Our high-throughput filtering of the proper insulating substrate materials for graphene starts from the 2D materials computational database~\cite{Haastrup-C2DB-2dmater-2018}. We only focus at those with bulk van der Waals structures which have been previously synthesized in laboratory. This ensures that it is experimentally feasible to exfoliate few layers from their bulk sample and then stack them on graphene to form heterostructures. The results are summarized in Table~\uppercase\expandafter{\romannumeral 1} of the main text.

\begin{figure*}[!htbp]
	\includegraphics[width=0.8\textwidth]{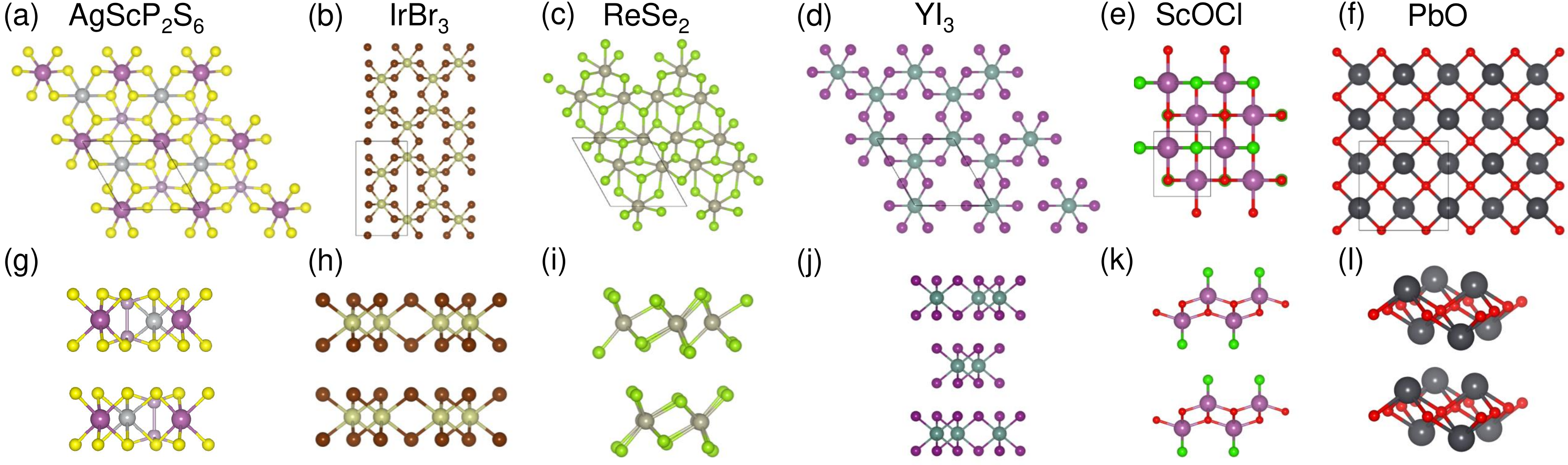}
	\caption{~\label{type1} (a)-(f): top views of the lattice structures of some candidate substrate materials in monolayer form. The primitive cells are remarked with black lines. (g)-(l): the side views of these substrate materials in few-layer form.}
\end{figure*}

\begin{figure*}[!htbp]
	\includegraphics[width=1.0\textwidth]{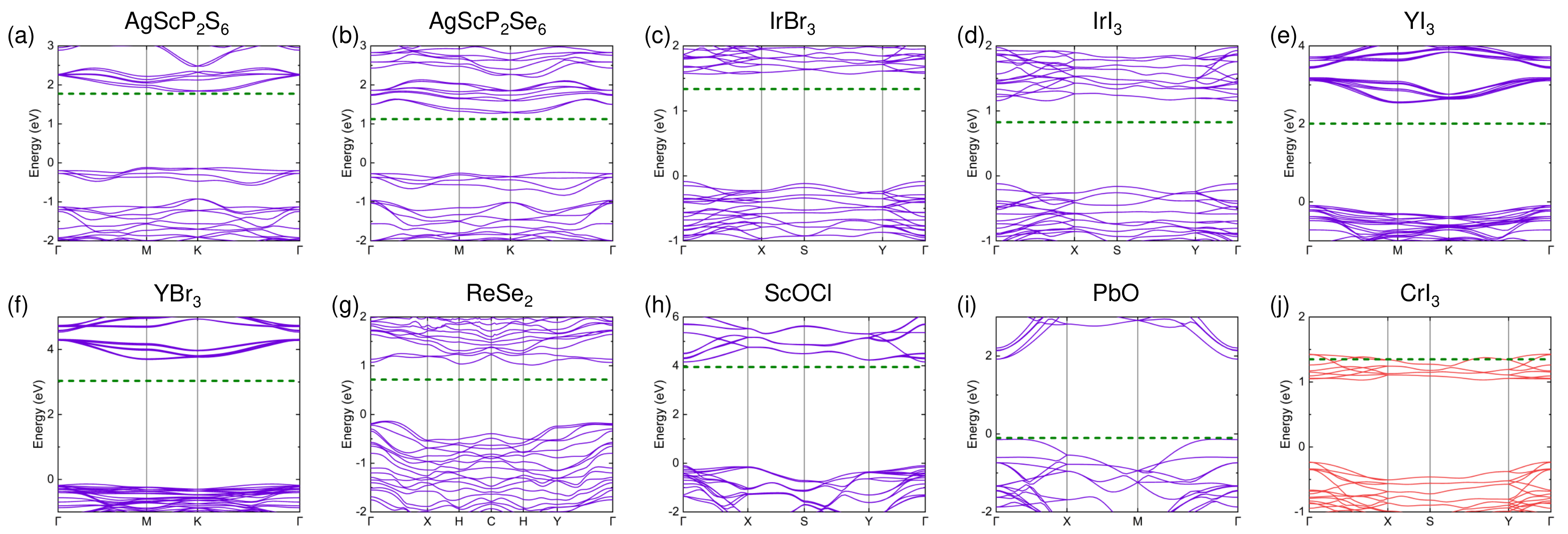}
	\caption{~\label{type2} The calculated energy bands of the candidate substrate materials, where the energy position of the Dirac point of graphene is marked by a dashed green line in the band structures. We single out CrI$_3$ by plotting its band structure using red solid lines. This is because CrI$_3$ is the only magnetic system among these ten materials so that it worth special attention on the magnetic nature of its ground state. Here, we suppose CrI$_3$ to have an intralayer ferromagnetic and interlayer antiferromagnetic ground state.}
\end{figure*}

The lattice structures of some of the substrate materials presented in Table~\uppercase\expandafter{\romannumeral 1} of the main text are shown in Fig.~\ref{type1}. They are either in the bilayer or trilayer structures. The lattice structure of CrI$_3$ is similar to that of YI$_3$ as shown in Fig.~\ref{type1}(d). 
Their band structures are presented in  Fig.~\ref{type2}, where the green dashed lines mark the energy position of the Dirac point in graphene. We note that the valence band maximum (VBM) of PbO bilayer is energetically close to the Dirac point of graphene; while for the other bilayer or trilayer substrate materials, their conduction band minima (CBM) are close to the Dirac point.
This indicates that charge transfer can easily occur between graphene and the substrates controlled by gate voltages. Moreover, we note that the conduction bands and valence bands of these materials are typically flat with large effective masses, which would be very susceptible to $e$-$e$ Coulomb interactions once these substrate materials are slightly charge doped, and may lead to Wigner-crystal-like state or long-wavelength ordered state as discussed in main text. Another important precondition for the Wigner-crystal state is that the screening effect of substrate materials can not be too strong. For example, the conduction band of ScOBr bilayer has a  large effective mass of 2.575$m_0$ ($m_0$ is the bare mass of a free electron), but the dielectric constant $\epsilon_r$ of ScOBr reaches $\sim$13, which makes it difficult to trigger the Wigner-crystal-like instability in this material under slight charge doping.

We note that all of these proposed substrate materials  all have been successfully synthesized  in laboratory as listed in Table.~\ref{host2}. Especially, few-layer of ReSe$_2$ as a highly anisotropic material \cite{Jariwala-ReSe2-CM-2016,Yang-ReSe2-NL-2015,Arora-ReSe2-NL-2017}, and few-layer CrI$_3$ system as a 2D magnetic material \cite{Huang-CrI3-nature-2017,Huang-CrI3-NN-2018,Jiang-CrI3-NN-2018,Klein-CrI3-science-2018}, have been extensively studied recently. Moreover, phonon spectra calculations have proved the dynamical stability of these substrate materials in monolayer from \cite{Haastrup-C2DB-2dmater-2018}. Thus the device fabrication of heterostructure consisting of graphene monolayer and one of these candidate substrate materials should be experimentally accessible.

There always exists  tension or compression in a heterostructure system. Under some lattice deformation, the variation of conduction band minimum (CBM) or valence band maximum (VBM) is  defined as deformation potential. We list the deformation potentials of the candidate substrate materials  in Table.~\ref{host2}. We note that the maximum value of the deformation potential is only 5.84\,eV for ScOCl, which means that the energy level of CBM of ScOCl  would move down by only 0.063\,eV under 1\% tensile strain. Therefore, even if strain is introduced in the graphene-insulator heterostructure proposed in this work, the band edges (with large effective masses) of those candidate substrate materials are still energetically close to the Dirac point of graphene.

In these candidate materials (except for CrOCl), CrI$_3$ bilayer is the only magnetic system. Previous theoretical studies reveal that the stacking configuration of CrI$_3$ bilayer plays an important role in the magnetic ground state \cite{Sivadas-bilayerCrI3-NL-2018}. Here we use the AB$^{\prime}$-type stacking in the bilayer structure, which is  consistent with the stacking configuration in the bulk phase of CrI$_3$. The AB$^{\prime}$-stacked CrI$_3$ bilayer is in an intralayer ferromagnetic and interlayer antiferromagnetic ground state.

\begin{figure*}[!htbp]
	\includegraphics[width=1.0\textwidth]{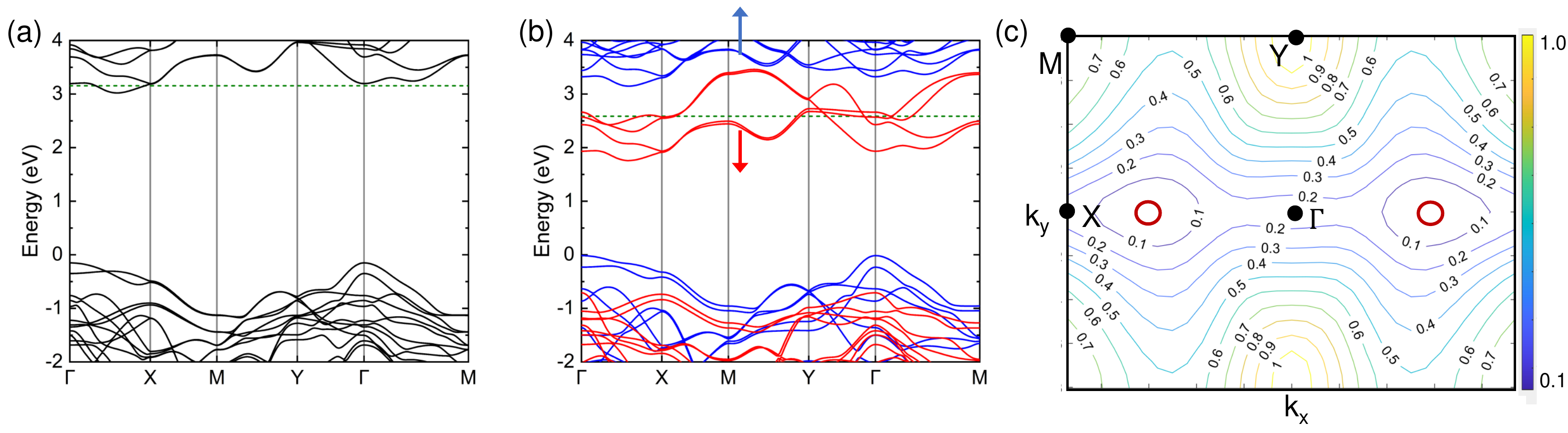}
	\caption{~\label{type3} The calculated energy bands of antiferromagnetic bilayer CrOCl: (a) without electric field, and (b) with an electric field 0.3\,V/nm. In (b), the energy bands from top and bottom layers are marked by red and blue lines, respectively. The energy position of the Dirac point of graphene are remarked with green dashed lines. (c) Fermi surface of bilayer CrOCl at different Fermi levels with respect to the conduction band minimum. The Fermi surface under 1/100 electron filling factor is remarked by red circles.}
\end{figure*}

\subsection{Band structure of graphene-CrOCl heterostructure}
We study graphene-CrOCl heterostructure using DFT calculations to show the small orbital overlap between graphene's carbon atom and CrOCl's Cr atom. Technically, we need to construct a commensurate supercell matching graphene and CrOCl's primitive unit-cell.

In our modeling, we use an $8\times \sqrt{7}$ supercell of graphene. We refer $\mathbf{a}_1$ and $\mathbf{a}_2$ to the lattice vectors along the short and long axis, respectively. Then, the corresponding supercell for CrOCl is a $(2 \mathbf{a}_1,  \mathbf{a}_1 + 5 \mathbf{a}_2)$ one. The average mismatch is $\sim 0.7\%$. In particular, there are $0.4\%$ tensile stress along long axis and $2.0\%$ compressive strain along short axis in the graphene supercell. 

\begin{figure}
	\centering
	\includegraphics[width=0.4\textwidth]{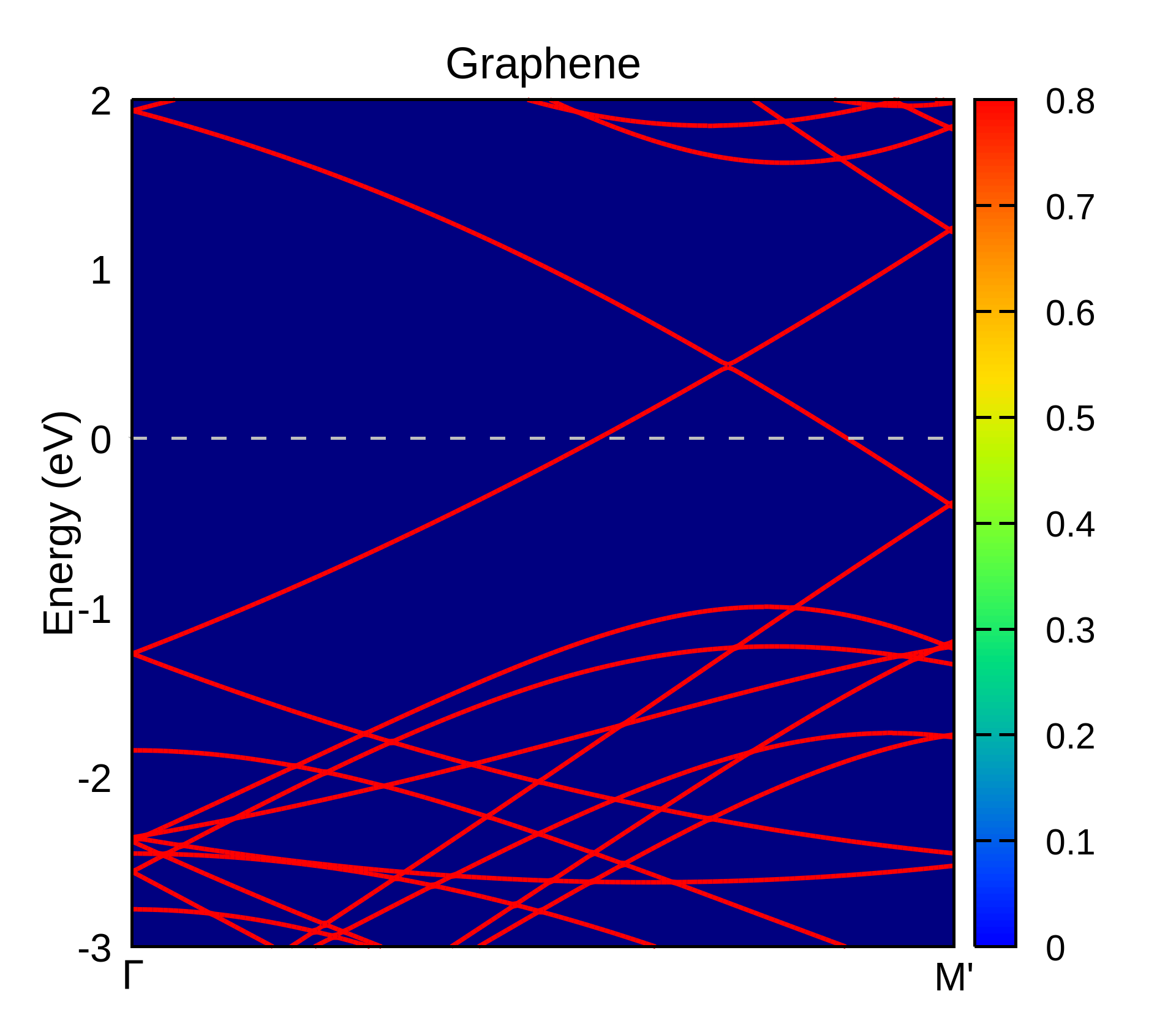}
	\includegraphics[width=0.4\textwidth]{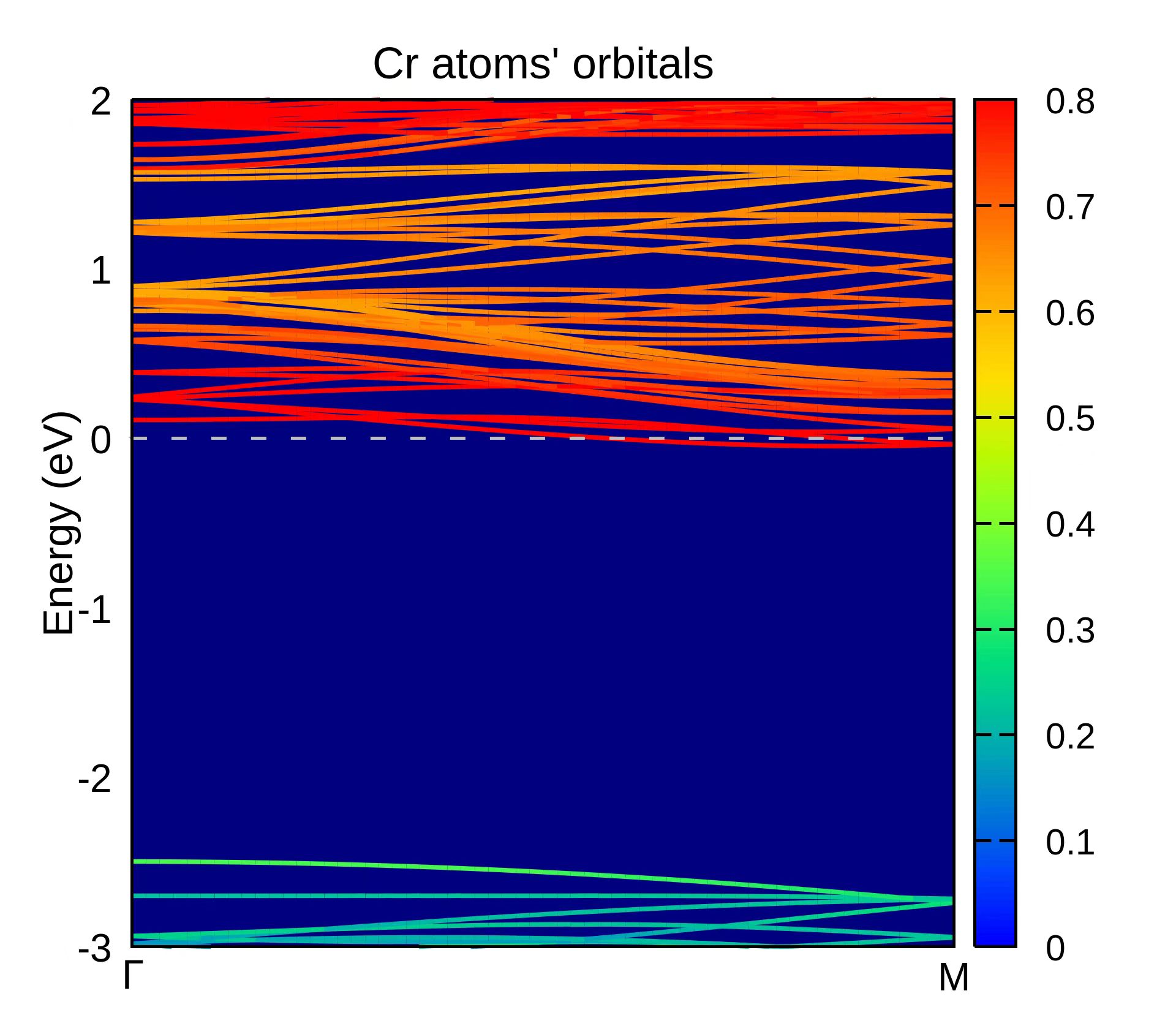}
	\caption{ Graphene-CrOCl heterostructure’s non-interacting bands, in which we project the orbital contribution on the two sides of the heterostructure: graphene’s C atoms (left) and CrOCl’s Cr atoms (right).}
	\label{fig:gr-crocl_dft}
\end{figure}

The band structure is shown in Fig.~\ref{fig:gr-crocl_dft}. The lattice distortion amounts to adding strain to graphene and thus moves the position of Dirac cone away from high-symmetry lines. Therefore, we choose the line (from $\Gamma$ to $M'$, which is not a high-symmetry point but near $M$) that passes the Dirac point in the Brillouin zone to plot the band structure. We see that the orbital overlap is small and thus negligible. 

\subsection{Electric-field tunable band structures of bilayer CrOCl}
Now we discuss  the electronic structure of CrOCl bilayer under vertical electrical fields. Here we consider an intralayer ferromagnetic and interlayer antiferromagnetic state for the bilayer configuration, which turns out to be one of competing low-energy magnetic states, and is the magnetic ground state when the on-site Hubbard $U$ value for the Cr $3d$ orbitals is large.  
\footnote{In our DFT+$U$ calculations, the on-site Hubbard $U\!=\!$ 5.48\,eV for the Cr $3d$ orbitals is used in the calculations, and the non-spherical contributions from the gradient corrections are taken into consideration. }
The calculated band gap of CrOCl bilayer with the DFT+$U$ calculation is 3.13\,eV, which is close to that of HSE06 calculation (3.12\,eV)~\cite{Haastrup-C2DB-2dmater-2018}. 
The band structure of antiferromagnetic CrOCl bilayer is shown in Fig.~\ref{type3}(a), where the green dashed line marks the energy position of the Dirac point of graphene. Without vertical electric field, the Dirac point is slightly above the CBM of bilayer CrOCl. Applying a vertical electric field of 0.03\,V/nm would push down the CBM as shown in Fig.~\ref{type3}(b). A closer inspection reveals that the top-layer conduction state (red lines) is pushed downwards while the bottom-layer state (blue lines) is pushed upward in energy as shown in Fig.~\ref{type3}(b), such that electron carriers in the graphene layer (if there is any) would be transferred to the top layer of CrOCl substrate, forming a Wigner-crystal-like state at the surface of CrOCl substrate given that the Wigner-Seitz radius of the CBM $\sim 55.7\rm{-}74.2$ (with a relative dielectric constant $\epsilon_r=3\rm{-}4$) is above the threshold value $\sim 31$ (see Table.~\uppercase\expandafter{\romannumeral 1}\ in the main text). Thus, our conjecture is supported by detailed first principles DFT calculations.  

In Fig.~\ref{type3}(c) we also present the Fermi surfaces at different Fermi energies above the CBM of bilayer CrOCl. At very low carrier densities with small Fermi energy (CBM is set to zero), the Fermi surface consists of two nearly isotropic circles. For example, at filling factor 1/100 (corresponding to a carrier density $\sim 8\times 10^{12}\,\rm{cm}^{-2}$), the Fermi surface is marked by the red  circles. Such isotropic Fermi surface with large effective mass ($\sim 1.308 m_0$) is likely to give rise to Wigner-crystal state as discussed in the main text. As the Fermi level further increases, the Fermi surfaces become more and more anisotropic.

\begin{table}[!htbp]
 \caption{The experimental works about the ten substrate materials, and the uni-axial deformation potentials of these materials~\cite{Haastrup-C2DB-2dmater-2018}.}
 \label{unitcell}
 \centering
\begin{tabular}{ccc} 
 \hline
 Materials & References & Deformation potentials \\ 
 \hline
 AgScP$_2$S$_6$ & Ref.~\cite{AgScP2S6-Lee1988} & --\\
 AgScP$_2$Se$_6$ & Ref.~\cite{AgScP2Se6-2006} & -- \\
 IrBr$_3$ & Ref.~\cite{IrBr3-1968} & -3.76\,eV \\
 IrI$_3$ & Ref.~\cite{IrI3-1968} & -2.17\,eV \\
 YI$_3$ & Ref.~\cite{YI3-1964} & 1.47\,eV \\
 YBr$_3$ & Ref.~\cite{YBr3-1980} & 1.43\,eV \\
 ReSe$_2$ & Ref.~\cite{ReSe2-1971} & -4.45\,eV \\
 ScOCl & Ref.~\cite{ScOCl-1985} & -5.84\,eV \\
 PbO & Ref.~\cite{PbO-1989} & -4.60\,eV \\
 CrI$_3$ & Ref.~\cite{Huang-CrI3-nature-2017,Huang-CrI3-NN-2018,Jiang-CrI3-NN-2018,Klein-CrI3-science-2018} & -2.20\,eV \\
 \hline
 \label{host2}
\end{tabular}
\end{table}

\subsection{Transition metal dichalcogenides as substrates for graphene}
\begin{table}[!htbp]
 \caption{Transition-metal dichalcogenides as substrate materials for graphene. The dielectric constants $\epsilon_r$ \cite{Fritz-dielectric-PhysRevB.93.115151-2016,Fritz-dielectric-Petousis2017,dielectric-choudhary2020joint}, conduction band minimum (CBM) position ($E_{\rm{CBM}}$), the corresponding  effective  mass $m^*$ at the CBM that is energetically close to the Dirac point, and the required critical doping concentration $n_c$ to realize the Wigner crystal state (with Wigner-Seitz radii $r_s=g_v m^*/\sqrt{\pi n_c}\epsilon_r m_0 a_{\rm{B}}=31$ \cite{wc-rs-prl09}) are tabulated. For clarity, the energy level of the Dirac point in graphene is set to zero. ``mono", ``bi", ``tri" and ``quad" stand for monolayer, bilayer, trilayer and quadruple-layer configurations, respectively.}
 \label{table:TMD}
 \centering
\begin{tabular}{cccccc} 
 \hline
 Materials & $\epsilon_r$ & $E_{\rm{CBM}}$  & $m^*$ & $g_v$  & $n_c$\\ 
 \hline
MoS$_2$ (mono) & 5.69 & -0.06\,eV & $0.425\,m_0$ & 2 & 2.84$\times$10$^{11}$\,cm$^{-2}$ \\ 
MoS$_2$ (bi) & 5.69 & -0.06\,eV & $0.446\,m_0$ & 2 & 3.10$\times$10$^{11}$\,cm$^{-2}$ \\
MoS$_2$ (tri) & 5.69 & -0.06\,eV & $0.467\,m_0$ & 2 & 3.40$\times$10$^{11}$\,cm$^{-2}$ \\
MoS$_2$ (quad) & 5.69 & -0.25\,eV & $0.484\,m_0$ & 2 & 3.64$\times$10$^{11}$\,cm$^{-2}$ \\
MoSe$_2$ (mono) & 7.29 & 0.36\,eV & $0.492\,m_0$ & 2 & 2.30$\times$10$^{11}$\,cm$^{-2}$ \\ 
MoSe$_2$ (bi) & 7.29 & 0.31\,eV & $0.773\,m_0$ & 6 & 5.10 $\times$10$^{12}$\,cm$^{-2}$ \\
MoSe$_2$ (tri) & 7.29 & 0.07\,eV & $0.739\,m_0$ & 6 & 4.65 $\times$10$^{12}$\,cm$^{-2}$ \\
MoSe$_2$ (quad) & 7.29 & -0.01\,eV & $0.730\,m_0$ & 6 & 4.56 $\times$10$^{12}$\,cm$^{-2}$ \\
MoTe$_2$ (mono) & 6.75 & 0.53\,eV & $0.471\,m_0$ & 2 & 2.46$\times$10$^{11}$\,cm$^{-2}$ \\ 
MoTe$_2$ (bi) & 6.75 & 0.42\,eV & $0.749\,m_0$ & 6 & 5.58$\times$10$^{12}$\,cm$^{-2}$ \\
MoTe$_2$ (tri) & 6.75 & 0.34\,eV & $0.711\,m_0$ & 6 & 5.04$\times$10$^{12}$\,cm$^{-2}$ \\
MoTe$_2$ (quad) & 6.75 & 0.31\,eV & $0.701\,m_0$ & 6 & 4.89$\times$10$^{12}$\,cm$^{-2}$ \\
WS$_2$ (mono) & 3.63 & 0.27\,eV & $0.468\,m_0$ & 2 & 8.38$\times$10$^{11}$\,cm$^{-2}$ \\ 
WS$_2$ (bi) & 3.63 & 0.25\,eV & $0.477\,m_0$ & 2 & 8.60$\times$10$^{11}$\,cm$^{-2}$ \\
WS$_2$ (tri) & 3.63 & 0.08\,eV & $1.155\,m_0$ & 6 & 4.59$\times$10$^{13}$\,cm$^{-2}$ \\
WS$_2$ (quad) & 3.63 & 0.00\,eV & $1.146\,m_0$ & 6 & 4.53$\times$10$^{13}$\,cm$^{-2}$ \\
WSe$_2$ (mono) & 4.07 & 0.53\,eV & $0.456\,m_0$ & 2 & 6.32$\times$10$^{11}$\,cm$^{-2}$ \\ 
WSe$_2$ (bi) & 4.07 & 0.52\,eV & $0.479\,m_0$ & 2 & 6.98$\times$10$^{11}$\,cm$^{-2}$ \\
WSe$_2$ (tri) & 4.07 & 0.47\,eV & $0.539\,m_0$ & 6 & 7.95$\times$10$^{12}$\,cm$^{-2}$ \\
WSe$_2$ (quad) & 4.07 & 0.27\,eV & $0.532\,m_0$ & 6 & 7.74$\times$10$^{12}$\,cm$^{-2}$ \\
PtS$_2$ (mono) & 12.34 & -0.29\,eV & $0.418\,m_0$ & 6 & 5.19$\times$10$^{11}$\,cm$^{-2}$ \\ 
PtS$_2$ (bi) & 12.34 & -0.51\,eV & $0.601\,m_0$ & 6 & 1.08$\times$10$^{12}$\,cm$^{-2}$ \\
PtSe$_2$ (mono) & 24.70 & -0.03\,eV & $0.327\,m_0$ & 6 & 7.95$\times$10$^{10}$\,cm$^{-2}$ \\
PtSe$_2$ (bi) & 24.70 & -0.31\,eV & $0.412\,m_0$ & 6 & 1.26$\times$10$^{11}$\,cm$^{-2}$ \\
 \hline
\end{tabular}
\end{table}

\begin{figure*}[!htbp]
\includegraphics[width=1.0\textwidth]{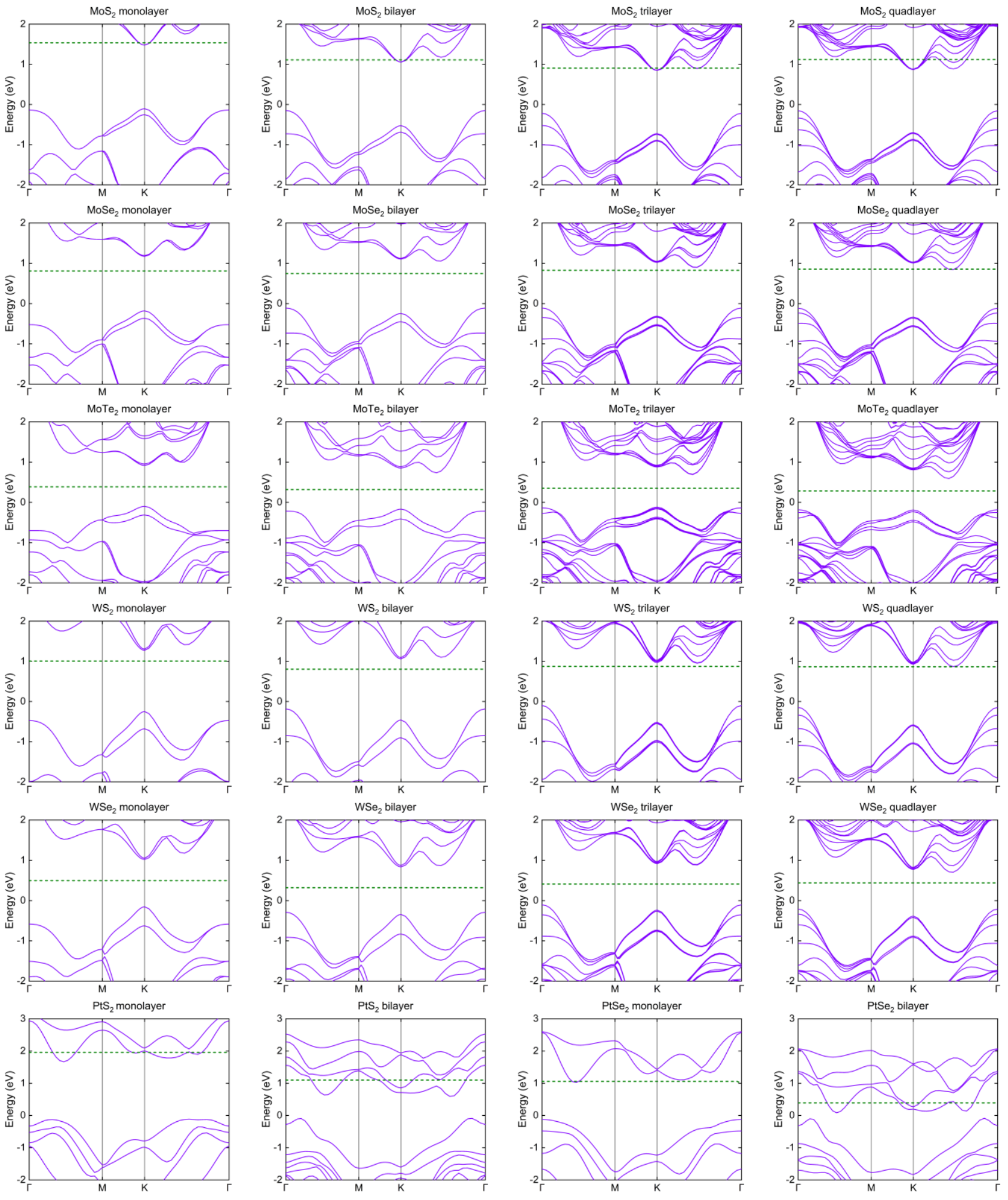}
\caption{~\label{type4} The calculated energy bands of different TMD system. Here the energy positions of Dirac point in graphene monolayer are remarked with green dashed lines. The structures of MoX$_2$ and WX$_2$ systems are trigonal prismatic 2H-phase, and those of PtX$_2$ systems are octahedral 1T-phase, which are common structural phase. The PtS$_2$ and PtSe$_2$ system has strong interlayer coupling, and the trilayer will become metallic.}
\end{figure*}

Now we focus on the few layers of transition metal dichalcogenides (TMD) which have attracted remarkable interest recently. 
Here we use the dielectric constants of the corresponding bulk TMD materials to estimate the condition for the onset of Wigner crystal states in these few-layer systems. We find that the conduction band minima  of MoX$_2$ (X = S, Se, Te), WY$_2$ and PtY$_2$ (Y = S, Se) few-layer systems are close to the Dirac point of graphene, as shown in the Table.~\ref{table:TMD}. Especially, the conduction band edges in the MoX$_2$ and WY$_2$ systems mainly originate from the transition metal Mo or W $d$ orbitals, which are localized and may have large effective masses. If the conduction bands of these TMD systems are slightly carrier doped, the systems may form Wigner crystal states as long as their Wigner-Seitz radii $r_s>31$.
We list the critical carrier concentration to realize the Wigner crystal states in various TMD few layers in the Table.~\ref{table:TMD}, and their energy bands are shown in  Fig.~\ref{type4}. The conduction band minima in the Pt-based TMD systems are  energetically close to the Dirac point of graphene, but the strong screening effect (large dielectric constants) may prevent the electrons to form Wigner crystal states in the Pt-based TMD system.
The out-of-plane electric field can easily tune the electronic structure in the MoX$_2$ and WY$_2$ systems. A previous theoretical work indicated that an electric field of 0.3\,V/nm can decrease the band gap by 0.2\,eV \cite{elect-filed-MX2-PhysRevB-2011}. Because of the excellent fabrication technology of the TMD system and their CBM close to the Dirac point, we believe that the graphene/Mo(W)-based TMD heterostructure can provide a feasible platform to realize gapped Dirac state concomitant with interaction-enhanced Fermi velocities.

\section{S8. Experimental measurements of the gaps in graphene-CrOCl heterostructure}
\setlabel{S8}{sec:exp}

To test our theory of the band reconstruction of Dirac fermions in graphene coupled with a long-wavelength charge order, we considered a few candidate substrates, among which CrOCl is suitable for device fabrication because of its high air-stability and easy-exfoliatable nature. By designing a dual-gated structure, we used few-layered CrOCl as an bottom dielectric while few-layered hexagonal boron nitride (h-BN) was served as top gate dielectric. The top and bottom gate voltages can then be converted into doping and displacement fields for further data analysis. In this section, we include below detailed experimental data on the quality of sample, the device configuration, the measurement setup as well as how we do the thermal gap measurements. 

\subsection{Quality of sample, device configuration and measurement setup}

\begin{figure}[!htbp]
	\includegraphics[width=1.0\textwidth]{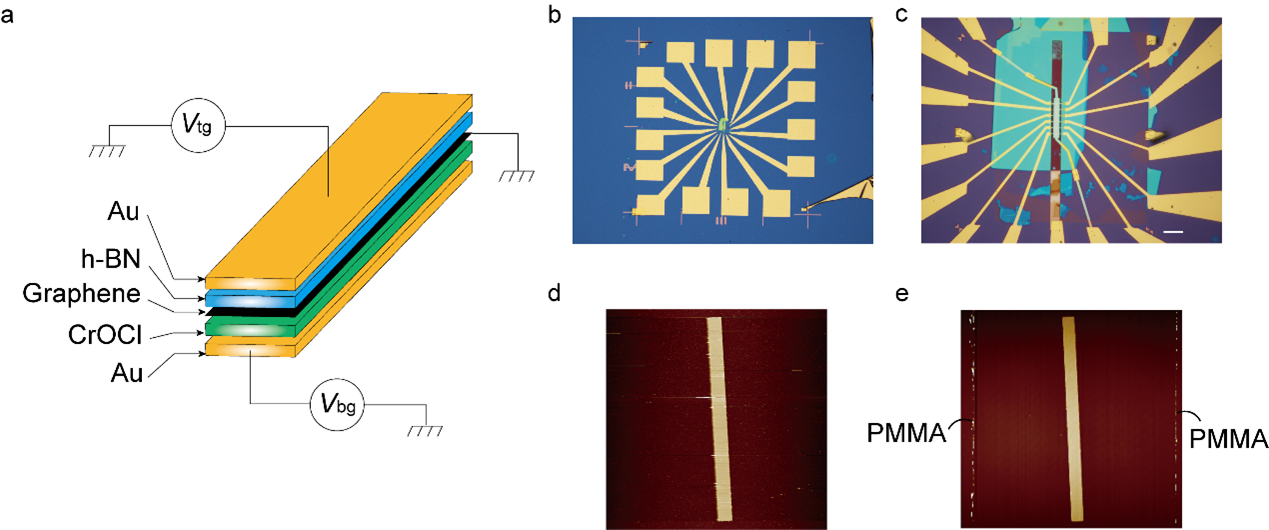}
	\caption{The characterizations of the experimental devices. (a) The cartoon illustration of a typical h-BN/Graphene/CrOCl heterostructure device, with its optical pictures shown in (b) and (c). Scale bar in the image is 10 $\mu$m. Images of AFM scans of the Au bottom gate before (d) and after (e) AFM Contact mode cleaning are also illustrated, where the cleaned window can be highlighted by the PMMA residues accumulated during the AFM ``sweeping'' cleaning. }
	\label{fig:device}
\end{figure}

The devices are made of graphene, h-BN, and CrOCl flakes, which are mechanically exfoliated from high quality bulk crystals. The vertical assembly of few-layered hBN, monolayer graphene and few-layered CrOCl were made using the polymer-assisted dry-transfer method. Electron beam lithography was done using a Zeiss Sigma 300 SEM with a Raith Elphy Quantum graphic writer. Top and bottom gates as well as contacting electrodes were fabricated with an e-beam evaporator, with typical thicknesses of Ti/Au $\sim$ 5/50\,nm. A cartoon illustration of the tested device is shown in Fig.~\ref{fig:device}(a), which includes h-BN/Graphene/CrOCl van der Waals heterostructure equipped with a top gate and a bottom gate. Fig.~\ref{fig:device}(b) and (c) show the $5\times$ magnification and $100 \times$ magnification optical pictures of the device, respectively. There are no visible bubbles or wrinkles on the heterostructure. To ensure the cleanliness of the sample, we will use an AFM (atomic force microscope) tip to clean the Au bottom gate in a Contact mode before landing heterostructures onto these local metallic gates. A significant amount of PMMA can be removed via such a process to make sure the homogeneity of the gate electrical fields (Fig.~\ref{fig:device}(d,e)\,). The device studied in this work is the same device S40 we investigated in our recent experimental paper published in \textit{Nat. Nanotechnol.} \textbf{17}, 1272–1279 (2022) \cite{wang_arxiv2021}. According to the high quality of quantum Hall plateaus (in the normal phase without charge transfer) seen at moderate magnetic field at a few or a few tens of Kelvin in these samples, we can claim that those samples, in term of cleanness, are of state-of-the-art quality.

\begin{figure}[!htbp]
	\includegraphics[width=1.0\textwidth]{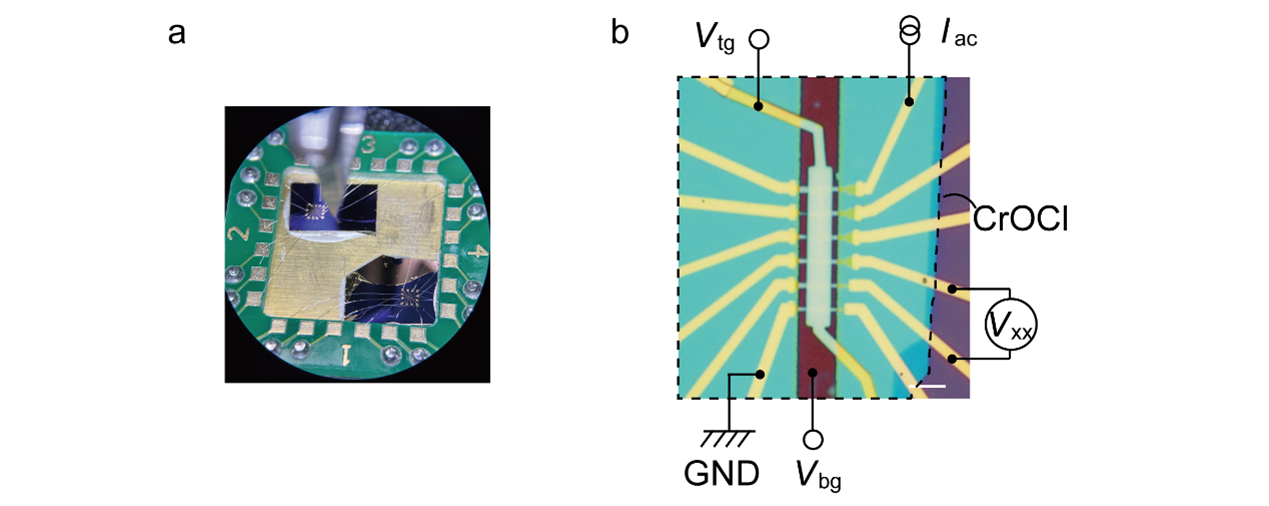}
	\caption{Measurement configurations. (a) The image of device bonded onto the holder. (b) Diagram of the 4-probe measurement method of Hall bar. The electrode labeled ``$I_{ac}$'' is applied 100nA AC signal by SR830 with a 10 M$\Omega$ bias resistor. The longitudinal voltage $V_{xx}$ is recorded by SR830, as well. The top gate and bottom gate electrodes are supplied voltage by Keithley 2400. Scale bar in the image is 5 $\mu$m.}
	\label{fig:measure_setup}
\end{figure}

For electrical transport measurements of the devices, we use a wire bonder to bond the device onto the sample holder (Fig.~\ref{fig:device}(a)). Samples are then loaded into a 1.5 K fridge (Oxford Tesla-Tron system), with a base temperature of 1.5K and a maximum magnetic field of 12 T. We adopt standard 4-probe low frequency (13.33 Hz) lock-in measurement method to perform electrical transport characterization of the devices, with the Standford SR830 lock-in amplifier in series with a 10 M$\Omega$ bias resistor, thus providing an AC signal at constant current of 100 nA (assume the sample resistance is way smaller than the bias resistor). Meanwhile, high precision voltage source (Keithley 2400) is used to regulate the top and bottom gate voltages. The longitudinal voltage $V_{xx}$ is measured by the lock-in amplifier. Pictures of the set-up are shown in Fig.~\ref{fig:measure_setup}. 

In addition, according to our experimental observations, we actually have two phases (see Fig.~4 in Ref.\cite{wang_arxiv2021}) in graphene-CrOCl heterostructure, revealed in the sample's resistance measurement in the parameter space of $V_{tg}$ and $V_{bg}$:
\begin{enumerate}
\item Phase-I is conventional graphene showing rather low Dirac peak at a few hundreds of Ohms; 
\item Phase-II is the interfacial-coupling phase, where the band structure of graphene is re-configured with a mild gap opening at the charge neutral.
\end{enumerate} 
In fact, we focus on the Phase-II since Phase-I is the trivial conventional graphene. In the latter phase, the Wigner crystal in the substrate is not activated since the charge transfer is not onset. Given that quantum Hall plateaus have been clearly seen under sub-Tesla magnetic field at tens of Kelvin, with the Shubnikov-de Haas quantum oscillations (see Supplementary Figure 36 in Ref.\cite{wang_arxiv2021}), the quality of sample is guaranteed. So, the measured resistivity peak has to do with gap opening in the Phase-II in CrOCl-contacted graphene. For even more experimental details, we would invite our referee to kindly refer to the Supplementary Information of \textit{Nat. Nanotechnol.} \textbf{17}, 1272–1279 (2022) \cite{wang_arxiv2021}.

\subsection{Measurement of the thermal gap}

\begin{figure}[htb]
    \centering
    \includegraphics[width=0.95\textwidth]{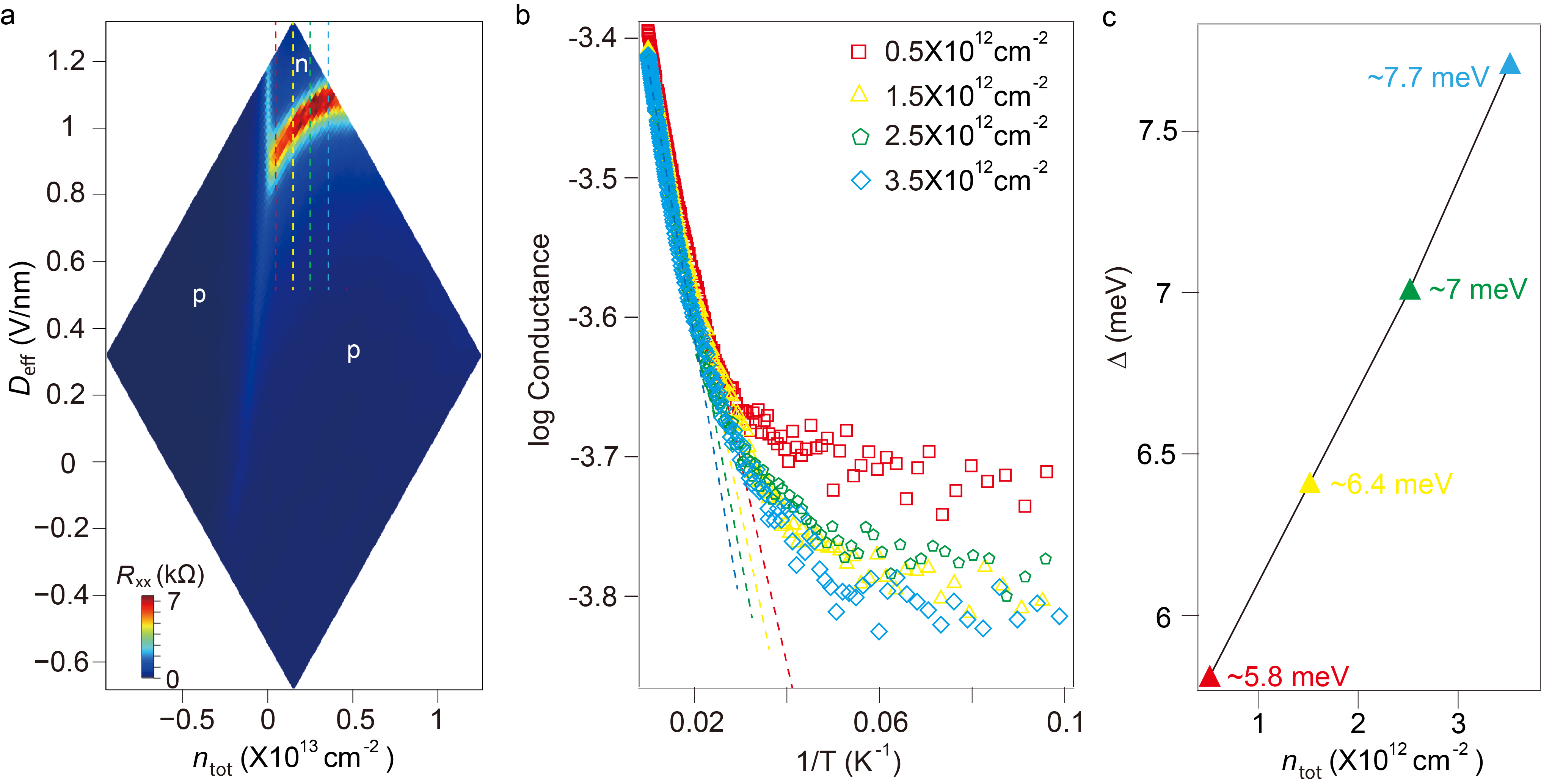}
    \caption{Gap size at the charge neutrality point (CNP) of monolayer graphene supported by CrOCl. (a) Channel resistance $R_{xx}$ measured in the space of $D_\mathrm{eff}$-$n_\mathrm{tot}$ at $T\!=\!1.5\,$K and $B\!=\!0\,$T. Here, $D_\mathrm{eff}$ is defined as $D_\mathrm{eff}=(C_\mathrm{tg}V_\mathrm{tg}-C_\mathrm{bg}V_\mathrm{bg})/2\epsilon_{0} - D_{0}$ and $n_\mathrm{tot}$ is defined as $n_\mathrm{tot}=(C_\mathrm{tg}V_\mathrm{tg}+C_\mathrm{bg}V_\mathrm{bg})/e-n_{0}$. $C_\mathrm{tg}$ and $C_\mathrm{bg}$ are the top and bottom gate capacitances per area, respectively. And $V_\mathrm{tg}$ and $V_\mathrm{bg}$ are the top and bottom gate voltages, respectively. $n_{0}$ and $D_{0}$ are residual doping and residual displacement field, respectively. A new phase driven by $e$-$e$ interaction is seen, with the CNP largely bent, and clearly much more resistive than the conventional vertical CNP at the fixed $n_{tot}=0$. This is consistent with the phase diagram reported in Ref.~\onlinecite{wang_arxiv2021}. (b) Log scale of the minimum conductance $\sigma_{xx}$ at the CNP with fixed $n_\mathrm{tot}$, i.e., $\log$(1/$R_{xx}$), plotted against $1/T$. The thermal excitation gap $\Delta$ can then be estimated from the linear part, using the formula $\sigma \sim$ $e^{-\Delta/2k_{B}T}$, where $k_{B}$ is the Boltzmann constant, $T$ is temperature. (c) The thermal activation gap extracted in each linear fit of the curve in (b) and plotted against $n_\mathrm{tot}$.} 
    \label{fig:S1}
\end{figure}

The gate dependencies of channel resistances are measured at various temperatures for the extraction of thermal gaps. In Fig.~\ref{fig:S1}(a), we present the detailed mapping of measured resistance in the space of displacement field $D_{\mathrm{eff}}$ and the nominal carrier density $n_{\mathrm{tot}}$. A resistivity peak which persists up to $n_{\mathrm{tot}}\lessapprox 5\times10^{12}\,\rm{cm}^{-2}$ is clearly seen, indicating gap opening at the charge neutrality point (CNP) of graphene. The gapped Dirac state persists up to $n_{\mathrm{tot}}\lessapprox 5\times10^{12}\,\rm{cm}^{-2}$ because the extra charge carriers are transferred from graphene to the surface of CrOCl, leaving graphene at the charge neutrality. As discussed in the main text, the charges transferred to the surface of CrOCl may form a long-wavelength ordered state through the Wigner-crystallization mechanism, which imposes a superlattice Coulomb potential to graphene and promotes the gap opening at the CNP. In order to determine the gap at the CNP of graphene at different $n_{\rm{tot}}$, we have further performed the temperature dependent conductance measurements at different $n_{\rm{tot}}$ as shown in Fig.~\ref{fig:S1}(b). The temperature dependence of the conductance $\sigma_{xx}(T)$ can be very well explained in a thermal excitation picture, with $\sigma_{xx}(T)\sim e^{-\Delta/2 k_{B} T}$ ($k_B$ is the Boltzmann constant), from which the gap at the CNP $\Delta$ can be extracted.  In Fig.~\ref{fig:S1}(c) we show the experimentally measured gaps at different nominal carrier density $n_{\rm{tot}}$. We see that the gap decreases linearly with $n_{\rm{tot}}$, consistent with theoretical calculations.

\end{document}